\RequirePackage{fix-cm}
\documentclass[twocolumn]{svjour3}          
\smartqed  
\usepackage{graphicx}
\usepackage{amsmath}
\usepackage{adjustbox}
\usepackage{soul}
\usepackage[final]{microtype}
\microtypesetup{protrusion=true, expansion=true}
\usepackage{color}
\usepackage{enumitem}
\usepackage{url}
\usepackage{graphicx}
\usepackage[compatibility=false]{caption}
\usepackage{subcaption}
\usepackage{subfig}
\PassOptionsToPackage{caption=false}{subfig}
\makeatletter
\renewcommand{\makeheadbox}{} 
\makeatother

\journalname{}  
\date{}         

\begin{document}

\title{Evaluating the Defense Potential of Machine Unlearning against Membership Inference Attacks
}


\author{Theodoros Tsiolakis \and Vasilis Perifanis \and Nikolaos Pavlidis \and Christos Chrysanthos Nikolaidis \and Aristeidis Sidiropoulos \and Pavlos S. Efraimidis
}


\institute{Theodoros Tsiolakis \at
              Democritus University of Thrace, Greece \\
              \email{ttsiolak@ee.duth.gr}           
           \and
}


\date{}

\maketitle

\begin{abstract}
Membership Inference Attacks (MIAs) pose a significant privacy risk by enabling adversaries to determine if a specific data point was part of a model's training set. This work empirically investigates whether Machine Unlearning (MU) algorithms can function as a targeted, active defense mechanism, in scenarios where a privacy audit identifies specific classes or individuals as highly susceptible to MIAs post-training. By 'dulling' the model's categorical memory of these samples, the process effectively mitigates the membership signal and reduces the MIA success rate for the most vulnerable users. We evaluate the defense potential of three MU algorithms, Negative Gradient (neg grad), SCalable Remembering and Unlearning unBound (SCRUB), and Selective Fine-tuning
and Targeted Confusion (SFTC), across four diverse datasets and three complexity-based model groups. Our findings reveal that MU can function as a countermeasure against MIAs, though its success is critically contingent on algorithm choice, model capacity, and a profound sensitivity to learning rates. While Negative Gradient often induces a generalized degradation of membership signals across both forget and retain set, SFTC identifies a critical ``divergence effect'' where targeted forgetting reinforces the membership signal of retained data. Conversely, SCRUB provides a more balanced defense with minimal collateral impact on MIA perspective.
\end{abstract}

\section{Introduction} 
\label{sec:intro}

As machine learning (ML) models continue to advance, their applications have become widespread across various industries, such as healthcare~\cite{Nikolaidis2024}, telecommunications~\cite{tsiolakis2024carbon}, biological sciences~\cite{pavlidis2023extensive} and beyond. These models have revolutionized how we approach tasks, making decision-making processes more efficient and accurate. However, the extensive use of ML raises concerns about the reliance on large volumes of user data, often containing sensitive or personal information. With the increasing number of privacy incidents where unauthorized parties gain access to this sensitive data, the need for robust privacy-preserving mechanisms has become critical. 

One of the primary privacy vulnerabilities of ML models is defending against MIAs~\cite{shokri2017membership}. In such attacks, adversaries aim to determine whether a specific data point or individual client was part of a model's training dataset~\cite{Nikolaidis2025}. For instance, if a social media model can be queried to confirm whether specific users' image were used during its training, it could expose confidential information about individuals. Therefore, understanding and mitigating the risks posed by MIAs is crucial for ensuring the privacy and safety of users. 
Beyond mere compliance, this study investigates \footnote{Code available at: \url{https://github.com/TheoTsio/MIA_vs_Unlearning.git}
} whether MU  ~\cite{Cao} can be repurposed as an active, targeted defense against such vulnerabilities. While both domains center on the residual influence of specific data points, MU traditionally allows service providers to erase data without retraining from scratch. While primarily utilized for regulatory compliance, such as fulfilling the ``right to be forgotten'' imposed by the General Data Protection Regulation (GDPR), MU has the potential to function as a latent security measure \cite{zhang2023review}. By strategically applying unlearning to ``patch'' high-risk data points, we can evaluate its efficacy in masking the traces of training membership and hardening the model against adversarial inference.

Despite its appeal, the relationship between Machine Unlearning and a model's vulnerability to MIAs remains largely unexplored, and these kind of attacks mostly used as an evaluation metric of the MU quality~\cite{tu2025}. While MU primarily focuses on the efficient removal of data to protect user privacy, it is unclear how this process influences the model's susceptibility to MIAs \cite{liu2024threats}. This knowledge gap is crucial because even after certain data has been ``unlearned'', models may still retain enough information about the deleted samples to be vulnerable to MIAs. As such, merely applying unlearning techniques may not suffice to mitigate these risks unless the process is carefully evaluated and tailored to address the specific threats posed by MIAs. 

This work aims to contribte to this research gap by assessing the impact of Machine Unlearning on a model's vulnerability against MIAs. We evaluate the efficacy of state-of-the-art Machine Unlearning algorithms across four distinct datasets, each representing different applications. Our analysis focuses on how unlearning methods affect a model's exposure to MIAs and whether certain unlearning strategies are more effective than others in reducing this risk. Our findings suggest several important insights. First, while Machine Unlearning can effectively remove targeted data, it does not inherently serve as a defense mechanism against MIAs. In some cases, the removal of data via unlearning may even introduce subtle vulnerabilities that could increase the model's susceptibility to adversarial attacks. Second, the effectiveness of unlearning techniques in reducing vulnerability to MIAs varies significantly depending on the unlearning algorithm used and the use-case scenario. Hence, designing a robust Machine Unlearning solution requires careful consideration of the the specific nature of the data being unlearned. 

The rest of this paper is structured as follows: Section~\ref{sec:problem} defines key concepts, including MIAs and Machine Unlearning, and explores the influence of unlearning algorithms on MIA vulnerabilities. Section~\ref{sec:related} presents related work, focusing on MIA, machine unlearning techniques, and the interplay between the two. In Section~\ref{sec:dataset}, we describe the datasets used in this study, covering both image and tabular data. Section~\ref{sec:experiments} details the experimental setup, discussing the models and unlearning algorithms. Section~\ref{sec:results} we present the results across different datasets, along with a sensitivity analysis. Finally, Section~\ref{sec:conclusions} concludes the paper with a summary of the findings and suggestions for future work.

\section{Key Takeaways of the paper}

\begin{itemize}
    \item[$\bullet$] Defense Focus: Explicitly frames Machine Unlearning (MU) as an active countermeasure against privacy threats that are recognized in the post-trained model.
    \item[$\bullet$] Comparative Analysis: Highlights the distinct performance profiles of SCRUB and SFTC, identifying a trade-off between the stability of the former and the high selectivity of the latter.
    \item[$\bullet$] Experimental Scope: Validates findings across a broad experimental matrix comprising four diverse datasets (image and tabular) and three distinct model complexity groups.
    \item[$\bullet$] Continuous Auditing Requirement: Establishes that evaluating MIA accuracy at every epoch is essential to capture unlearning dynamics and identify the optimal ``stopping point'' before ``over-unlearning'' degrades model utility.
    \item[$\bullet$] MIA Accuracy Trade-offs: Reveals that a successful reduction in MIA Forget Accuracy can paradoxically lead to an increase in MIA Retain Accuracy. While the model becomes less vulnerable regarding the forgotten set, intensive fine-tuning can reinforce the membership signal of the retain set, potentially compromising its privacy.
\end{itemize}

\section{Related Work} 
\label{sec:related}

\subsection{MIA}
Shokri et al.~\cite{shokri2017membership} introduced the concept of MIAs, demonstrating that adversaries could determine whether a particular data point was part of the training dataset by observing the model's outputs. Their work laid the groundwork for subsequent research in this domain.

Salem et al.~\cite{salem2018ml} extended the MIA concept to more practical scenarios, showing that even with limited knowledge of the target model, adversaries could perform effective MIA. They explored different attack models, including black-box and white-box scenarios, highlighting the pervasive risk of MIA.

A comprehensive survey by Hu et al.~\cite{hu2021membership} details the variety of MIA techniques across different ML models, including classification and generative models. Their work also outlines the defences proposed to mitigate these attacks, emphasizing the ongoing need for robust privacy-preserving mechanisms. Other studies, such as \cite{Zhong2022Understanding} explore the privacy unfairness in MIA and how certain defence mechanisms, like differential privacy and dropout, affect different subgroup data. 

Further advancements in MIA, such as the MIA-BAD framework proposed by Banerjee et al~\cite{banerjee2023miabad} leverage federated learning as a potential defence mechanism against batch-wise MIAs. Similarly, Shi et al.~\cite{shi2023scalemia} introduce Scale-MIA, a scalable attack capable of recovering training data even in secure federated learning environments, demonstrating the persistent risks to user privacy. 

\subsection{Machine Unlearning}

Cao et al.~\cite{Cao} introduced the foundational concept of machine unlearning, where they focused on the efficient removal of data from ML models. Their method transforms models like Naive Bayes into a summation-based structure, enabling the deletion of specific data points without requiring full retraining of the entire model.

Building on this foundational work, Bourtoule et al.~\cite{Bourtoule} introduced a more scalable unlearning framework called SISA in 2019. The core idea behind SISA training is to divide the training dataset into several disjoint shards and train sub-models on these separate shards.

Ginart et al.~\cite{Ginart} tackled unlearning in the context of unsupervised learning, specifically k-means clustering. Their method quantizes data points into discrete clusters, enabling efficient unlearning by allowing modifications or deletions of the clusters containing the data points in question, without needing to retrain the entire model.

While these approaches focus on unlearning efficiency, it is equally important to minimize the negative impact on model accuracy. Addressing this concern, Perifanis et al.~\cite{perifanis2024sftc} proposed SFTC, an approach that fine-tunes the model on the remaining instances while intentionally confusing the model about the forget set, thus balancing unlearning with preserving predictive performance.

Moving to the context of logit-based classifiers, Baum\-hauer et al.~\cite{Baumhauer} explored machine unlearning by leveraging the structure of logit-based models. This approach provides an efficient method for removing data points, avoiding full retraining while ensuring data removal.

To tie all these efforts together, Xu et al.~\cite{Xu} conducted a comprehensive survey of various machine unlearning techniques. Their survey covers the broad range of methods introduced so far, analyzing the trade-offs between accuracy, efficiency, and scalability, and highlighting the ongoing challenges in this evolving field.

\subsection{Machine Unlearning and MIA}

Several studies have explored various aspects of MU with different ideas of adversarial attacks and more specific to MIAs, shedding light on both its potential and its limitations.

For instance, Sommer et al.~\cite{Sommer} investigate probabilistic verification methods to defend against MIA in unlearning systems, emphasizing the need for strong verification protocols to prevent data leakage. Their work underscores the necessity of robust mechanisms to ensure the integrity of unlearning processes.

Also, Doughan and Itani~\cite{DoughanItani} provide a comparative analysis of various machine unlearning methods, focusing on their effectiveness in hindering MIAs. They emphasize the importance of future research in this area, particularly in the context of different models and datasets. In their study, they examine machine unlearning techniques applied to Resnet-18 models trained with the CIFAR-10 dataset. By comparing Fine-Tuning and Fisher Noise-based Impair-Repair methods, they demonstrate that the latter significantly reduces MIA scores while maintaining model accuracy.

However, even with such advancements, new challenges continue to emerge. Lu et al.~\cite{Lu} introduce label-only MIAs, highlighting a novel threat to machine unlearning systems. These attacks enable adversaries to infer membership without relying on posterior probabilities, thus complicating existing defense strategies.

In addition to these concerns, Chen et al.~\cite{Chen} delve deeper into the privacy risks associated with unlearning. They show that MIAs can still exploit unlearning processes, even when sensitive data is expected to be removed. The authors propose a novel MIA that targets models post-unlearning by leveraging differences in model outputs before and after the unlearning process, demonstrating how attackers can infer which data points were likely part of the original training set.

Finally, Niu et al.~\cite{NIU2024404} offer a comprehensive review of the landscape of MIAs. Surveying over 200 publications, they systematically examine various defense strategies, including machine unlearning, to mitigate these attacks. Their findings highlight the importance of careful evaluation in the unlearning process, noting that incomplete or ineffective unlearning can leave models vulnerable to residual traces of removed data.

Although the recent advances in the connection of unlearning to MIAs have expanded our understanding of defense mechanisms, most efforts focus mainly on image classification tasks, with limited attention given to other types of data, such as tabular datasets. In addition, most recent approaches tend to evaluate the effectiveness of MIAs by comparing only two snapshots of the model: one before unlearning and one after. While this evaluation provides a basic measure of improvement or degradation in MIA resistance, it neglects to offer insights into the dynamics of the unlearning process itself and how it incrementally impacts the model's vulnerability to MIAs. To address these gaps, we conduct a comprehensive experimental study using both image and tabular datasets, evaluating MIAs not only before and after unlearning but throughout the entire unlearning process.

\section{Problem Definition} 
\label{sec:problem}
\subsection{Membership Inference Attacks (MIAs)}
\begin{figure}[t!]
    \centering
    \includegraphics[width=\columnwidth]{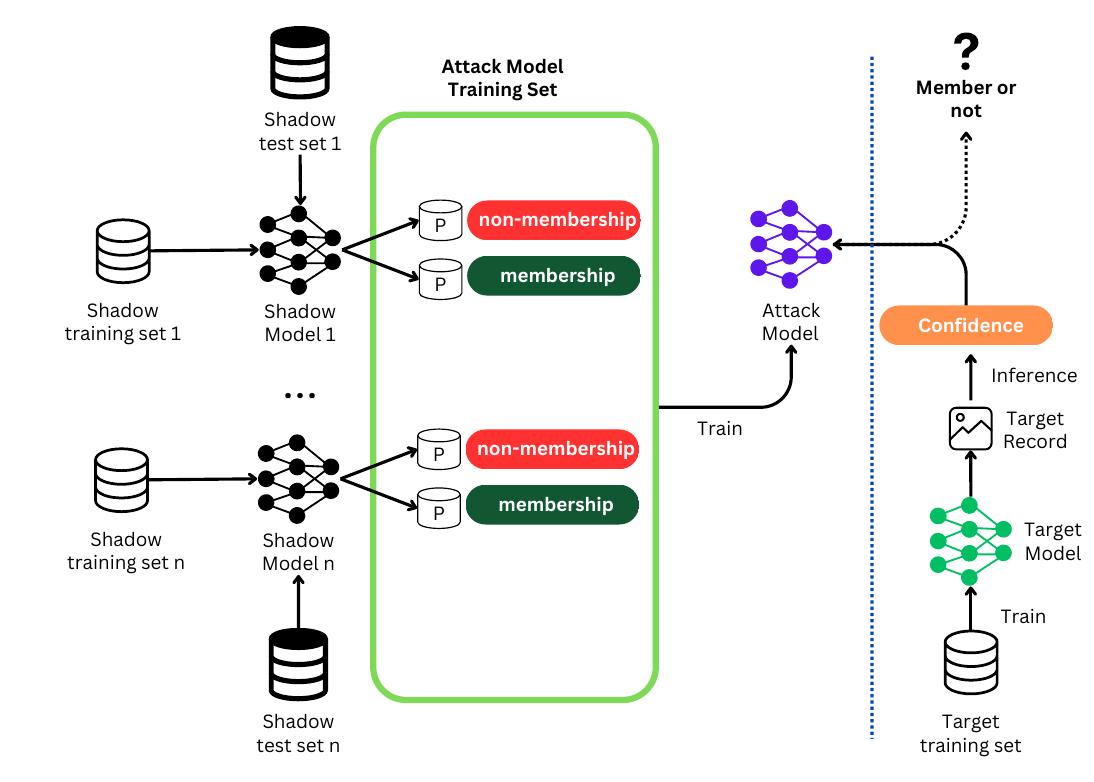}
    \caption{MIA - Flow Diagram}
    \label{fig:mia-flow}
\end{figure}

MIAs present a significant privacy challenge in ML models. Specifically, a MIA involves an adversary trying to determine if a specific data point was part of the model's training dataset \cite{hu2021membership}. Given access to the model's outputs, the adversary can exploit differences in the model's behavior when given data that was part of the training set versus data that was not (see Fig.~\ref{fig:mia-flow}). This capability enables the adversary to deduce whether specific instances are in the training set, potentially revealing sensitive information about individuals.

Formally, let $\mathcal{D}_{train}$ represent the training dataset used to train a model $f$, and $\mathcal{D}_{test}$ represent a separate test dataset. The goal of an adversary is to design a membership inference attack algorithm, $A$, that, given a data point $x$ and model $f$, can output a binary label:
\[
A(f(x)) =
\begin{cases} 
      1 & \text{if } x \in \mathcal{D}_{train} \\
      0 & \text{if } x \notin \mathcal{D}_{train}
\end{cases}
\]

The success of a MIA depends on how effectively the adversary can differentiate between members of $\mathcal{D}_{train}$ and non-members based on the model's predictions. Such attacks pose significant privacy risks as it could potentially leak sensitive information, especially in models trained on personal or confidential data, such as medical or financial records. In this work, we implement a white-box attack, where the adversary has knowledge of the target model's architecture and the distribution of the dataset \cite{niu2024survey}. This type of attack is more powerful, as it exploits deeper insights into the model, making it easier to infer membership.

\subsection{Shadow Models}

To execute MIAs, we employ shadow models to simulate the target model's behavior and establish a ground truth for membership. The adversary trains multiple proxy models on datasets with known membership, allowing them to observe the distinct output probability distributions, often characterized by higher confidence or lower entropy, associated with training data versus unseen data. These observations are used to train an attack classifier that learns to recognize the ``membership signal''. In the context of machine unlearning, shadow models are critical for auditing; they allow for a direct comparison between a model that has ``unlearned'' a specific data point and a ``retrained-from-scratch model. If an attack model can still distinguish a forgotten sample from a truly unseen one, it indicates a privacy residual, suggesting that the unlearning algorithm has failed to completely erase the influence of the deleted data.

\subsection{Machine Unlearning}
\begin{figure}[t]
    \centering
    \includegraphics[width=\columnwidth]{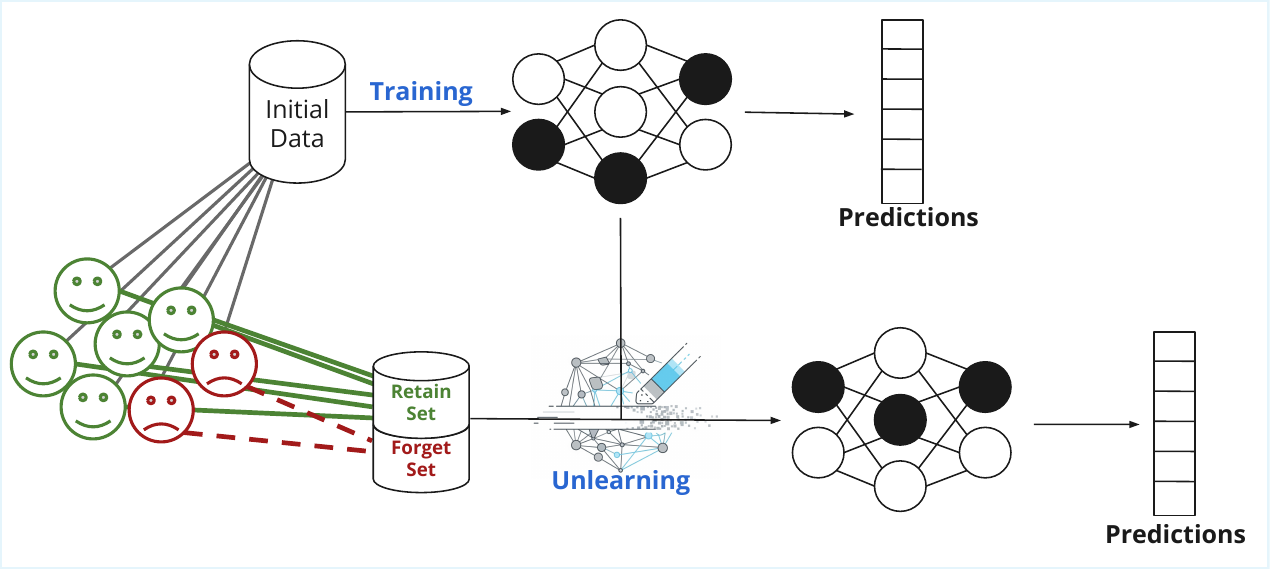}
    \caption{Machine Unlearning - Flow Diagram}
    \label{fig:unlearning-flow}
\end{figure}

The operational workflow of Machine Unlearning (as illustrated in Fig.~\ref{fig:unlearning-flow}) involves identifying a target subset of data, defined here as a specific percentage of the original training set, to be removed from a pre-trained model. Once the target data is selected, the unlearning algorithm is applied to adjust the model's parameters. A critical challenge throughout this procedure is the preservation of the model's utility; if the predictive capacity is substantially degraded following the unlearning process, the procedure loses its practical and functional significance. Therefore, an effective unlearning execution must achieve a state where specific data influence is erased without inducing a total predictive collapse.

Let $\mathcal{D}_{train} = \mathcal{D}_{retain} \cup \mathcal{D}_{forget}$, where $\mathcal{D}_{forget}$ represents the data that needs to be unlearned. Given a model $f$ trained on $\mathcal{D}_{train}$, the goal of Machine Unlearning is to produce a modified model $f'$ such that:
\[
f'(x) \approx f(x) \quad \forall x \in \mathcal{D}_{retain}
\]
and the influence of $\mathcal{D}_{forget}$ is effectively removed. In other words, for any data point $x \in \mathcal{D}_{forget}$, the model $f'$ should behave as though $x$ was never part of the original training dataset. Achieving this efficiently, without retraining the model from scratch, is the core challenge of Machine Unlearning.

\subsection{Influence of Machine Unlearning on Membership Inference Attacks (MIAs)}
While Machine Unlearning is designed to remove data points from a model's training set, its influence on the model's vulnerability to MIA is not straightforward. The process of unlearning aims to erase the impact of the data to be forgotten ($\mathcal{D}_{forget}$), but this can either reduce or exacerbate the model's susceptibility to MIAs depending on how the unlearning is implemented \cite{Chen}.

The key question is whether the model $f'$ after unlearning becomes less distinguishable with respect to the forget data points, thus mitigating the effectiveness of MIAs, or whether the unlearning process itself leaves traces that adversaries can exploit. In practice, if the model retains subtle information about $\mathcal{D}_{forget}$ even after unlearning, it might still be vulnerable to MIAs, as adversaries could use this residual information to determine whether specific data points were in the original training set.

Formally, we are concerned with the impact of Machine Unlearning on the adversary's success rate, $S(A, f')$, where:
\begin{align}
S(A, f') = & \, P(A(f'(x)) = 1 \mid x \in \mathcal{D}_{forget}) \nonumber \\
          + & \, P(A(f'(x)) = 0 \mid x \notin \mathcal{D}_{forget}), 
\end{align}
where $P(A(f'(x)) = 1)$ denotes the probability that the adversary's MIA algorithm $A$ predicts that the data point $x$ was part of the original training dataset $\mathcal{D}_{train}$, based on the modified model $f'$ after the unlearning process. Specifically, the adversary queries the model $f'$ with $x$ and, based on the model's output and attempts to infer whether $x$ was included in the forget dataset $\mathcal{D}_{forget}$. A prediction of $1$ indicates that the adversary believes $x$ was in the original training set.

Similarly, $P(A(f'(x)) = 0)$ represents the probability that the adversary predicts that $x$ was not part of the original training set, indicating that $x$ was not in $\mathcal{D}_{forget}$. Together, these probabilities reflect the adversary's success in conducting a membership inference attack on the unlearned model $f'$.

The goal is for $S(A, f')$ to be no better than random guessing for adversaries after the unlearning process, ideally reducing their success rate to that of flipping a coin (i.e., $S(A, f') \approx 0.5$). However, depending on the unlearning algorithm, $S(A, f')$ could remain higher, indicating that the model is still vulnerable to MIAs. Understanding and mitigating this influence is critical for designing robust unlearning mechanisms that not only comply with data deletion requests but also enhance the overall privacy and security of the model.

\subsection{White box and Black box MIA}

Membership Inference Attacks (MIA) are categorized based on the information available to the adversary regarding the target model's parameters and architecture.
\begin{itemize}
    \item[$\bullet$] \textbf{Black-box MIA}: In this restricted setting, the adversary can only interact with the target model via an API, providing an input x and receiving the corresponding output y, which is typically a vector of prediction probabilities (posteriors). The attacker exploits the statistical differences in these confidence scores, as models often exhibit higher confidence and lower entropy for members of the training set compared to unseen data. This study utilizes shadow models to simulate this behavior and train an attack classifier without internal model access.

    \item[$\bullet$] \textbf{Grey-box MIA}: This intermediate setting assumes the adversary possesses partial information about the target model or its training process. This may include knowledge of the model's architecture (e.g., specific ResNet or MLP structures) or the underlying data distribution, without having direct access to the model's weights. By aligning the shadow model's architecture with that of the target, the attacker can more accurately replicate the target model's decision boundaries, leading to a more effective inference than a strictly black-box approach.

    \item[$\bullet$] \textbf{White-box MIA}: In a white-box scenario, the adversary has full access to the target model's internal components, including its weights, specific architecture, and intermediate layer activations. This allows for more sophisticated attacks that exploit the model's internal gradients or ``memorization'' patterns within specific neurons. White-box attacks generally provide an upper bound on privacy leakage as they utilize the maximum possible information available.
\end{itemize}

\subsection{Unlearning Algorithms Families}

Machine unlearning algorithms are generally classified into two broad categories based on their approach to removing the influence of specific data points from a trained model.

\begin{itemize}
    \item[$\bullet$] \textbf{Exact Unlearning}: This paradigm aims to produce a model that is statistically indistinguishable from one trained from scratch without the ``forget set.'' While retraining is the most straightforward example, more efficient exact methods (e.g., SISA) leverage specialized data partitioning and partial retraining to reduce computational costs. Because these methods provide a theoretical guarantee of data removal, they serve as the gold standard for privacy compliance.
    \item[$\bullet$] \textbf{Unlearning}: This more common category seeks to efficiently approximate the ideal unlearned model through heuristics and optimization without full retraining. These methods typically involve targeted parameter updates to reduce the impact of the forgotten data to a practical level. Our study focuses on three representative approximate methods:
    \begin{itemize}
        \item[$\bullet$] \textbf{NegGrad (Negative Gradient)}: Serves as a fundamental gradient-based baseline. It is the standard approach for direct parameter manipulation and is essential for establishing a lower bound on unlearning performance.
        \item[$\bullet$] \textbf{SCRUB}: Represents the state-of-the-art in teacher - student frameworks. It is widely cited \cite{kurmanji2023towards} for its ability to balance the removal of ``forget'' information with the preservation of ``retain'' utility via knowledge distillation.
        \item[$\bullet$] \textbf{SFTC}: A recent and promising approach that optimizes unlearning fidelity \cite{perifanis2024sftc}. represents the ``selective update'' family of algorithms, which are currently favored for their efficiency and ability to handle complex data distributions.
    \end{itemize}
\end{itemize}

\subsection{Metrics}

\subsubsection{Forget Accuracy}
Forget Accuracy is a metric used to evaluate the effectiveness of a machine unlearning algorithm. It is defined as the accuracy of the unlearned model when evaluated on the forget set, the specific subset of data that was targeted for removal. A successful unlearning process aims to minimize this accuracy, ideally reducing it to a level equivalent to that of a model that was never trained on this data in the first place (i.e., no better than random guessing or its baseline generalization performance on unseen data). A low Forget Accuracy indicates that the model no longer retains the specific patterns learned from the forgotten data.

\subsubsection{Retain Accuracy}
Retain Accuracy is a metric that measures the utility preservation of a machine unlearning algorithm. It is defined as the accuracy of the unlearned model when evaluated on the retain set—the portion of the original training data that was not targeted for removal. The primary goal is to maintain a Retain Accuracy that is as high as possible, and ideally, statistically indistinguishable from the accuracy of the original model on this same set. This metric ensures that the unlearning process, while removing specific information, does not cause ``catastrophic forgetting'' or degrade the model's performance on the remaining, valid data.

\subsubsection{MIA Accuracy}
MIA Accuracy refers to the accuracy of a Membership Inference Attack (MIA). Meaning the number of samples that correctly identified as part of the training dataset. So if we want to quantify it, this metric is the number of samples that correctly identified as part or not of the original dataset compared to total number of samples queried.

\begin{equation}
MIA\ Acc = \frac{Correct\ Predictions}{Total\ Samples}
\end{equation}
where Total Samples are all the samples we are testing, both members and non members.

\subsubsection{MIA Forget Accuracy}
MIA Forget Accuracy is a privacy-verification metric used to quantify the residual influence of deleted data by evaluating the success rate of a Membership Inference Attack on the forget set. It measures the proportion of samples within the forget set that the adversary can still correctly identify as having been part of the original training distribution. By applying the MIA specifically to the forget set, we can empirically determine the extent to which ``deleted'' samples continue to leave recognizable traces or unique signatures within the model's parameters. A successful unlearning process aims to reduce this accuracy to a baseline level, confirming that the model's behavior on the forgotten data is indistinguishable from its behavior on samples it has never encountered.

\begin{equation}
MIA\ Accuracy_{forget} = \frac{C_{F} + C_{U}}{|D_{forget}| + |D_{unseen}|}
\end{equation}
where $C_{F}$ and $C_{U}$ represent correct predictions for forget and unseen samples, respectively. A successful unlearning process aims to reduce this accuracy to a baseline of 50 \%, confirming that the model's behavior on the forgotten data is indistinguishable from its behavior on samples it has never encountered.

\subsubsection{MIA Retain Accuracy}
MIA Retain Accuracy serves as a measure of how the unlearning process affects the membership privacy of the data intended to be kept. It is defined as the accuracy of an MIA when evaluated on the retain set—the samples that were not targeted for removal. This metric tracks how many samples in the retain set continue to be correctly identified as members of the training distribution after the model has been modified. A significant deviation in MIA Retain Accuracy compared to the original model's baseline suggests that the unlearning process may have overly perturbed the model's parameters, potentially reducing its confidence or changing its behavior on the valid remaining data in a way that alters its privacy profile.

\begin{equation} MIA\ Accuracy_{retain} = \frac{C_{R} + C_{U}}{|D_{retain}| + |D_{unseen}|} \end{equation}
where $C_{R}$ and $C_{U}$ represent correct predictions for retain and unseen samples, respectively.

\section{Datasets} 
\label{sec:dataset}
To evaluate the defense potential of machine unlearning against Membership Inference Attacks (MIAs), we conduct experiments on four widely utilized datasets: CIFAR-10, Machine Unlearning Faces (MuFac), Purchase-100, and Texas-100. These datasets offer a diverse representational scope, with CIFAR-10 and MuFac representing the image domain, while Purchase-100 and Texas-100 consist of tabular data.

The Purchase and Texas datasets are benchmarks within the privacy domain; they are integrated into frameworks such as MLPrivacyMeter \cite{murakonda2020mlprivacymeteraiding} and were utilized in the seminal work that introduced MIAs \cite{shokri2017membership}. Furthermore, these datasets, along with versions of CIFAR, are frequently employed in broader membership privacy research \cite{nasr2018machinelearningmembershipprivacy}. Similarly, the MuFac dataset has emerged as a standard benchmark specifically for machine unlearning experiments \cite{perifanis2024sftc} \cite{spartalis2025}. Below, we provide a description of each dataset and its specific relevance to our evaluation of unlearning efficacy.

\subsection{Image Data}
The CIFAR-10 dataset~\cite{cifar10} is a foundational benchmark for image classification, consisting of 60,000 32x32 color images across 10 distinct classes. While the dataset's categories are general, it serves as a critical proxy for user-contributed image repositories where individuals may assert their 'right to be forgotten.' In this context, each image represents a unique data contribution that a model may inadvertently memorize during training.

Evaluating unlearning on CIFAR-10 allows us to measure how effectively an algorithm can remove the specific visual signature of a data point—preventing an adversary from confirming its presence in the training set. Furthermore, as a widely adopted baseline in privacy research, its use ensures that our evaluation of an algorithm's ability to mitigate fundamental memorization is both reproducible and directly comparable with other methods.

The MuFac Dataset~\cite{choi2023towards} is a large-scale collection featuring over 13,000 Asian facial images. The dataset is comprehensive, with images capturing multiple factors influencing appearance, such as different angles, accessories (like glasses or hats), and emotional expressions. While the data is annotated with both age groups and personal identities, we clarify that the specific classification task used in our experiments is multi-class age classification.

The MuFac dataset serves as a high-fidelity proxy for privacy-sensitive biometric systems, where the right to be forgotten' is particularly critical due to the immutable nature of facial features. Its detailed annotations, specifically the explicit link between individual images and personal identities, provide a rigorous framework for evaluating an algorithm's ability to erase a specific biometric fingerprint. By targeting individual identities for removal while maintaining the model's capacity to classify age for the remaining population, we can simulate real-world requirements for facial analysis technologies. This ensures that a model can comply with a user's deletion request by successfully masking their membership traces without compromising the system's overall functional utility.

\subsection{Tabular Data}

The Purchase-100 dataset~\cite{purchase100}, derived from Kaggle's ``Acquire Valued Shoppers Challenge'', consists of 197,324 customer records represented by 600-dimensional binary feature vectors. Each vector indicates whether a customer purchased products from specific category-company pairs. The classification task is to predict which of 100 behavioral segments—derived through clustering similar purchasing patterns, a given customer belongs to.

Purchase-100 serves as a critical benchmark for machine unlearning due to the high sensitivity of the transaction categories, which include pharmaceuticals, health products, and lifestyle choices. These features can reveal intimate details about an individual's health conditions and personal circumstances. Despite the dataset's overall size, the high-dimensional and sparse nature of the features makes individual purchase patterns potentially unique. This creates a realistic scenario where organizations must comply with deletion requests to mask these sensitive behavioral signatures, balancing the need for accurate market segmentation with the obligation to prevent adversarial re-identification through MIAs.

The Texas-100 dataset~\cite{shokri2017membership} is a hospital discharge dataset consisting of patient records from Texas hospitals. Each record contains over 6,000 features, including demographics, treatment details, and hospital-related metadata. The classification task in our experiments is to predict the patient's primary diagnosis across 100 distinct categories.

Due to the presence of highly sensitive information, such as discharge status and surgical procedures, Texas-100 is a standard benchmark for privacy-preserving machine learning. Its high dimensionality makes individual patient records highly unique, creating a significant risk for membership inference. In the context of machine unlearning, this dataset represents a critical healthcare application where a model must be able to erase specific clinical fingerprints upon request to prevent the unauthorized disclosure of a patient's medical history, all while maintaining high accuracy for general diagnostic research.

\section{Experiments}
\label{sec:experiments}

In this section, we present the experimental setup and results of our study regarding the defense potential of machine unlearning algorithms against Membership Inference Attacks (MIAs). Our experiments are designed to systematically evaluate the performance of state-of-the-art unlearning algorithms across diverse model architectures. Specifically, we investigate how effectively these algorithms remove the influence of target data points while monitoring the resulting changes in the model's vulnerability to MIAs.

The experiments were conducted on a workstation running Ubuntu 24.04.3 LTS, equipped with an \textbf{AMD Ryzen™ 7 6800HS} processor (16 logical cores) and an \textbf{AMD Radeon™ 680M} integrated graphics unit. For accelerated model training and unlearning, we utilized an \textbf{NVIDIA GeForce RTX™ 3050 Laptop GPU} \textbf{(4GB VRAM)} and a \textbf{512.1 GB SSD} for high-speed data access and storage. All unlearning procedures were implemented using Python-based deep learning frameworks, ensuring that hardware utilization was optimized for the high-dimensional datasets described previously.

\subsection{Target Model Methodology}

The total training dataset was bifurcated equally using a 50-50 split between the target and shadow training sets. This partitioning strategy ensures that the shadow models are trained on data drawn from the same underlying distribution as the target model, while remaining strictly disjoint from the target model's actual training samples.

All target models were trained using an 80/20/20 train-validation-test split, derived from the 50\% of the total dataset allocated to the target model's environment. We implemented various architectures for each dataset to observe how different levels of model complexity influence both predictive capacity and the 'shielding' efficacy provided by unlearning against Membership Inference Attacks (MIAs).

In this study, we employed a spectrum of target models, ranging from simple baselines to complex, real-world architectures. Table~\ref{tab:model_archs} summarizes the architectures used for each dataset, with Multi-Layer Perceptrons (MLPs) serving as a common baseline across all data types. We categorized these models into three groups: simple, intermediate, and complex architectures. In this classification, the \textit{TargetModelID} denotes the complexity group, while the suffixes \textit{a, b, c,} and \textit{d} correspond to the CIFAR-10, MUFAC, Purchase-100, and Texas-100 datasets, respectively.

The evaluated models include standard fully connected neural networks, Convolutional Neural Networks (CNNs), and more complex residual architectures such as ResNet. ResNet \cite{He2016resnet}  a deep CNN architecture that utilizes ``skip connections'' (or shortcut connections) within its residual blocks. These connections allow gradients to bypass certain layers, effectively mitigating the vanishing gradient problem and enabling the training of extremely deep networks. By learning residual mappings rather than direct transformations, ResNet facilitates the optimization of deep models, achieving superior performance in computer vision tasks such as image classification.

In all experiments, the target models were trained using hyperparameters that yielded the optimal test accuracy. To ensure the validity and fairness of our comparative analysis, we utilized the Optuna Python library \cite{akiba2019optuna}, which employs a Tree-structured Parzen Estimator (TPE) to tune key parameters, including the number of hidden-layer neurons, the number of layers, the learning rate, the epoch count, and the batch size. Once the optimal hyperparameters were identified, the models were deliberately trained for additional epochs to induce an approximate 30\% generalization gap between the training and test accuracy. This controlled overfitting was implemented to increase model susceptibility to attacks, thereby allowing us to evaluate whether unlearning algorithms can effectively protect even the most vulnerable, worst-case models.

\subsection{Membership Inference Attack Configuration}

\subsubsection{Attack Model}

For the implementation of the Attack Model of the MIA, we utilized a Random Forest classifier as the meta-model (attack model). The classifier was configured with 100 estimators to ensure a robust and stable decision boundary. The selection of this architecture was motivated by the following factors:

\begin{itemize}
    \item[$\bullet$] \textbf{Non-linear Relationship Mapping:} The prediction vectors (softmax probabilities) from target models often exhibit complex, non-linear distributions. Random Forests are inherently capable of capturing these high-dimensional interactions without requiring feature scaling or assuming a specific underlying distribution.
    
    \item[$\bullet$] \textbf{Variance Reduction via Bagging:} By aggregating the predictions of 100 individual decision trees through bootstrap aggregating (bagging), the model significantly reduces variance. This prevents the attack model from overfitting to specific artifacts of the shadow models, ensuring better generalization when attacking the unseen target model.
    
    \item[$\bullet$] \textbf{Dimensionality Handling:} In datasets such as Purchase-100 or Texas-100, where the output vector consists of 100 classes, the attack model must process high-dimensional input. Random Forests handle multi-modal probability scores efficiently, identifying subtle differences in confidence levels between members and non-members.
    
    \item[$\bullet$] \textbf{Stability and Efficiency:} The choice of 100 estimators provides an optimal balance between computational overhead and predictive stability.
\end{itemize}

\subsubsection{Shadow Model}
For the shadow model architecture, as proposed by \cite{shokri2017membership}, we utilized the same architecture as the target model in each scenario. This approach assumes prior knowledge of the target model's architecture by the adversary.While assuming knowledge of the target architecture represents a 'gray-box' framing, this choice is justified by several real-world considerations. First, in the modern ML ecosystem, many organizations utilize standardized, state-of-the-art architectures (e.g., ResNet, VGG, or standard MLP bottlenecks) which are easily guessable by an informed adversary. Second, a system's security should not rely on the secrecy of the algorithm but on the secrecy of the data or keys; therefore, evaluating the 'worst-case' scenario where the attacker knows the architecture provides a more robust upper bound of the privacy risk. Finally, research has shown that Membership Inference Attacks are effective even when there is an architectural mismatch \cite{zou2020privacyanalysisdeeplearning}, but using the same architecture ensures that the observed privacy leakage, ot lack of leackage, is attributable to the unlearning algorithm's performance rather than structural differences between models. In our experiments we used 5 shadow models.

\begin{table*}[htbp]
\begin{adjustbox}{width=1\textwidth,center}
\begin{tabular}{ccccl}
\textbf{TargetModelID} & \textbf{Dataset} & \textbf{Epochs} & \textbf{Learning Rate} & \textbf{Layers and Units} \\  \hline \\
1a & CIFAR-10         & 30     & 0.001            & 3 FC layers (512, 256, 128 units) \\ \\
1b & MuFac            & 60     & 0.0007            & 3 FC layers (512, 256, 128 units) \\ \\
1c & Purchase-100     & 100     & 0.01            & 1 FC layers (128 units) \\ \\
1d & Texas-100        & 100     & 0.01            & 1 FC layers (128 units) \\ \\
\hline \\
2a & CIFAR-10         & 30     & 0.001            & 2 Conv Blocks  (16, 32 filters)+1FC layers (128 units) \\ \\
2b & MuFac            & 60     & 0.0007            & 2 Conv Blocks  (16, 32 filters)+1FC layers (128 units) \\ \\
2c & Purchase-100     & 100     & 0.01            & 2 FC layers (256, 128 units) \\ \\
2d & Texas-100        & 100     & 0.01            & 2 FC layers (256, 128 units) \\ \\ \hline \\ \\
3a & CIFAR-10         & 30     & 0.001            & Resnet \\ \\
3b & MuFac            & 60     & 0.0007           & Resnet \\ \\
3c & Purchase-100     & 100     & 0.01            & 4-Layer MLP (1024, 512, 256 units) + Dropout \\ \\
3d & Texas-100        & 100     & 0.01            & 4-Layer MLP (512, 128, 128 units) + Dropout

\end{tabular}
\end{adjustbox}
\caption{Target Model Architectures}
\label{tab:model_archs}
\end{table*} 

\subsection{Overfitting as a Worst-Case Privacy Benchmark}

A notable characteristic of our experimental setup is the significant generalization gap observed in the target models, which reaches up to 30\%. While such levels of overfitting are unrepresentative of high-performing production models, this setting was intentionally chosen to maximize the Membership Inference (MI) signal.

Since MIA success is fundamentally bounded by the degree of empirical risk minimization and memorization, utilizing overfitted models serves as a ``stress test'' for the unlearning algorithms under evaluation. By amplifying the signal that the unlearning process is intended to erase, we can more precisely measure the privacy residual, the traces of training data that remain even after an unlearning request is processed.

This 'worst-case' framing establishes a conservative upper bound on privacy risk ; if an algorithm can successfully mitigate a strong membership signal in an overfitted model, its defensive efficacy is expected to be even more robust in well-regularized production environments where the baseline signal is naturally weaker. Furthermore, this controlled gap ensures that the observed privacy results are attributable to the unlearning algorithm's performance rather than a failure of the attack to detect members due to low baseline memorization.

\subsection{Unlearning Algorithms}

In this study, we focus on a comparative evaluation of three representative machine unlearning algorithms: NegGrad \cite{golatkar2020eternal} , SCRUB \cite{kurmanji2023towards}, and SFTC \cite{perifanis2024sftc}. These methods were carefully selected to cover distinct and widely recognized families of unlearning approaches. NegGrad exemplifies gradient-based parameter manipulation techniques aimed at directly removing the influence of target data. SCRUB uses a teacher-student framework that leverages knowledge distillation and divergence maximization to balance retention and unlearning. SFTC, a recent state-of-the-art method, integrates selective fine-tuning with targeted confusion to optimize unlearning effectiveness. Together, these algorithms represent a diverse spectrum of strategies addressing the unlearning challenge. While many machine unlearning algorithms exist in the literature, incorporating dozens of variants and hybrids, this study focuses on these three due to their strong empirical performance and methodological variety, which allows for a meaningful yet tractable comparison consistent with practices in recent benchmarking and survey works. This approach aligns with the field's current standards, where in-depth analysis typically centers on a carefully chosen subset of leading algorithms to provide clear insights and rigorous evaluation.

We include the SFTC algorithm \cite{perifanis2024sftc} as our main focus, given its novel approach to balancing utility preservation and targeted forgetting. In addition we use NegGrad, the most straightforward baseline which modifies model parameters by applying gradient ascent on the forget set, aiming to erase the influence of the forget set. We also consider SCRUB, a teacher-student approach, which minimizes the divergence between the student and original models on the retain set while maximizing divergence on the forget set. Each algorithm is tested under similar conditions to ensure a fair comparison of their unlearning and defense capabilities.

\subsection{Experimenting Methodology}

In these experiments, we apply unlearning algorithms to the target models and perform an MIA at each epoch to measure both MIA Forget Accuracy and MIA Retain Accuracy. This approach provides a high-granularity estimation of the attack's efficacy throughout the unlearning process, enabling a deeper understanding of the temporal unlearning dynamics.

Our primary objective is to observe whether the MIA accuracy on the forget set diminishes over time keeping the MIA Retain accuracy stable. Ideally, a model should unlearn specific data such that it yields a low MIA forget accuracy, behaving as if those samples had never been part of the training set. Simultaneously, it is essential to maintain a stable MIA retain accuracy to ensure that the model's knowledge of the remaining data is preserved. A decrease in MIA retain accuracy may indicate that the unlearning method is starting to understand after unlearning the Retain Data with a different way. Conversely, a significant increase in retain accuracy may suggest that the retain set is becoming more vulnerable to privacy leaks as the unlearning process progresses.

Maintaining a stable MIA Retain Accuracy alongside a significantly diminished MIA Forget Accuracy indicates that the unlearning algorithm has the potential to effectively mitigate privacy risks for the forgotten cohort without compromising the privacy profile of the remaining data.

To ensure a rigorous and comparative analysis, a uniform data partitioning strategy was applied across all evaluated algorithms. Specifically, 20\% of the training samples were allocated to the forget set, while the remaining 80\% served as the retain set.

\section{Results}

In this section, we present the results of our experiments based on the graphs we have created to demonstrate the MIA Forget and the MIA Retain accuracies throughout the unlearning epochs. We categorize the experiments in this section based on the unlearning algorithms, and for each unlearning algorithm, we present the results for the first, second, and third groups of architectures.

\label{sec:results}
\subsection{Negative Gradient Algorithm}

\subsubsection{First Group}

\begin{figure*}[t]
    \centering
    \begin{subfigure}[b]{0.49\textwidth}
        \centering
        \includegraphics[width=\textwidth]{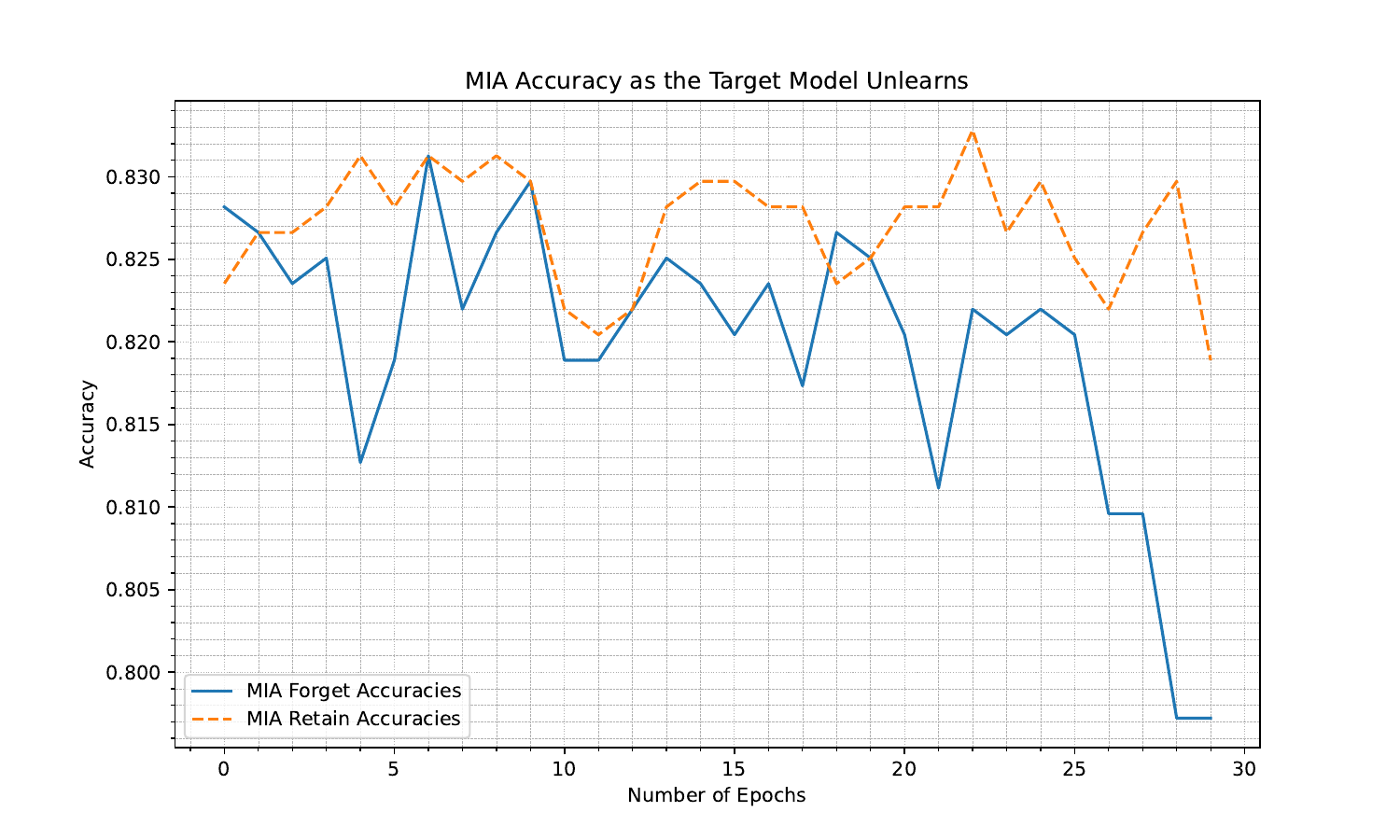}
        \caption{MIA Forget and Retain Accuracy \textbf{neg grad} for \textbf{Cifar}}
        \label{fig:mia_accuracy_1a_neg_grad}
    \end{subfigure}
    \hfill
    \begin{subfigure}[b]{0.49\textwidth}
        \centering
        \includegraphics[width=\textwidth]{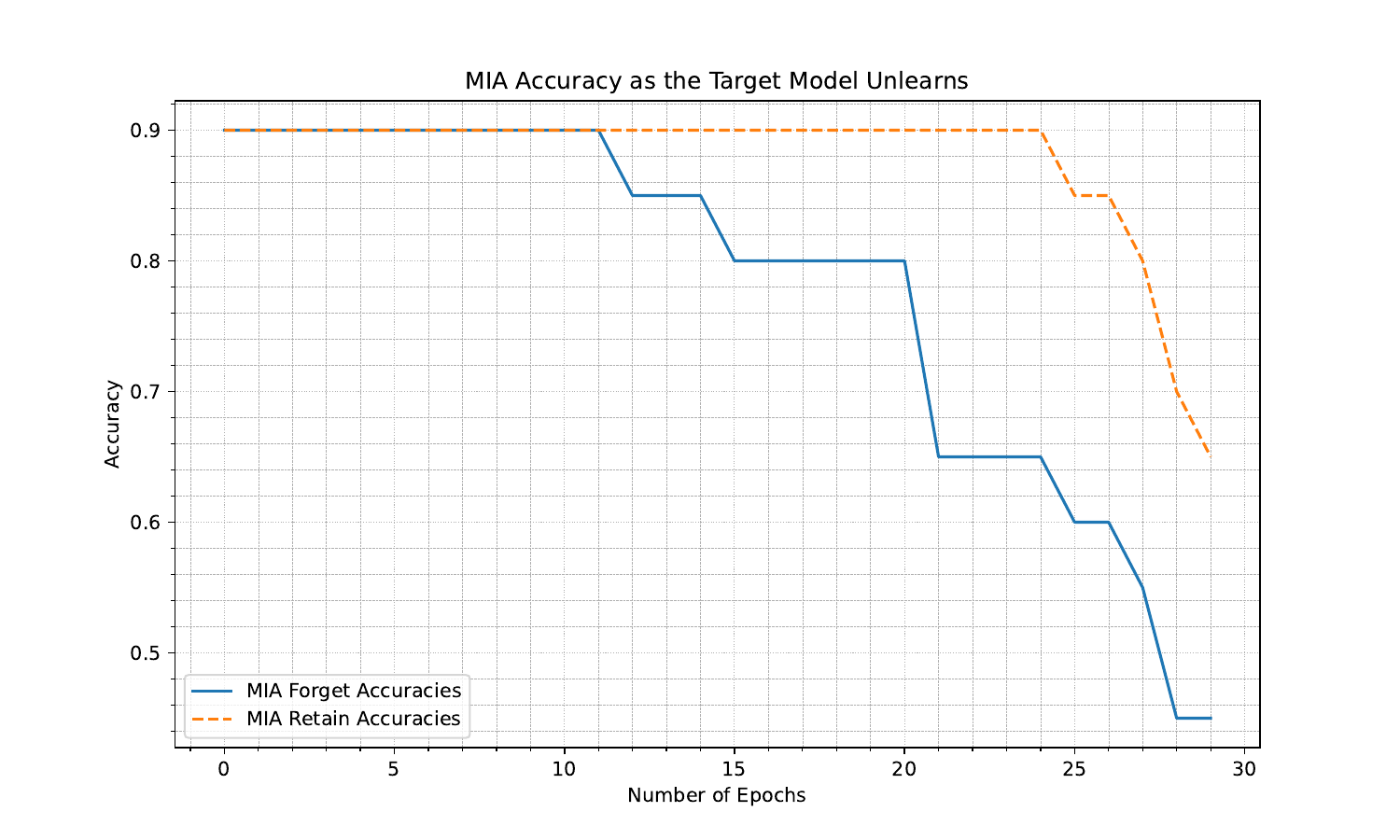}
        \caption{MIA Forget and Retain Accuracy \textbf{neg grad} for \textbf{Mufac}}
        \label{fig:mia_accuracy_1b_neg_grad}
    \end{subfigure}

    \medskip 
    
    \begin{subfigure}[b]{0.49\textwidth}
        \centering
        \includegraphics[width=\textwidth]{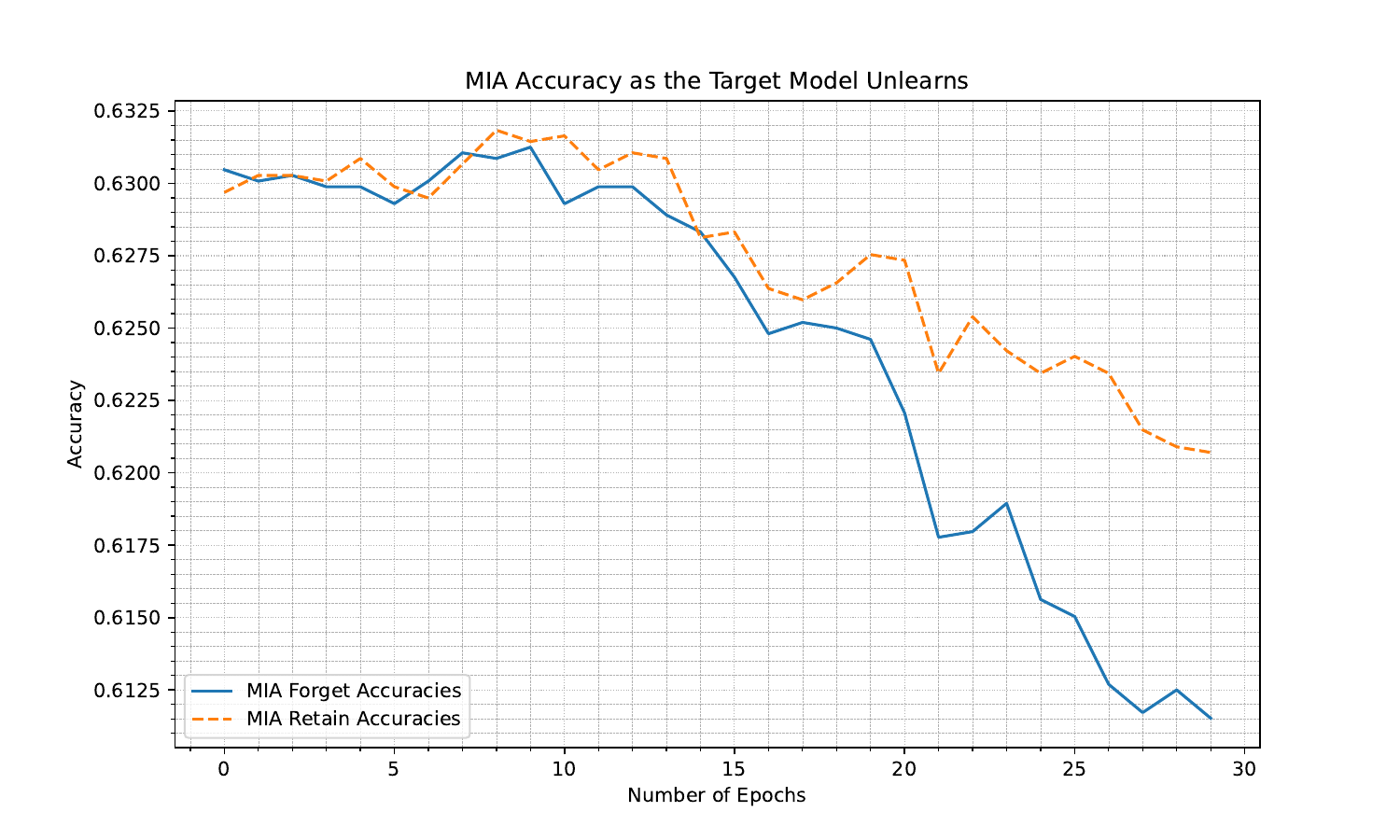}
        \caption{MIA Forget and Retain Accuracy \textbf{neg grad} for \textbf{Purchase}}
        \label{fig:mia_accuracy_1c_neg_grad}
    \end{subfigure}
    \hfill
    \begin{subfigure}[b]{0.49\textwidth}
        \centering
        \includegraphics[width=\textwidth]{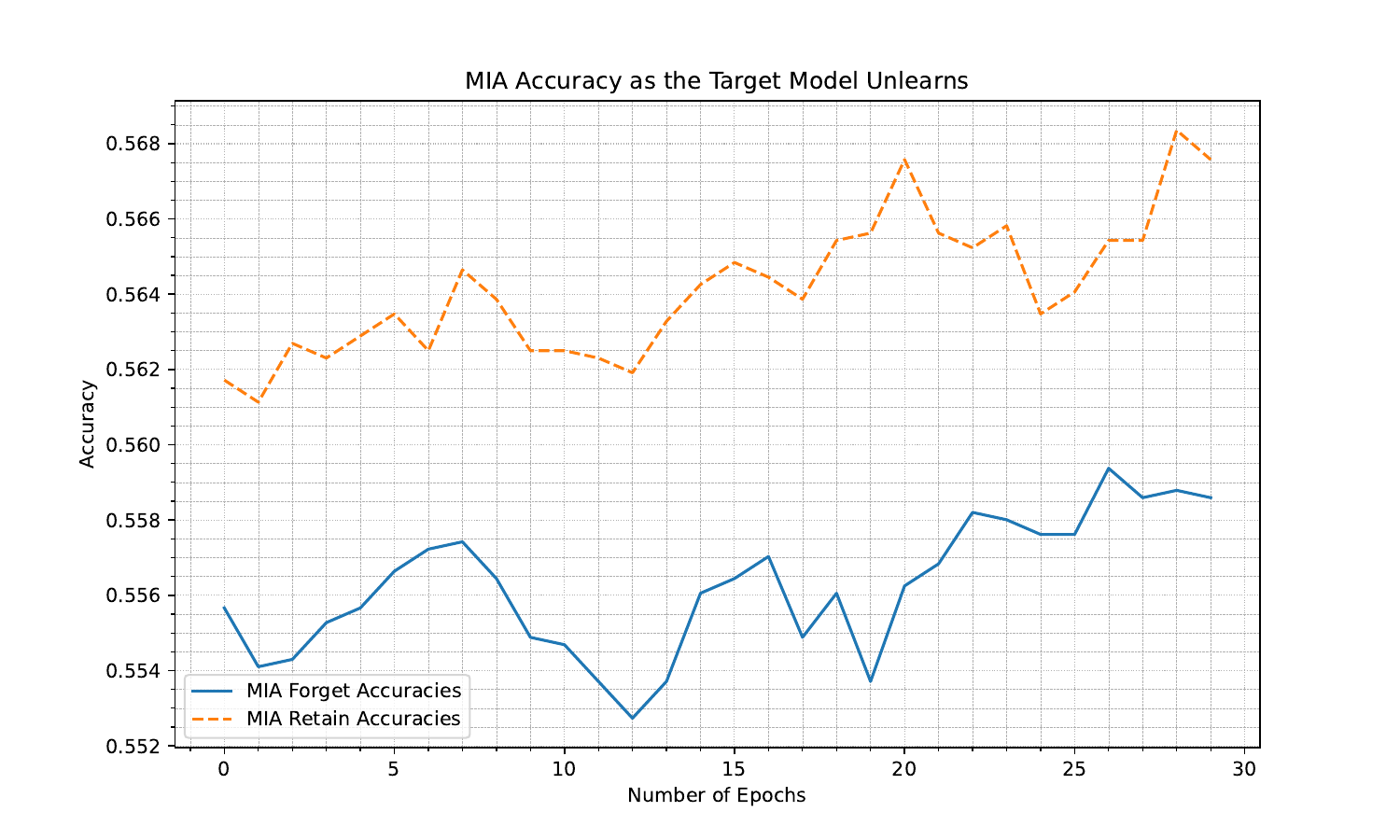}
        \caption{MIA Forget and Retain Accuracy \textbf{neg grad} for \textbf{Texas}}
        \label{fig:mia_accuracy_1d_neg_grad}
    \end{subfigure}
    
    \caption{These graphs illustrate the Membership Inference Attack (MIA) performance at each epoch of the unlearning process. This specific set of results corresponds to the \textbf{first group} of experiments, utilizing the Negative Gradient unlearning algorithm.}
    \label{fig:neg_grad_1_group}
\end{figure*}

Based on observations, (see Figure~\ref{fig:neg_grad_1_group}) the Negative Gradient algorithm demonstrated favorable performance on the CIFAR-10 dataset within the first group of target models.

Following the unlearning process (see Figure~\ref{fig:mia_accuracy_1a_neg_grad}), the MIA Retain Accuracy decreased slightly from 0.823 to 0.819, effectively maintaining the desired stability of the retain set. Simultaneously, the MIA Forget Accuracy dropped from 0.828 to 0.77. These results align with our primary objective: diminishing the MIA effectiveness on the forget set while preserving the integrity of the retain set. However, it should be noted that the MIA remained highly effective against this simple architecture, maintaining a 77\% accuracy on the forget set even after unlearning. This suggests that while the Negative Gradient method provides a degree of defense against membership inference, its protective capacity in this specific scenario is relatively limited.

Regarding the unlearning process illustrated in Figure~\ref{fig:mia_accuracy_1b_neg_grad}, we observe a simultaneous decline in both MIA Retain and MIA Forget accuracies. Both metrics initially plateau at 0.9, indicating a highly effective MIA attack prior to unlearning. By epoch 29, the Retain accuracy falls to 0.65 while the Forget accuracy reaches 0.478. At this stage, the model has undergone excessive unlearning; the significant degradation of the Retain accuracy suggests that the model's predictive capacity has been partially compromised. Had the unlearning process been terminated at epoch 22, the results would have more closely aligned with our initial objectives: the Retain accuracy remained stable at 0.9, while the Forget accuracy decreased to 0.75. This indicates that while the MIA originally demonstrated high precision in detecting membership, the unlearning process successfully bolstered the protection of the target model by reducing the MIA Forget accuracy by 0.15.

Regarding the unlearning process for the Purchase-100 dataset (Figure~\ref{fig:mia_accuracy_1c_neg_grad}), both the MIA Forget and MIA Retain accuracies exhibit a downward trend. We observe a reduction of 0.075 in retain accuracy and 0.175 in forget accuracy. While this trajectory aligns with our initial hypothesis, the magnitude of these changes may not be considered a substantial enough improvement for critical real-world production environments.

Finally, regarding the Texas-100 dataset (Figure~\ref{fig:mia_accuracy_1d_neg_grad}), the MIA initially failed to achieve a high degree of effectiveness, with a baseline Retain accuracy of 0.561 and a Forget accuracy of 0.555. Following the unlearning process, we observed a slight increase in both metrics. This highlights a critical edge case: an initially ineffective MIA became paradoxically more potent due to the artifacts or 'traces' left by the unlearning algorithm. While the increase is marginal (approximately 0.006), it demonstrates a trend where the unlearning process inadvertently enhances the model's susceptibility to membership inference. This phenomenon suggests that unlearning can sometimes act as a signal rather than a mask, a dynamic that warrants further investigation in future work.

\subsubsection{Second Group}

\begin{figure*}[t]
    \centering
    \begin{subfigure}[b]{0.49\textwidth}
        \centering
        \includegraphics[width=\textwidth]{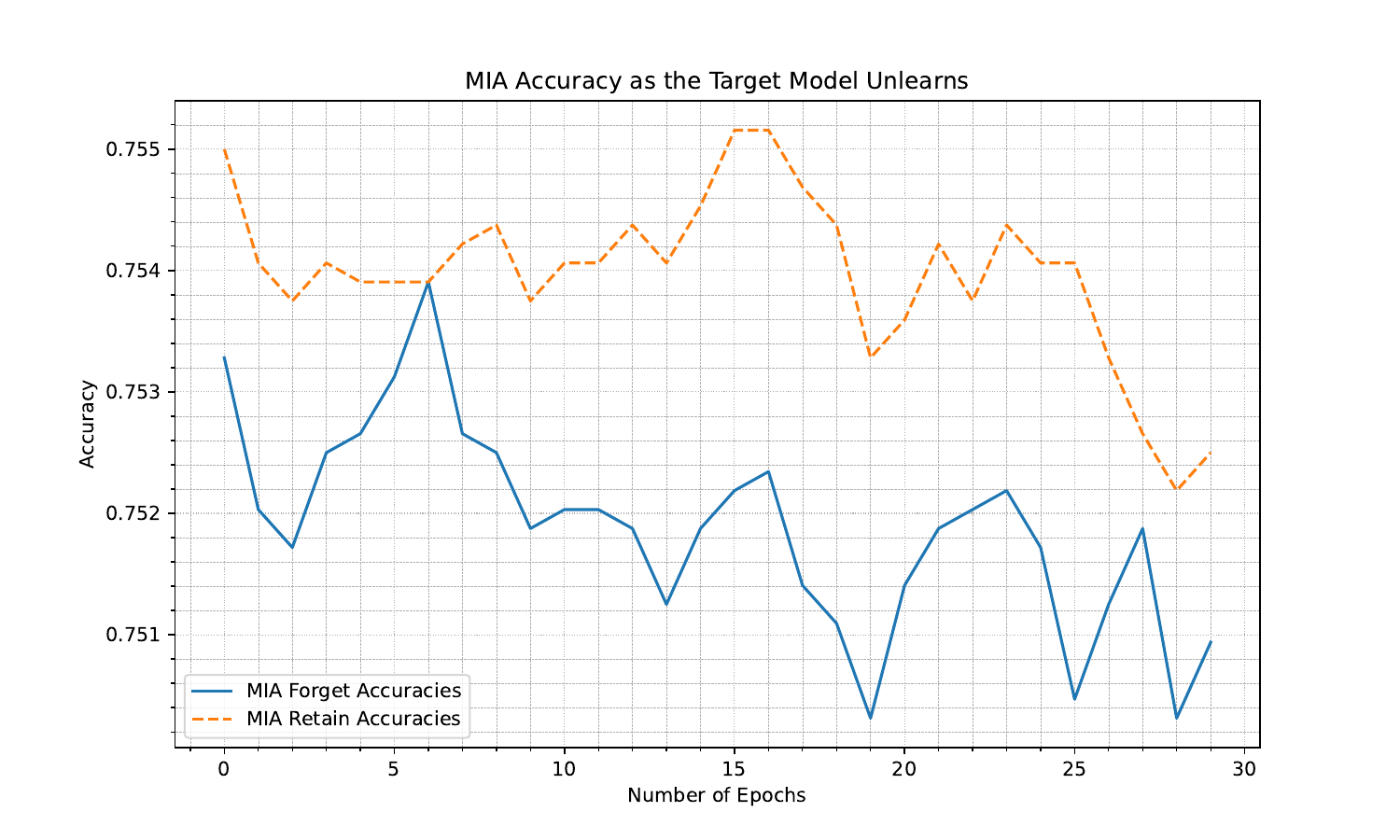}
        \caption{MIA Forget and Retain Accuracy \textbf{neg grad} for \textbf{Cifar}}
        \label{fig:mia_accuracy_2a_neg_grad}
    \end{subfigure}
    \hfill
    \begin{subfigure}[b]{0.49\textwidth}
        \centering
        \includegraphics[width=\textwidth]{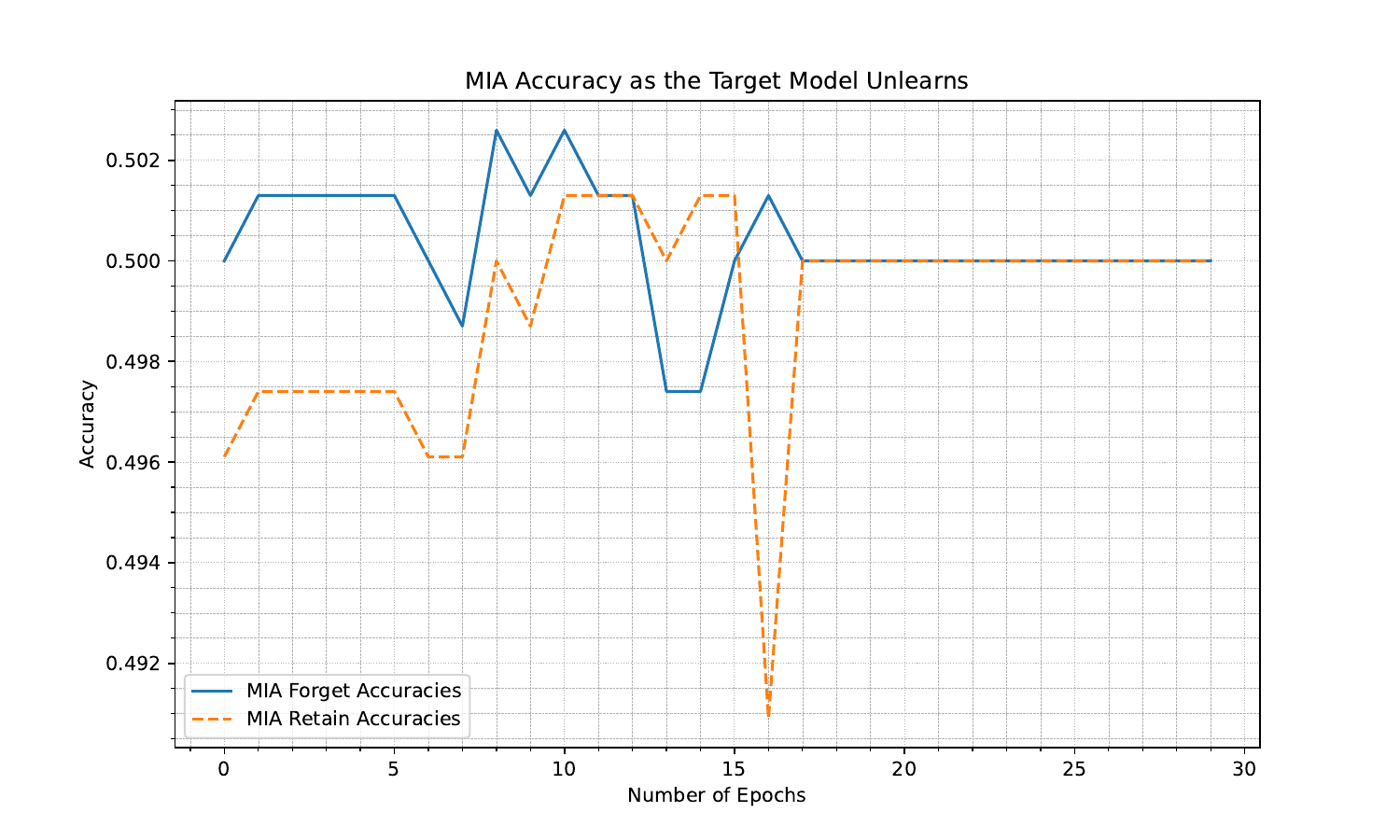}
        \caption{MIA Forget and Retain Accuracy \textbf{neg grad} for \textbf{Mufac}}
        \label{fig:mia_accuracy_2b_neg_grad}
    \end{subfigure}

    \medskip 
    
    \begin{subfigure}[b]{0.49\textwidth}
        \centering
        \includegraphics[width=\textwidth]{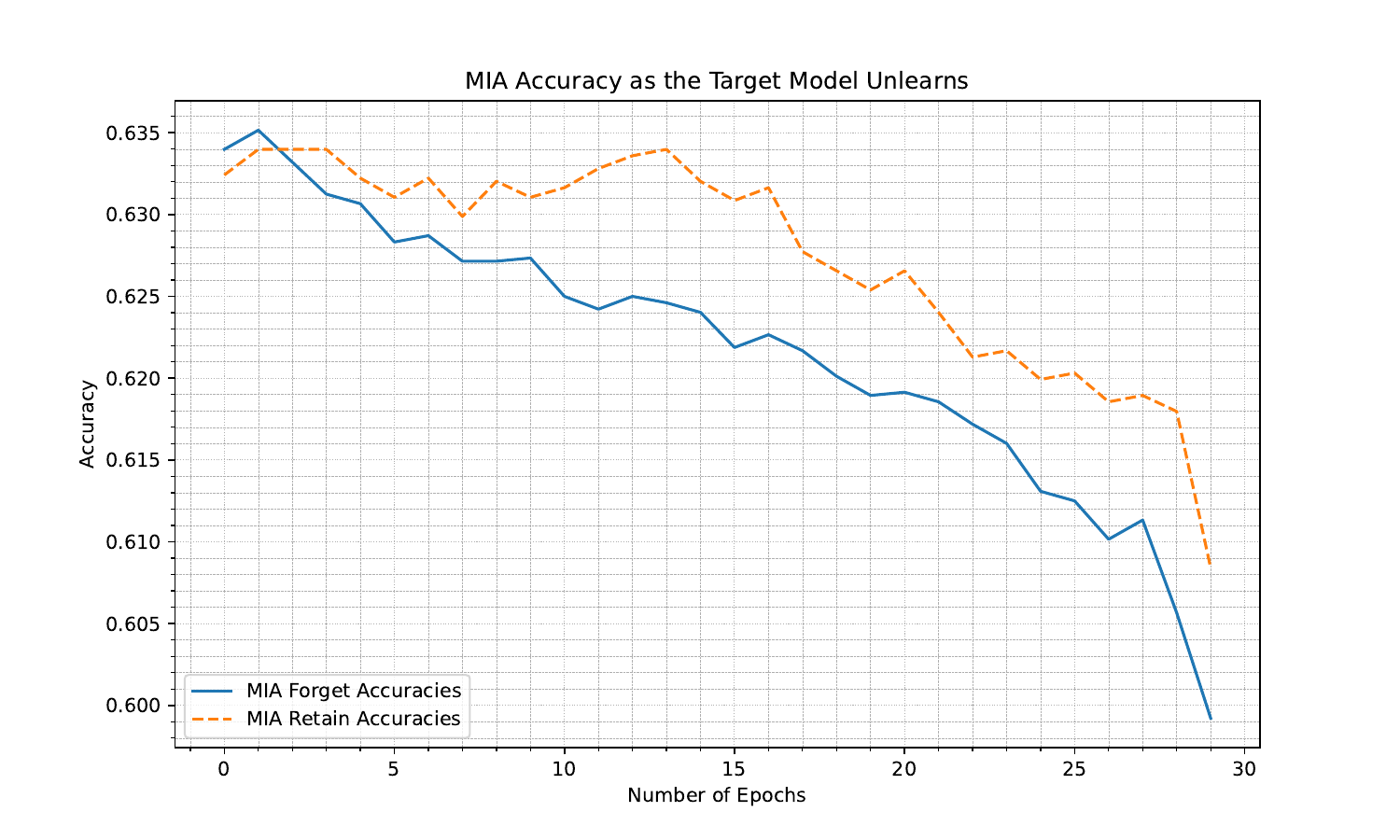}
        \caption{MIA Forget and Retain Accuracy \textbf{neg grad} for \textbf{Purchase}}
        \label{fig:mia_accuracy_2c_neg_grad}
    \end{subfigure}
    \hfill
    \begin{subfigure}[b]{0.49\textwidth}
        \centering
        \includegraphics[width=\textwidth]{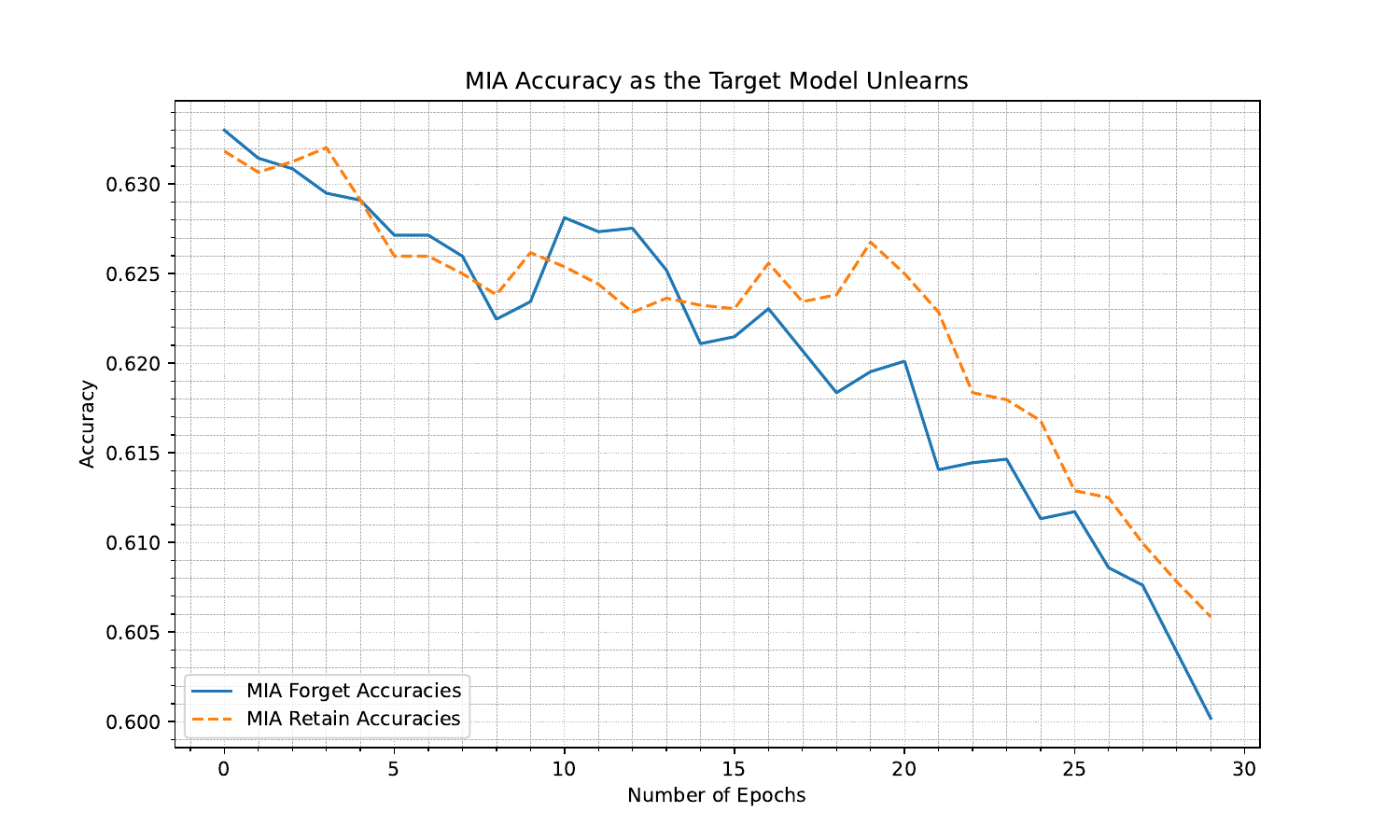}
        \caption{MIA Forget and Retain Accuracy \textbf{neg grad} for \textbf{Texas}}
        \label{fig:mia_accuracy_2d_neg_grad}
    \end{subfigure}
    
    \caption{These graphs illustrate the Membership Inference Attack (MIA) performance at each epoch of the unlearning process. This specific set of results corresponds to the \textbf{second group} of experiments, utilizing the Negative Gradient unlearning algorithm.}
    \label{fig:neg_grad_2_group}
\end{figure*}

In this subsequent series of experiments (refer to Figure~\ref{fig:neg_grad_2_group}), we replicate the epoch-wise Membership Inference Attack (MIA) evaluation. However, in this instance, the analysis is conducted on more complex model architectures to assess how increased capacity influences the unlearning process.

Regarding the CIFAR-10 dataset (see Figure~\ref{fig:mia_accuracy_2a_neg_grad}), 
we observe a clear declining trend in the MIA Forget Accuracy. By the 30th unlearning epoch, a simultaneous diminution of the MIA Retain Accuracy becomes apparent; however, if the unlearning process were terminated at epoch 24, we would achieve a comparatively stable MIA Retain Accuracy alongside a reduced MIA Forget Accuracy. In this instance, while the directional trend of the MIA performance aligns with our hypotheses, the absolute impact is marginal, yielding an accuracy reduction of only 0.004. Thus, while the qualitative behavior of the algorithm is correct, the quantitative impact on privacy remains limited.

Regarding the MuFac dataset (see Figure~\ref{fig:mia_accuracy_2b_neg_grad}), 
we observe that the MIA is largely ineffective, as the accuracy for both the forget and retain sets remains at approximately 50\%. This baseline performance, equivalent to random guessing, persists throughout the unlearning process. Such results indicate that the target model does not exhibit the necessary patterns of memorization or overfitting required for a successful membership inference attack, rendering the unlearning algorithm's impact on privacy metrics negligible in this specific scenario.

For the Purchase-100 dataset (see Figure~\ref{fig:mia_accuracy_2c_neg_grad}), the MIA accuracies for the forget and retain sets initiate at 0.634 and 0.631, respectively. Throughout the unlearning process, we observe a simultaneous diminishing trend across both datasets. This suggests that as the model's parameters are adjusted to forget the target data, there is a collateral impact on the retain set's membership signal, indicating that the unlearning process affects the overall posterior distribution of the model rather than isolated samples.
Regarding the Texas-100 dataset (see Figure~\ref{fig:mia_accuracy_2d_neg_grad}), the results exhibit a trajectory highly similar to that of the Purchase-100 dataset. The observed trends are consistent, characterized by a significant collateral impact on the retain set. Specifically, the decline in MIA accuracy is not isolated to the forget set; rather, it extends to the retained data, suggesting that the unlearning process induces a generalized degradation of the model's membership signal across all training samples.

\subsubsection{Third Group}

\begin{figure*}[t]
    \centering
    \begin{subfigure}[b]{0.49\textwidth}
        \centering
        \includegraphics[width=\textwidth]{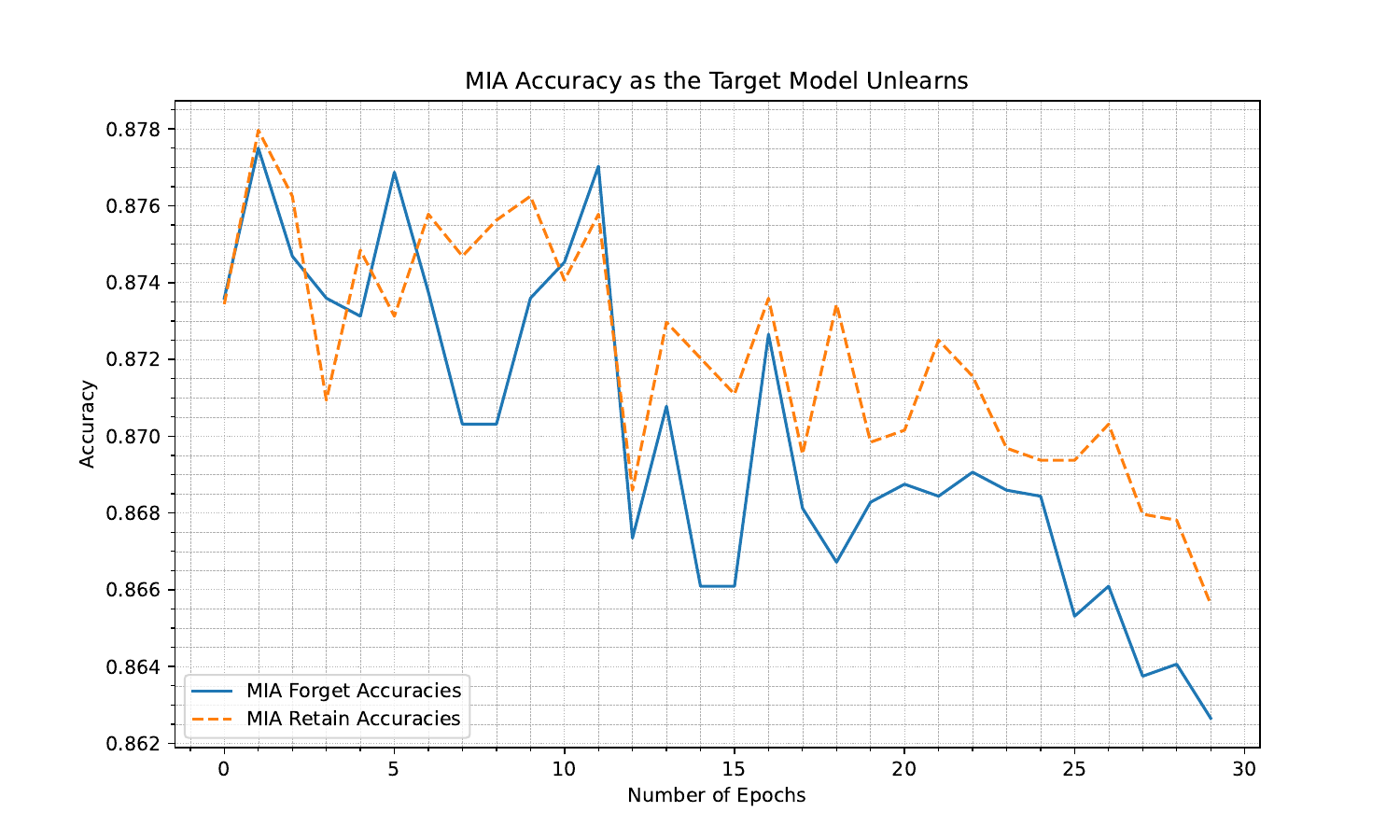}
        \caption{MIA Forget and Retain Accuracy \textbf{neg grad} for \textbf{Cifar}}
        \label{fig:mia_accuracy_3a_neg_grad}
    \end{subfigure}
    \hfill
    \begin{subfigure}[b]{0.49\textwidth}
        \centering
        \includegraphics[width=\textwidth]{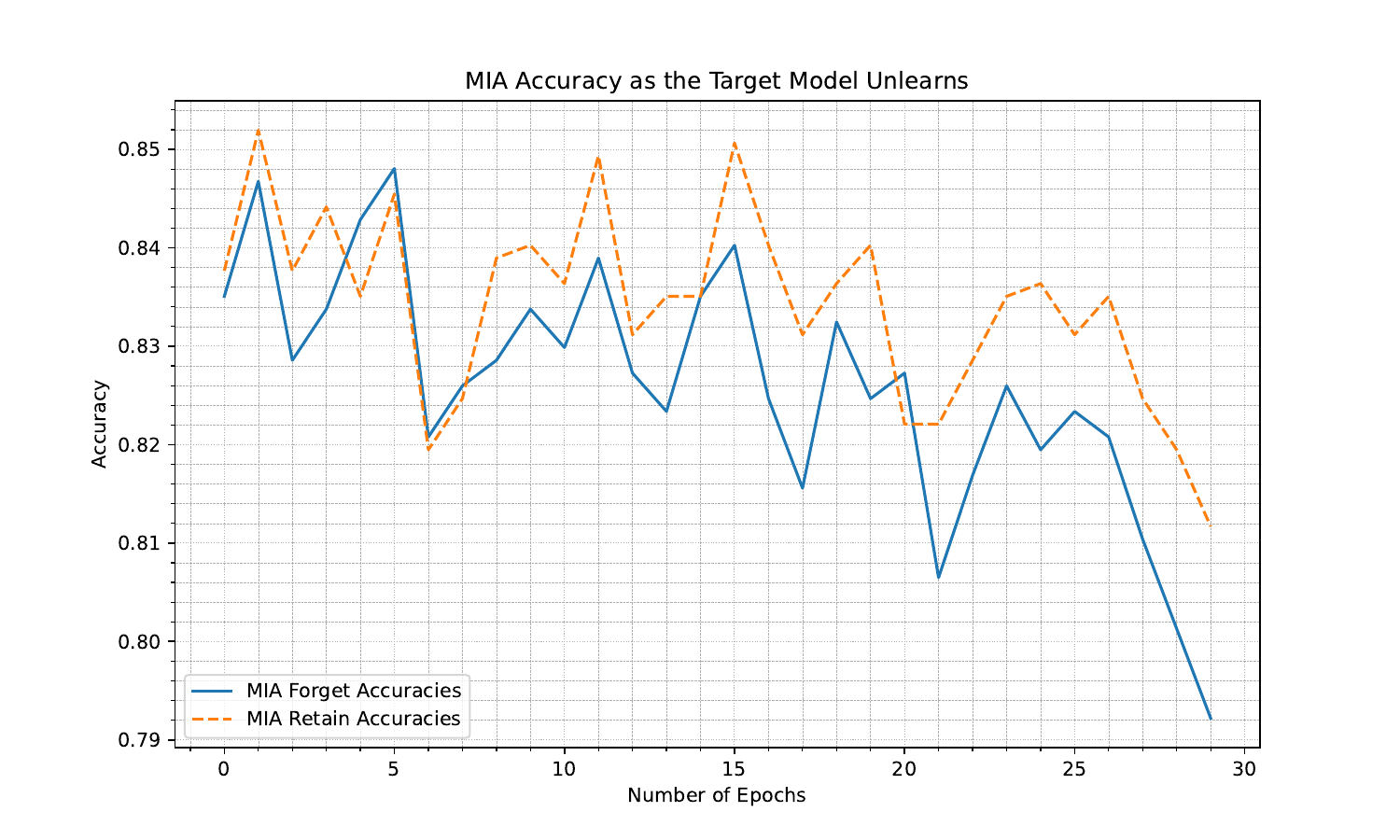}
        \caption{MIA Forget and Retain Accuracy \textbf{neg grad} for \textbf{Mufac}}
        \label{fig:mia_accuracy_3b_neg_grad}
    \end{subfigure}

    \medskip 
    
    \begin{subfigure}[b]{0.49\textwidth}
        \centering
        \includegraphics[width=\textwidth]{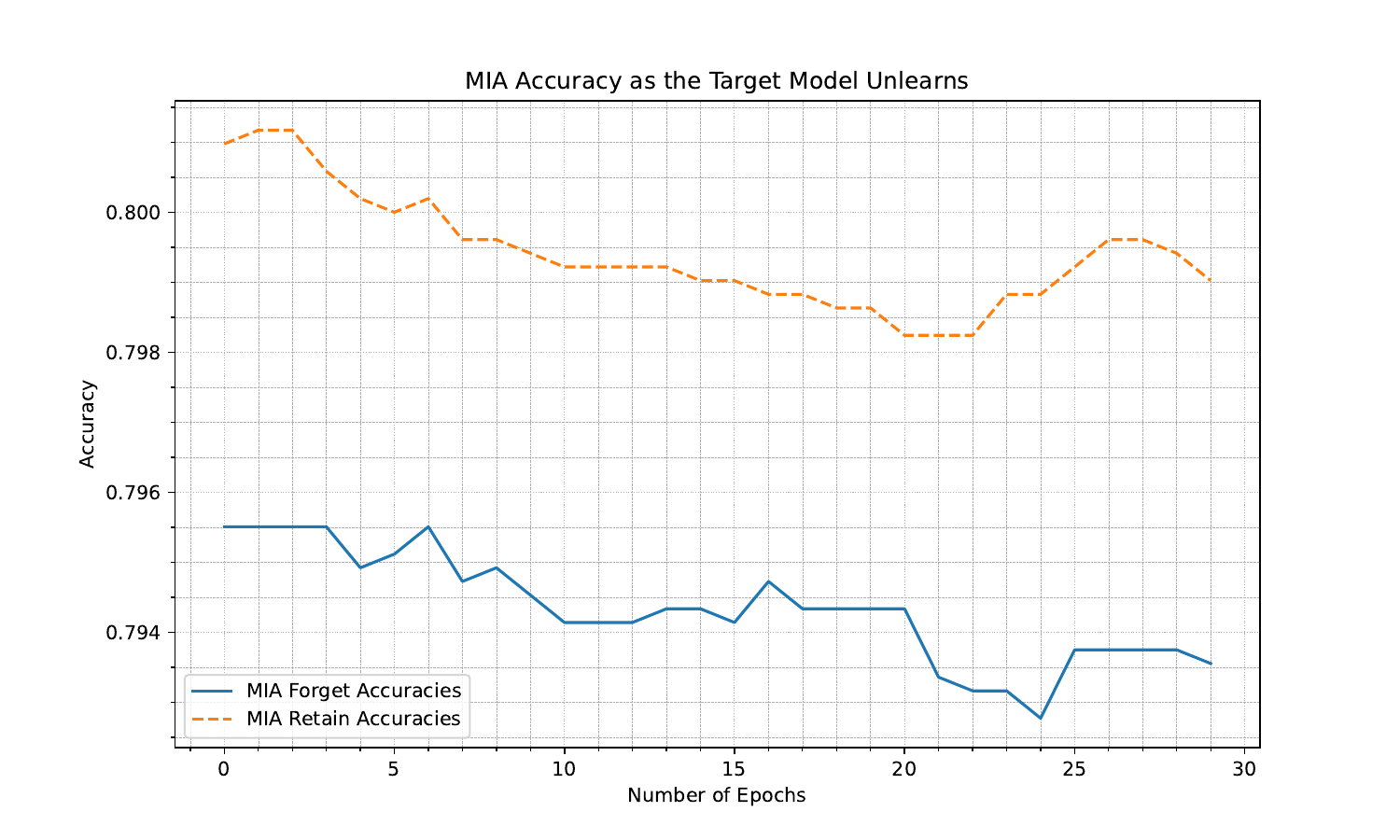}
        \caption{MIA Forget and Retain Accuracy \textbf{neg grad} for \textbf{Purchase}}
        \label{fig:mia_accuracy_3c_neg_grad}
    \end{subfigure}
    \hfill
    \begin{subfigure}[b]{0.49\textwidth}
        \centering
        \includegraphics[width=\textwidth]{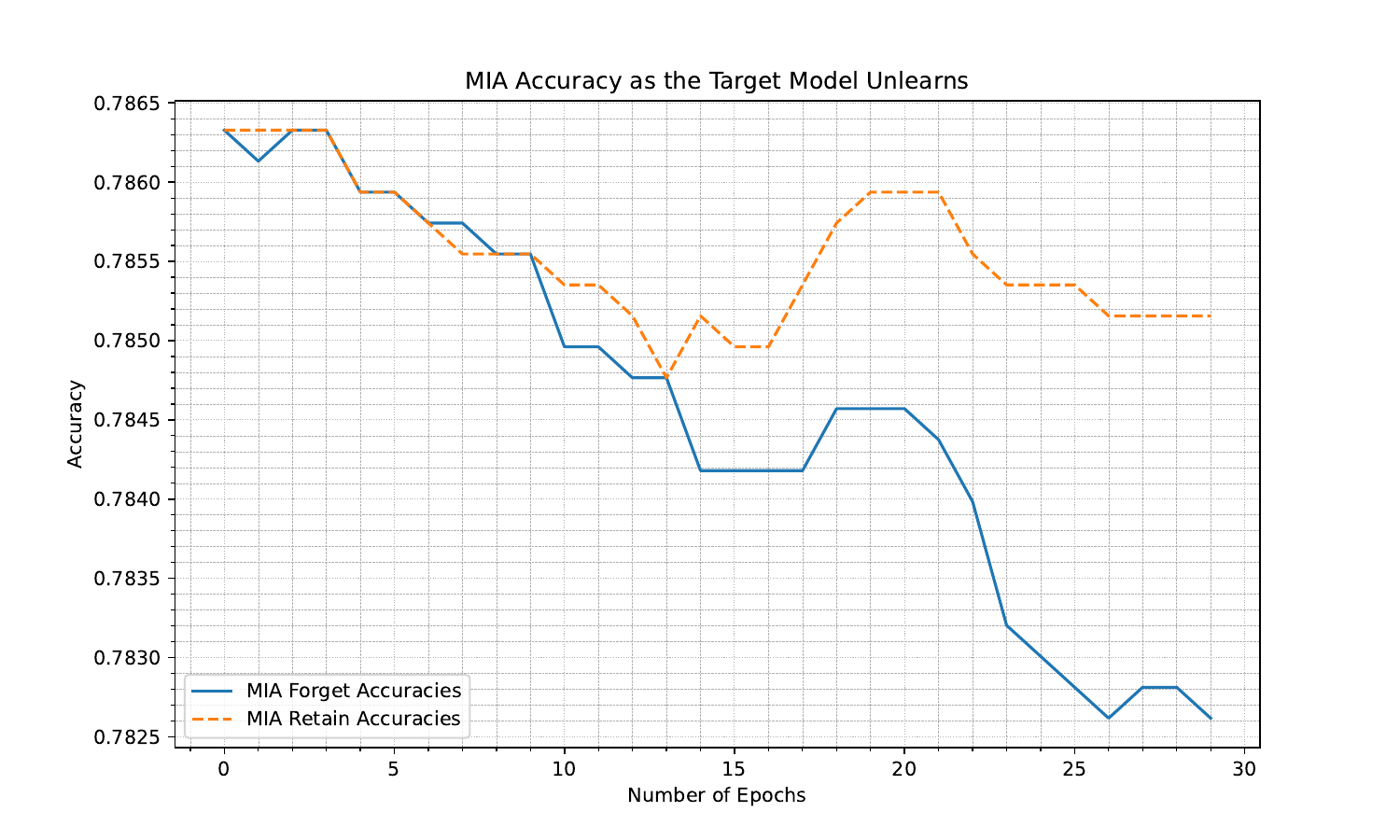}
        \caption{MIA Forget and Retain Accuracy \textbf{neg grad} for \textbf{Texas}}
        \label{fig:mia_accuracy_3d_neg_grad}
    \end{subfigure}
    
    \caption{These graphs illustrate the Membership Inference Attack (MIA) performance at each epoch of the unlearning process. This specific set of results corresponds to the \textbf{third group} of experiments, utilizing the Negative Gradient unlearning algorithm.}
    \label{fig:neg_grad_3_group}
\end{figure*}

In this section, we present the experimental results for the third group of target models, which utilize the most complex architectures evaluated in this study. The following analysis focuses on the performance of the Negative Gradient unlearning algorithm across these high-capacity models (see Figure~\ref{fig:neg_grad_3_group}).

Regarding the CIFAR-10 dataset (Figure~\ref{fig:mia_accuracy_3a_neg_grad}), we observe a high degree of volatility in both the MIA Forget and MIA Retain accuracies during the early stages of unlearning. However, as the process progresses, a distinct downward trend emerges. Notably, the MIA Forget accuracy experiences a more pronounced decline compared to the Retain set, particularly after epoch 25. This suggests that in complex architectures, the Negative Gradient method can achieve a differentiated unlearning effect, though the overall accuracy remains high (above 0.86), indicating substantial residual memorization.

For the MUFAC dataset (Figure~\ref{fig:mia_accuracy_3b_neg_grad}), the trends for both forget and retain accuracies remain tightly coupled for the majority of the unlearning trajectory. While there is a general decrease in MIA effectiveness, the gap between the two metrics is marginal until the final epochs. By epoch 30, we see a sharp divergence where the MIA Forget accuracy drops more significantly than the Retain accuracy. This implies that for this specific dataset and architecture, extended unlearning epochs are required to achieve any meaningful separation between the membership signals of the two sets.

In the Purchase-100 experiment (Figure~\ref{fig:mia_accuracy_3c_neg_grad}), the model demonstrates high stability in its membership signal. Interestingly, the MIA Retain accuracy remains consistently higher than the Forget accuracy throughout the process. We observe a gradual, slight decline in both metrics; however, the impact is minimal (shifting less than 0.01 in accuracy). This suggests that the complex architecture utilized here is highly robust against the Negative Gradient algorithm, maintaining a nearly constant privacy profile for both sets despite the unlearning iterations.

The results for the Texas-100 dataset (Figure~\ref{fig:mia_accuracy_3d_neg_grad}) highlight a significant divergence in the latter half of the experiment. Initially, the Retain and Forget accuracies decline in tandem. However, after epoch 15, the MIA Retain accuracy begins to recover and stabilize, while the MIA Forget accuracy continues to drop sharply toward epoch 30. This represents an ideal unlearning outcome for complex models: the model successfully reduces the privacy risk associated with the forgotten data while effectively ``protecting'' or restoring the membership signal stability of the retained data.

In these complex architectures, the Negative Gradient algorithm shows a more nuanced behavior than in simpler models. While Texas-100 shows the most ``surgical'' unlearning (forgetting without harming the retain signal), Purchase-100 appears almost entirely resistant to the process. This confirms that model complexity and dataset characteristics significantly dictate the efficiency of gradient-based unlearning

\subsection{Scrub Algorithm}

\subsubsection{First Group - Simple Target}

\begin{figure*}[t]
    \centering
    \begin{subfigure}[b]{0.49\textwidth}
        \centering
        \includegraphics[width=\textwidth]{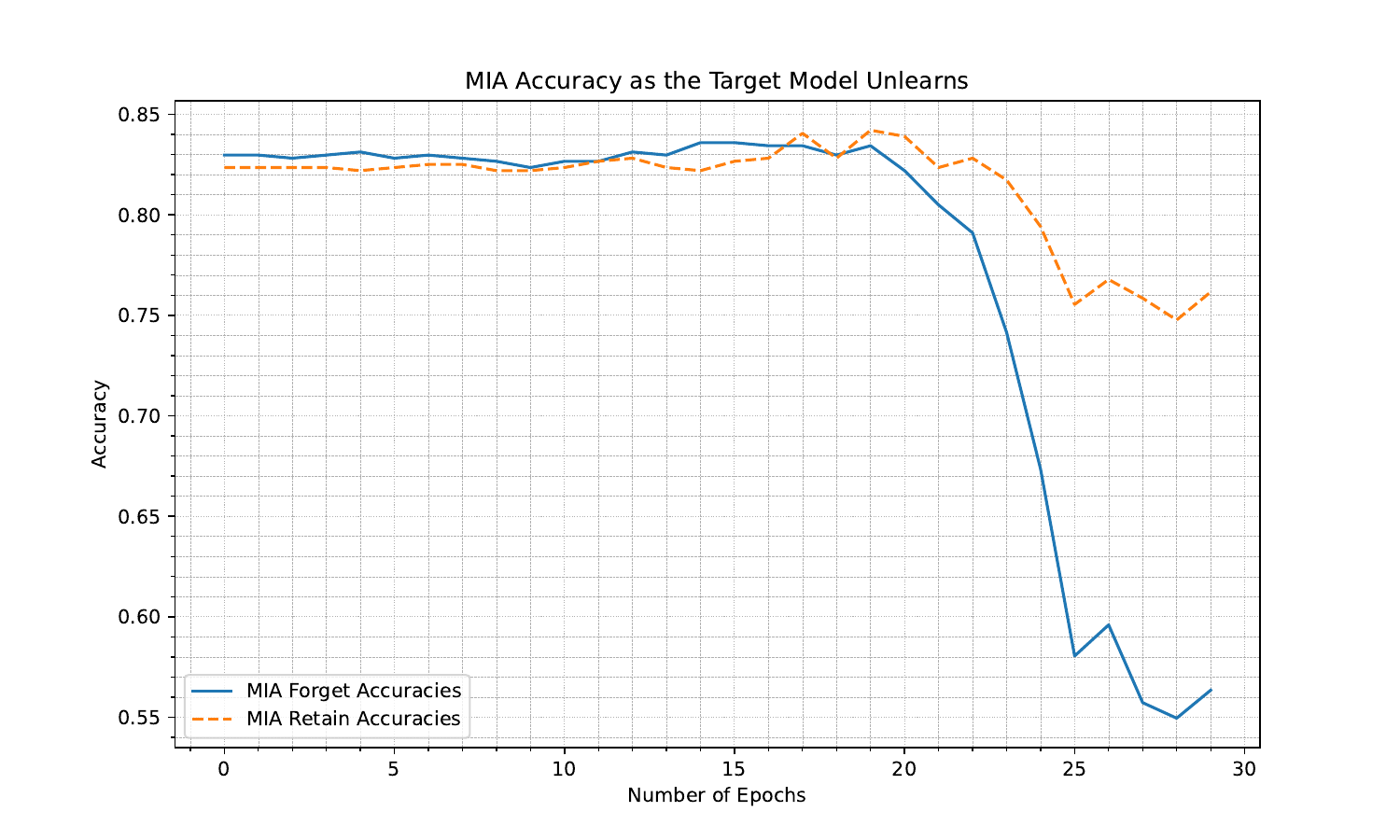}
        \caption{MIA Forget and Retain Accuracy \textbf{scrub} for \textbf{Cifar}}
        \label{fig:mia_accuracy_1a_scrub}
    \end{subfigure}
    \hfill
    \begin{subfigure}[b]{0.49\textwidth}
        \centering
        \includegraphics[width=\textwidth]{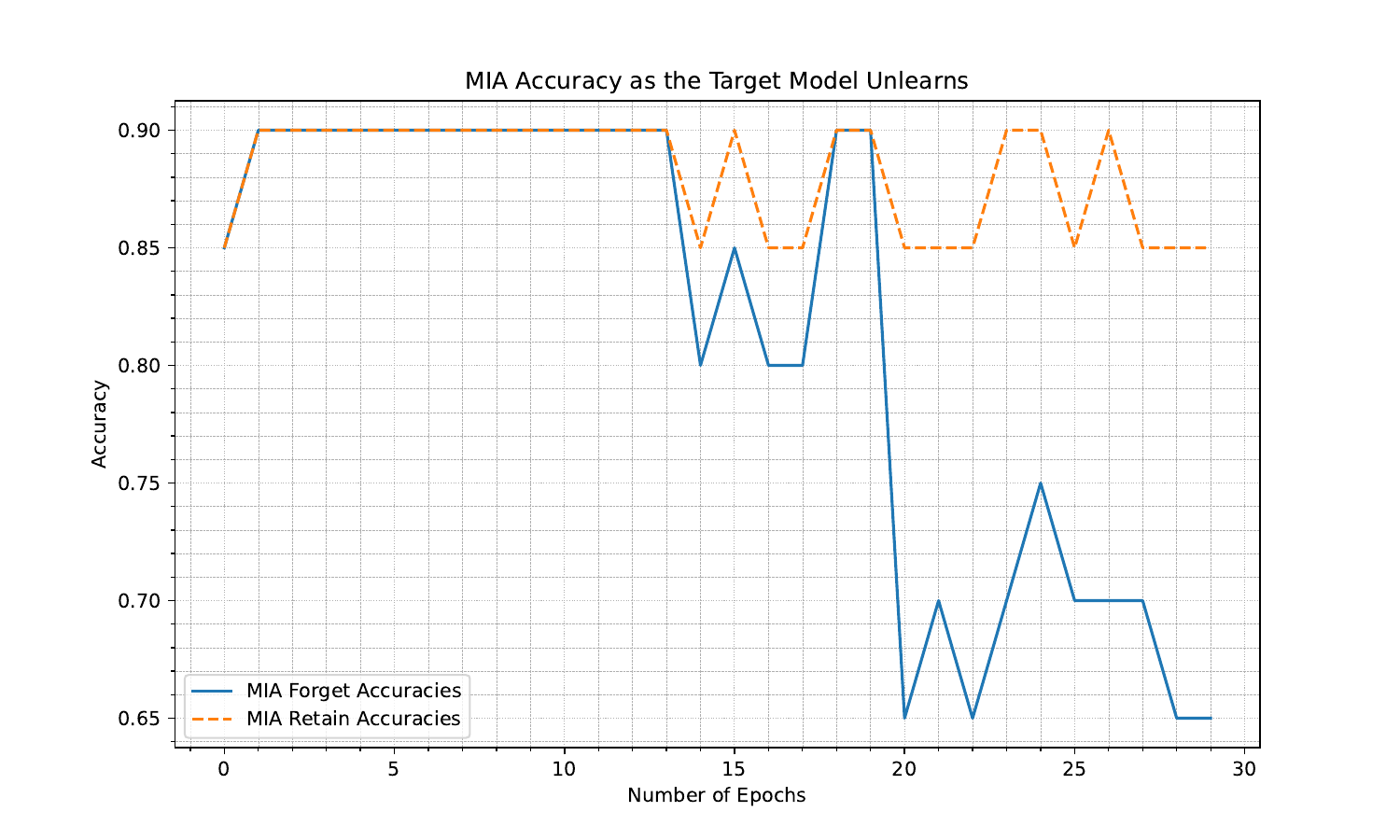}
        \caption{MIA Forget and Retain Accuracy \textbf{scrub} for \textbf{Mufac}}
        \label{fig:mia_accuracy_1b_scrub}
    \end{subfigure}

    \medskip 
    
    \begin{subfigure}[b]{0.49\textwidth}
        \centering
        \includegraphics[width=\textwidth]{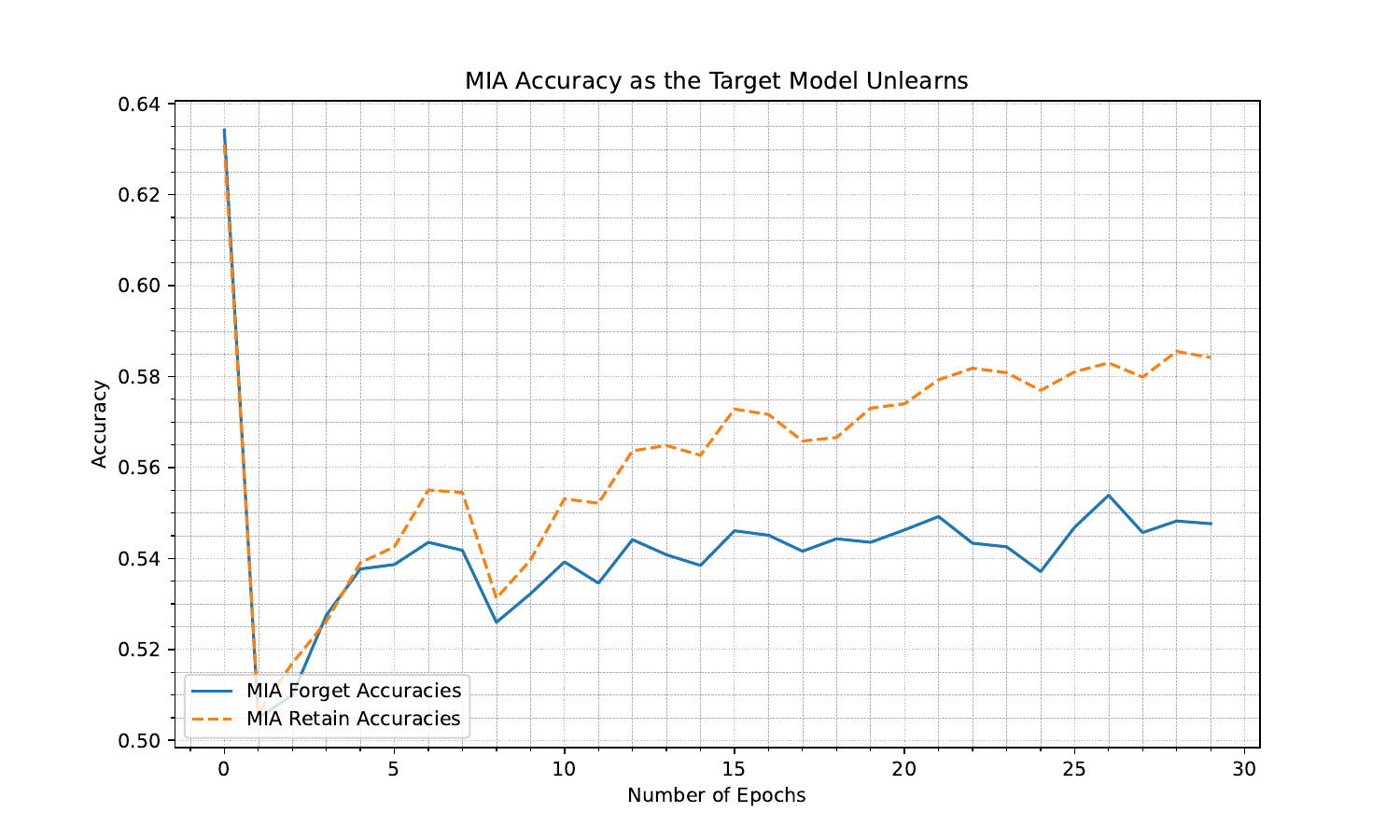}
        \caption{MIA Forget and Retain Accuracy \textbf{scrub} for \textbf{Purchase}}
        \label{fig:mia_accuracy_1c_scrub}
    \end{subfigure}
    \hfill
    \begin{subfigure}[b]{0.49\textwidth}
        \centering
        \includegraphics[width=\textwidth]{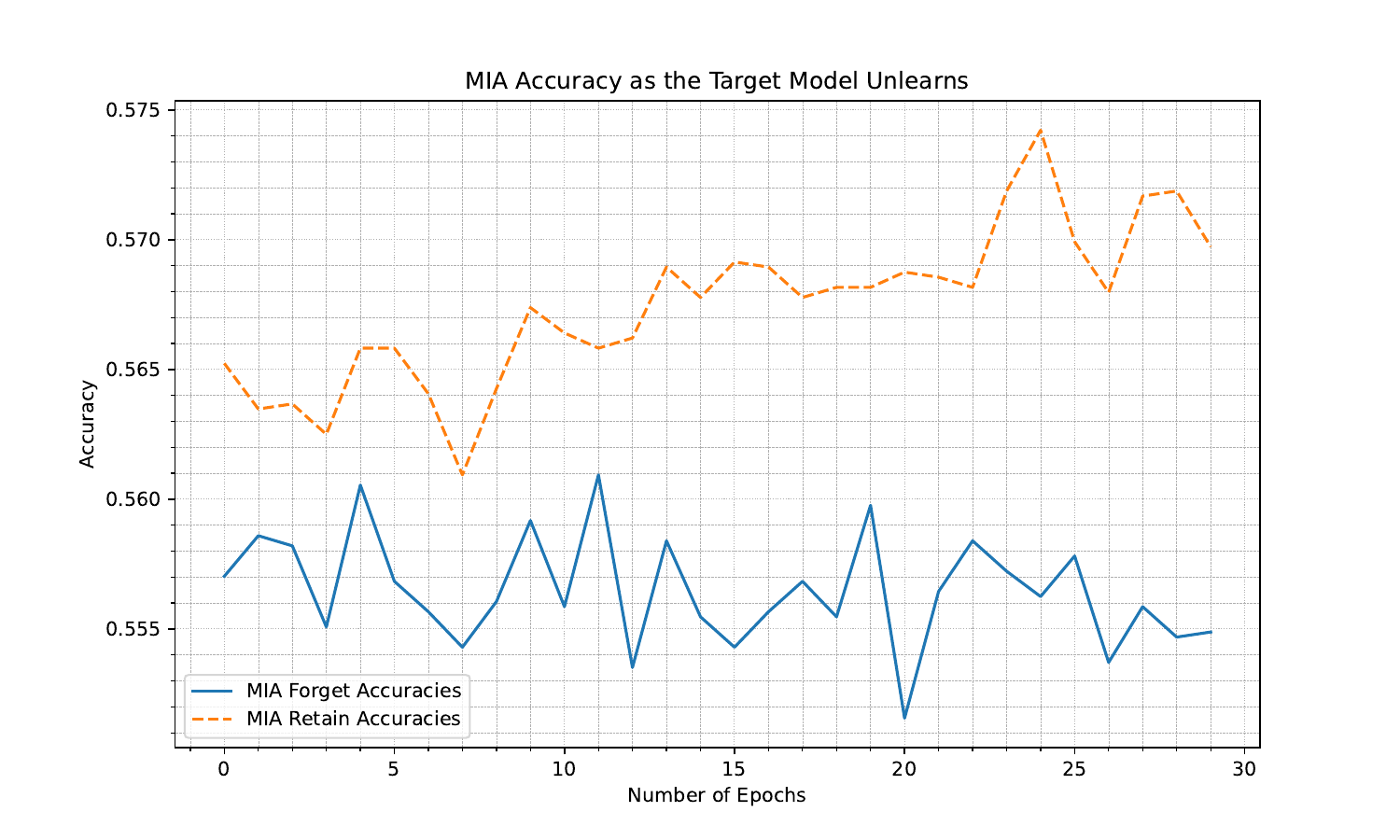}
        \caption{MIA Forget and Retain Accuracy \textbf{scrub} for \textbf{Texas}}
        \label{fig:mia_accuracy_1d_scrub}
    \end{subfigure}
    
    \caption{These graphs illustrate the Membership Inference Attack (MIA) performance at each epoch of the unlearning process. This specific set of results corresponds to the \textbf{first group} of experiments, utilizing the scrub unlearning algorithm.}
    \label{fig:scrub_1_group}
\end{figure*}

For the CIFAR-10 dataset (Figure~\ref{fig:mia_accuracy_1a_scrub}), the Scrub algorithm exhibits a highly stable MIA accuracy for both sets during the initial 20 epochs, maintaining a baseline of approximately 0.83. However, after epoch 20, we observe a sharp and significant decline in the MIA Forget accuracy, which drops precipitously to nearly 0.55. While the MIA Retain accuracy also experiences a reduction, it stabilizes around 0.75. This indicates that Scrub is particularly effective at ``breaking'' the membership signal for the forget set in later stages, though the widening gap suggests a clear divergence between the privacy protection of the two sets.

The MUFAC results (Figure~\ref{fig:mia_accuracy_1b_scrub}) reveal a highly volatile unlearning trajectory. Both metrics remain at a maximum ceiling of 0.90 for the first half of the process, indicating severe vulnerability to MIA. Post-epoch 13, the MIA Forget accuracy becomes extremely unstable, fluctuating sharply before settling near 0.65. Interestingly, the Retain accuracy also undergoes significant oscillations but recovers to 0.85. This ``spiky'' behavior suggests that the Scrub algorithm may induce temporary structural instabilities in simpler models before reaching a state of partial unlearning.

In the Purchase-100 dataset (Figure~\ref{fig:mia_accuracy_1c_scrub}), we observe a distinct ``reboun'' effect. Immediately following the start of unlearning, both accuracies drop to a baseline of 0.50 (random chance). However, as the epochs progress, both accuracies begin to climb steadily. By epoch 30, the Retain accuracy reaches approximately 0.58, while the Forget accuracy stabilizes around 0.54. This suggests that while Scrub initially successfully masks membership, the continued unlearning process inadvertently allows the model to re-learn or expose features that slightly increase its susceptibility to MIA over time.

The results for Texas-100 (Figure~\ref{fig:mia_accuracy_1d_scrub}) highlight a failure to achieve the primary unlearning objective. Throughout the 30 epochs, the MIA Forget accuracy remains consistently lower than the Retain accuracy, but both show a gradual upward trend. The Retain accuracy increases from 0.56 to roughly 0.57, while the Forget accuracy oscillates around 0.555. This upward trajectory in both metrics suggests that the Scrub algorithm, when applied to this specific architecture and dataset, may actually be increasing the model's transparency to an attacker rather than providing protection.

Compared to the Negative Gradient method, the Scrub algorithm appears more aggressive in its later stages (especially in CIFAR-10) but exhibits significant instability in MUFAC. The most concerning trend is observed in Purchase-100 and Texas-100, where unlearning leads to an increase in MIA accuracy over time. This highlights that for simple architectures, Scrub can inadvertently ``leak'' information as it attempts to remove it, a phenomenon that emphasizes the importance of early stopping at the ``dip'' (e.g., epoch 2 in Purchase-100).

\subsubsection{Second Group}

\begin{figure*}[t]
    \centering
    \begin{subfigure}[b]{0.49\textwidth}
        \centering
        \includegraphics[width=\textwidth]{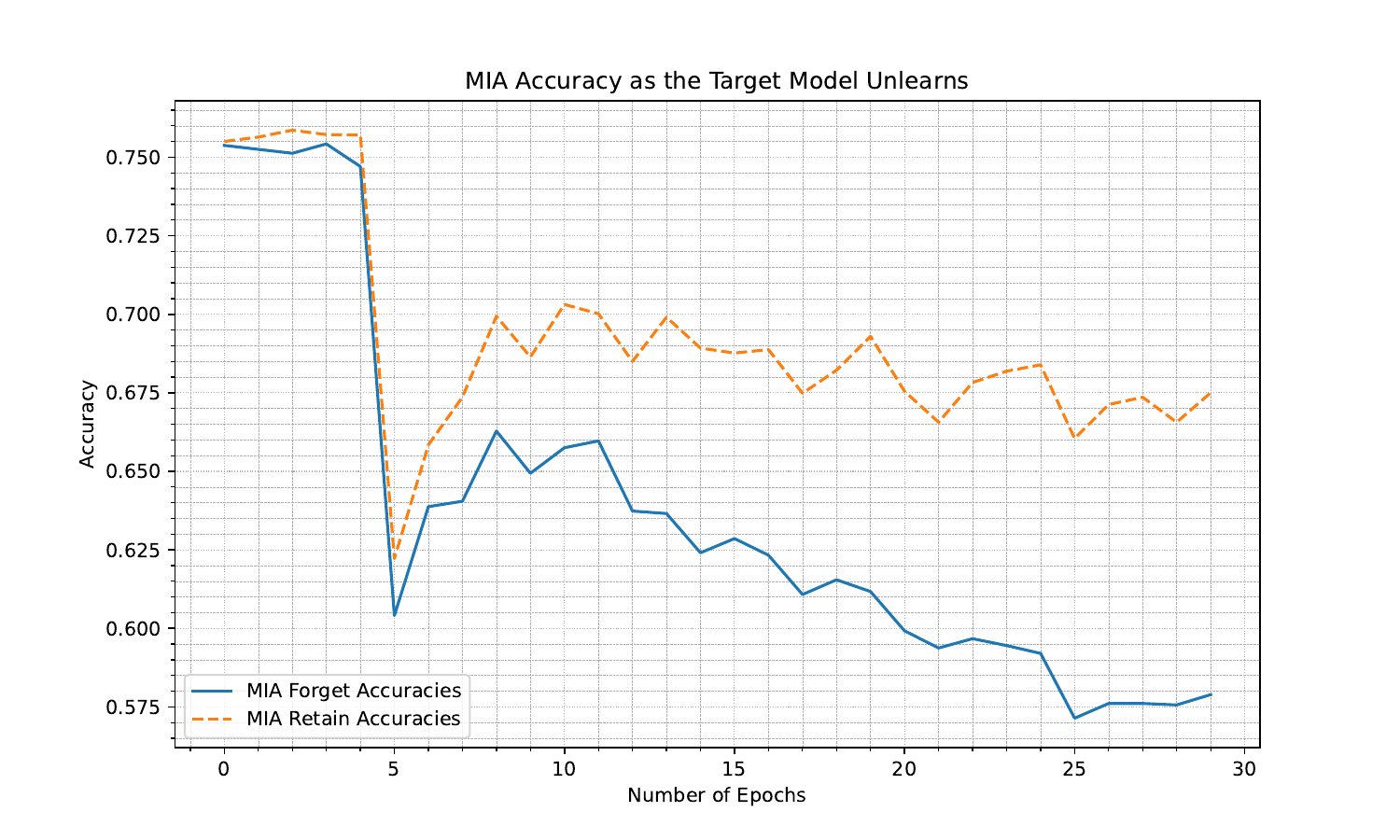}
        \caption{MIA Forget and Retain Accuracy \textbf{scrub} for \textbf{Cifar}}
        \label{fig:mia_accuracy_2a_scrub}
    \end{subfigure}
    \hfill
    \begin{subfigure}[b]{0.49\textwidth}
        \centering
        \includegraphics[width=\textwidth]{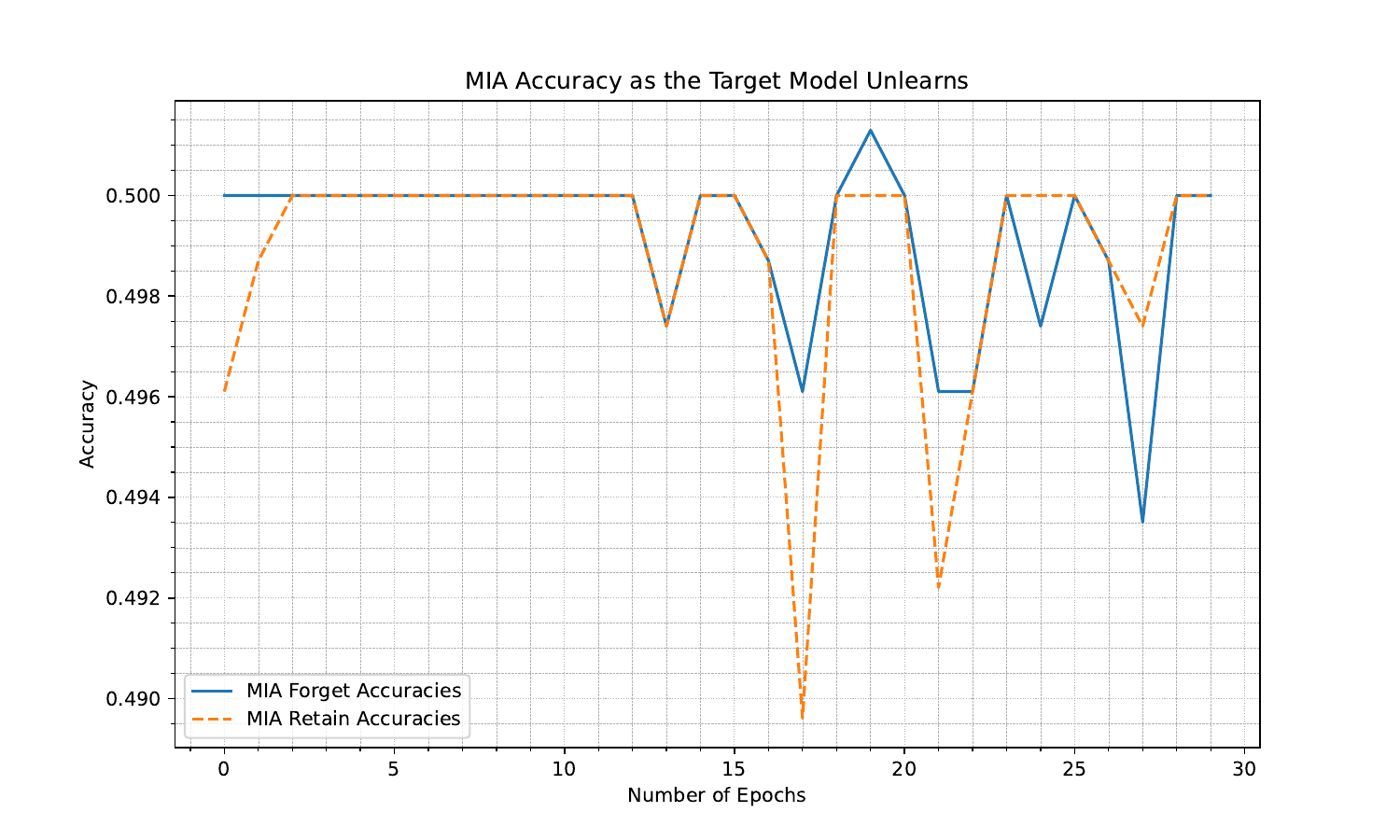}
        \caption{MIA Forget and Retain Accuracy \textbf{scrub} for \textbf{Mufac}}
        \label{fig:mia_accuracy_2b_scrub}
    \end{subfigure}

    \medskip 
    
    \begin{subfigure}[b]{0.49\textwidth}
        \centering
        \includegraphics[width=\textwidth]{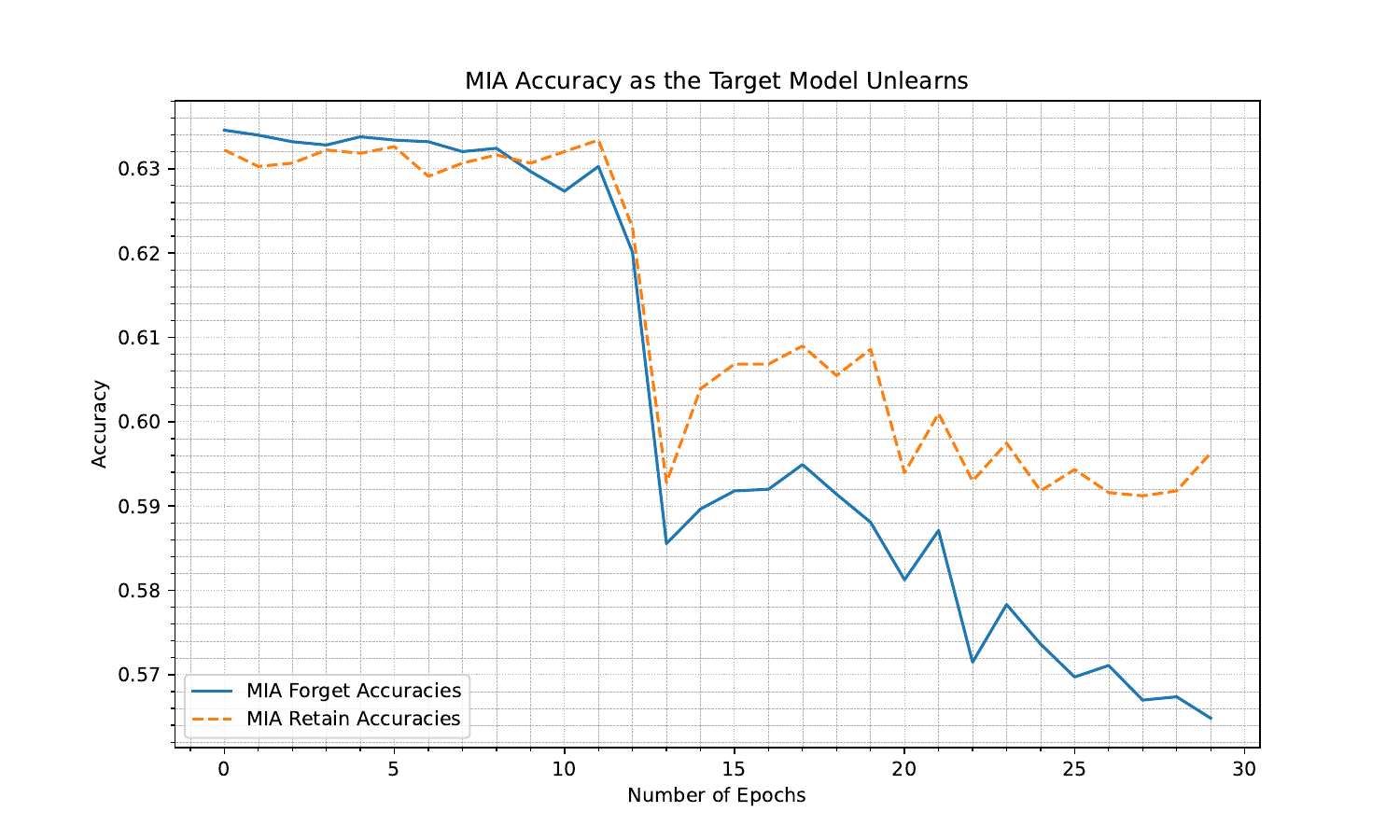}
        \caption{MIA Forget and Retain Accuracy \textbf{scrub} for \textbf{Purchase}}
        \label{fig:mia_accuracy_2c_scrub}
    \end{subfigure}
    \hfill
    \begin{subfigure}[b]{0.49\textwidth}
        \centering
        \includegraphics[width=\textwidth]{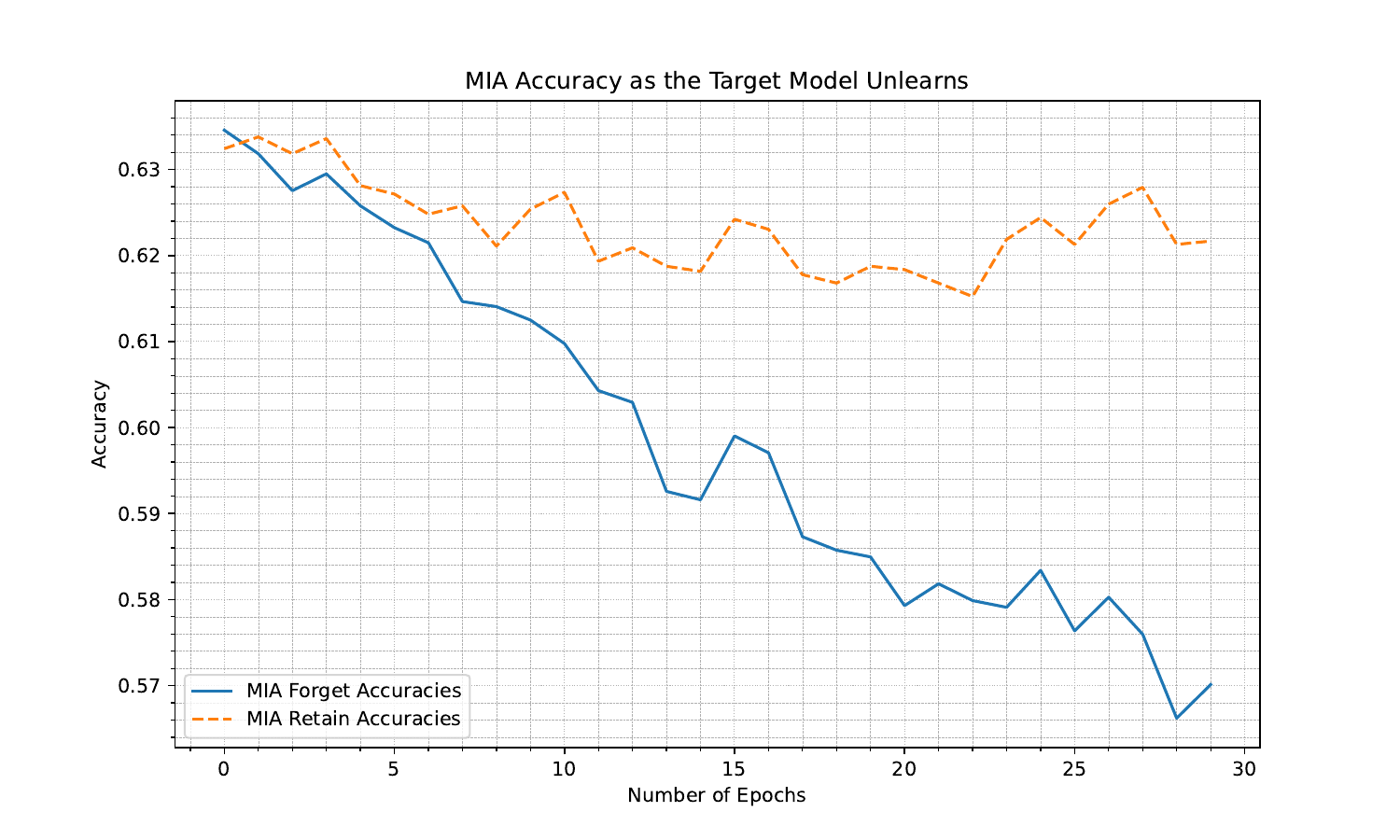}
        \caption{MIA Forget and Retain Accuracy \textbf{scrub} for \textbf{Texas}}
        \label{fig:mia_accuracy_2d_scrub}
    \end{subfigure}
    
    \caption{These graphs illustrate the Membership Inference Attack (MIA) performance at each epoch of the unlearning process. This specific set of results corresponds to the \textbf{second group} of experiments, utilizing the scrub unlearning algorithm.}
    \label{fig:scrub_2_group}
\end{figure*}

In the CIFAR-10 experiment (Figure~\ref{fig:mia_accuracy_2a_scrub}), the Scrub algorithm demonstrates a notable ``plunge'' in MIA accuracy during the initial five epochs. Both the Forget and Retain accuracies drop sharply from a baseline of 0.75, with the Forget accuracy reaching a local minimum near 0.60. Following this initial phase, the Retain accuracy recovers and stabilizes around 0.675, while the Forget accuracy continues a steady, gradual decline toward 0.575. This represents an effective unlearning trajectory, as the algorithm successfully creates a consistent gap between the membership signals of the two sets after the initial instability.

The results for the MUFAC dataset (Figure~\ref{fig:mia_accuracy_2b_scrub}) highlight a scenario where the MIA is initially ineffective, with both accuracies hovering at a baseline of 0.50 (random chance). However, as the unlearning process continues, the model exhibits significant instability. We observe frequent ``spikes'' and fluctuations in both metrics, particularly between epochs 15 and 30. Despite these oscillations, the accuracies generally return to the 0.50 baseline. This suggests that while Scrub does not increase overall privacy risk here, it introduces transient structural changes that temporarily make certain samples more identifiable before they are re-masked.

For the Purchase-100 dataset (Figure~\ref{fig:mia_accuracy_2c_scrub}), the Scrub algorithm shows a distinct ``delayed reaction.'' For the first 12 epochs, both MIA Forget and Retain accuracies remain high and stable at approximately 0.635. At epoch 13, there is a sudden and significant drop in both metrics. While the Retain accuracy partially recovers to approximately 0.595, the Forget accuracy continues to decline, ending near 0.565. This indicates that the unlearning mechanism for this architecture requires a specific threshold of iterations before the membership signal is successfully disrupted.

The Texas-100 results (Figure~\ref{fig:mia_accuracy_2d_scrub}) demonstrate a more linear and ``surgical'' unlearning process compared to the other datasets in this group. We observe a consistent downward trend for both metrics from the outset. Crucially, the MIA Forget accuracy declines at a steeper rate than the Retain accuracy throughout the 30 epochs. By the end of the process, the Forget accuracy reaches a low of 0.57, while the Retain accuracy remains higher at roughly 0.62. This suggests that for intermediate architectures on the Texas dataset, Scrub effectively prioritizes the removal of the forget set's membership information while preserving a stronger signal for the retained data.

For this second group of models, the Scrub algorithm appears more stable than it was with simpler architectures, particularly in the Texas-100 and CIFAR-10 cases. The most prominent feature of this group is the initial volatility (CIFAR-10) and delayed disruption (Purchase-100), suggesting that the timing of the ``unlearning shock'' is highly dependent on the dataset's characteristics. Unlike the first group, we do not see a significant ``rebound'' where accuracies increase at the end, implying that higher model complexity may help in maintaining the privacy gains achieved during the unlearning process.

\subsubsection{Third Group}

\begin{figure*}[t]
    \centering
    \begin{subfigure}[b]{0.49\textwidth}
        \centering
        \includegraphics[width=\textwidth]{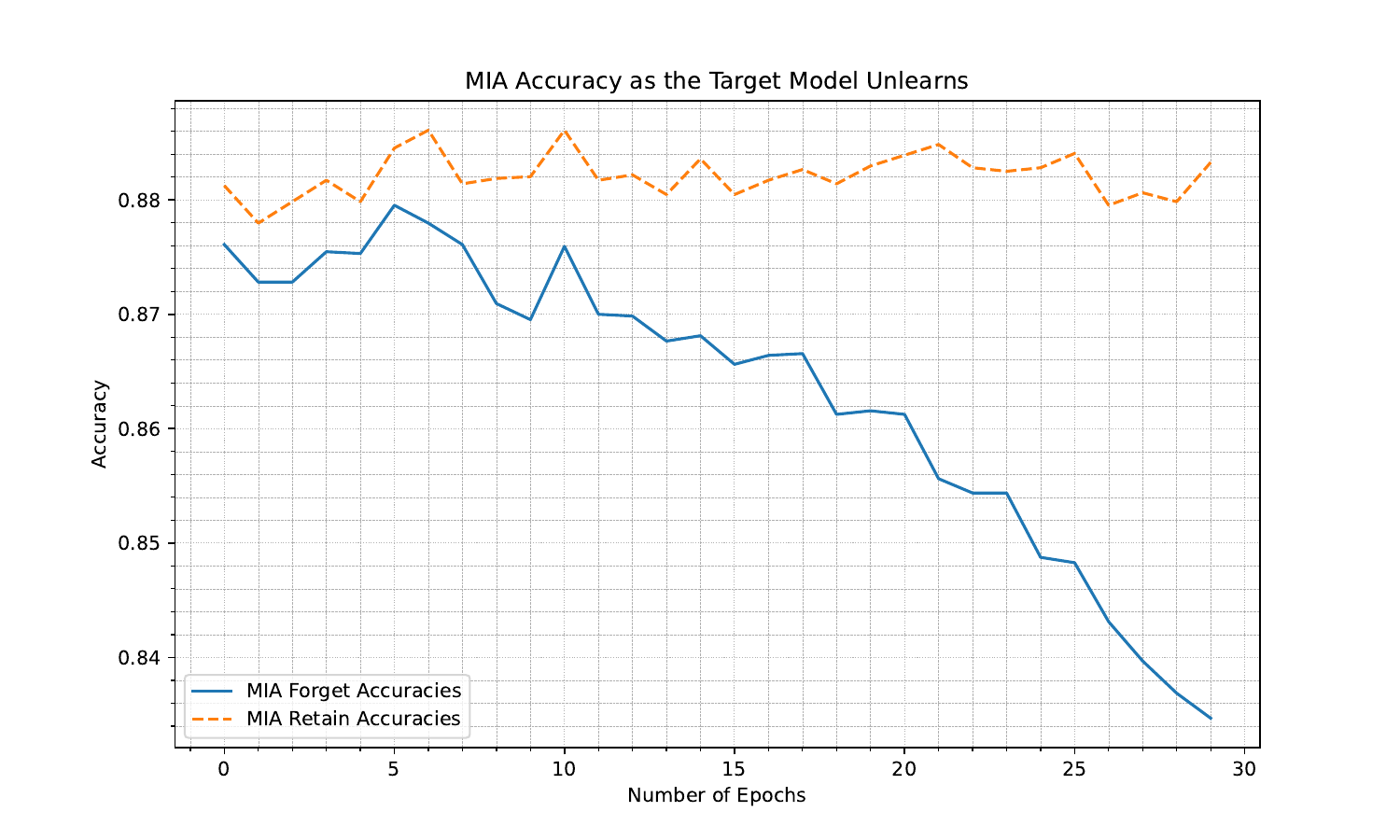}
        \caption{MIA Forget and Retain Accuracy \textbf{scrub} for \textbf{Cifar}}
        \label{fig:mia_accuracy_3a_scrub}
    \end{subfigure}
    \hfill
    \begin{subfigure}[b]{0.49\textwidth}
        \centering
        \includegraphics[width=\textwidth]{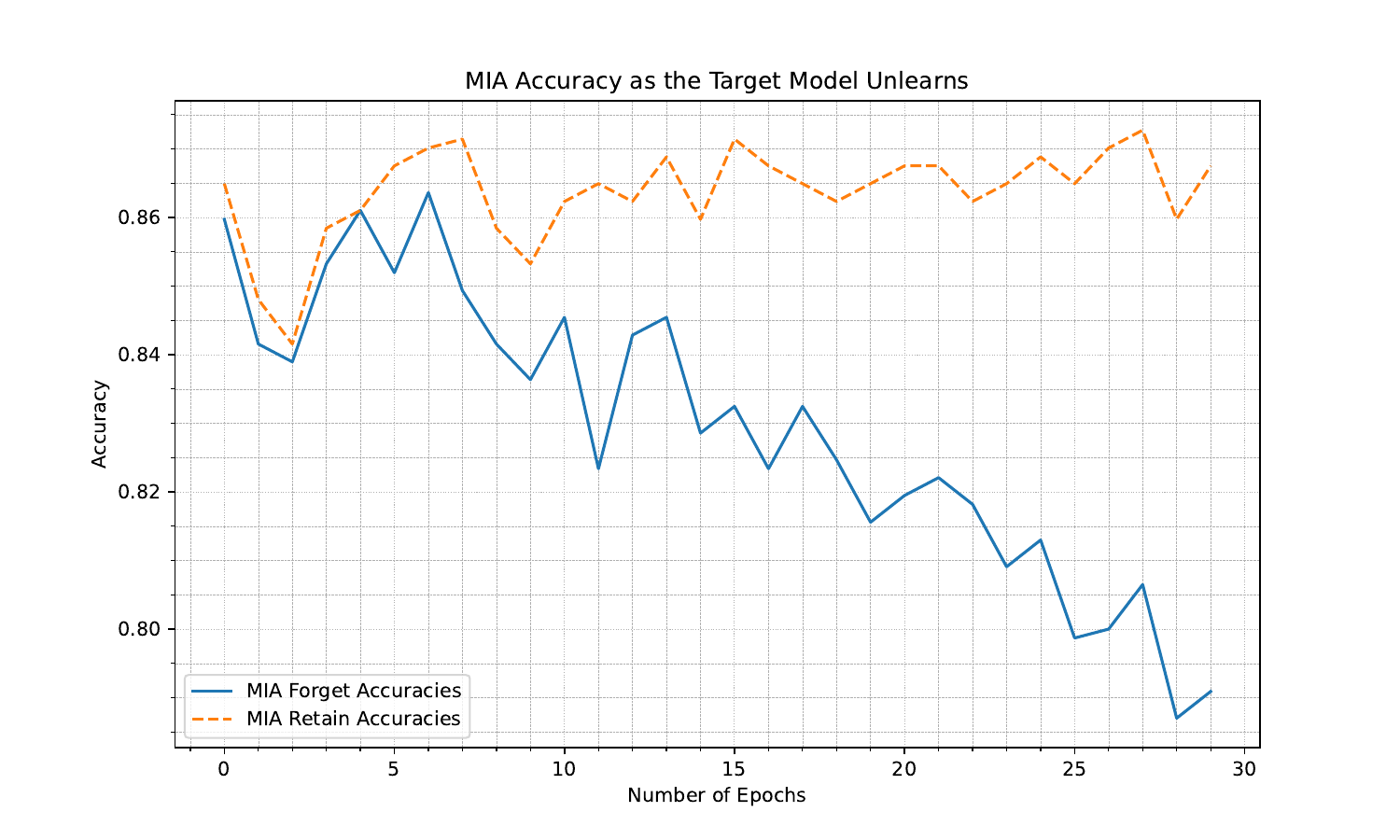}
        \caption{MIA Forget and Retain Accuracy \textbf{scrub} for \textbf{Mufac}}
        \label{fig:mia_accuracy_3b_scrub}
    \end{subfigure}

    \medskip 
    
    \begin{subfigure}[b]{0.49\textwidth}
        \centering
        \includegraphics[width=\textwidth]{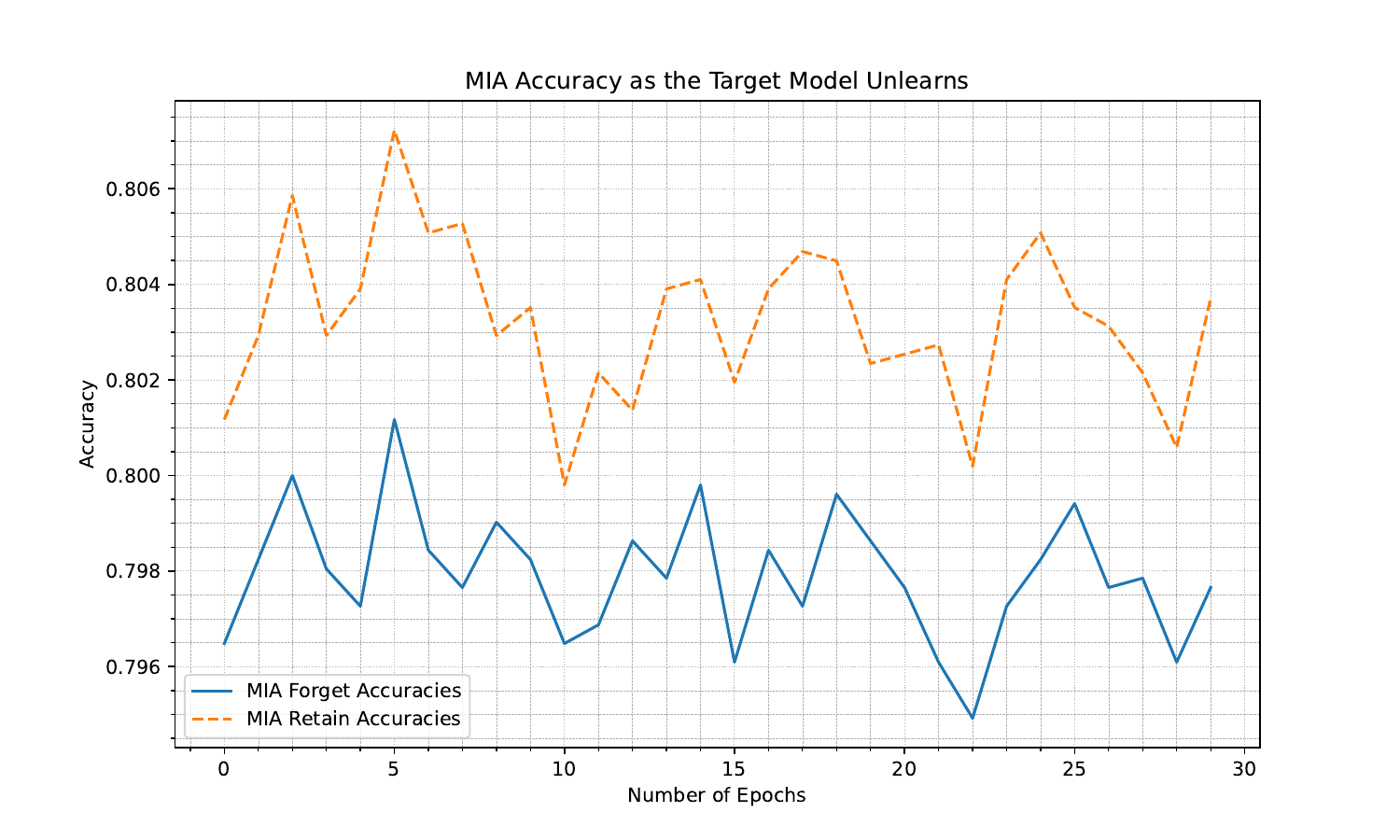}
        \caption{MIA Forget and Retain Accuracy \textbf{scrub} for \textbf{Purchase}}
        \label{fig:mia_accuracy_3c_scrub}
    \end{subfigure}
    \hfill
    \begin{subfigure}[b]{0.49\textwidth}
        \centering
        \includegraphics[width=\textwidth]{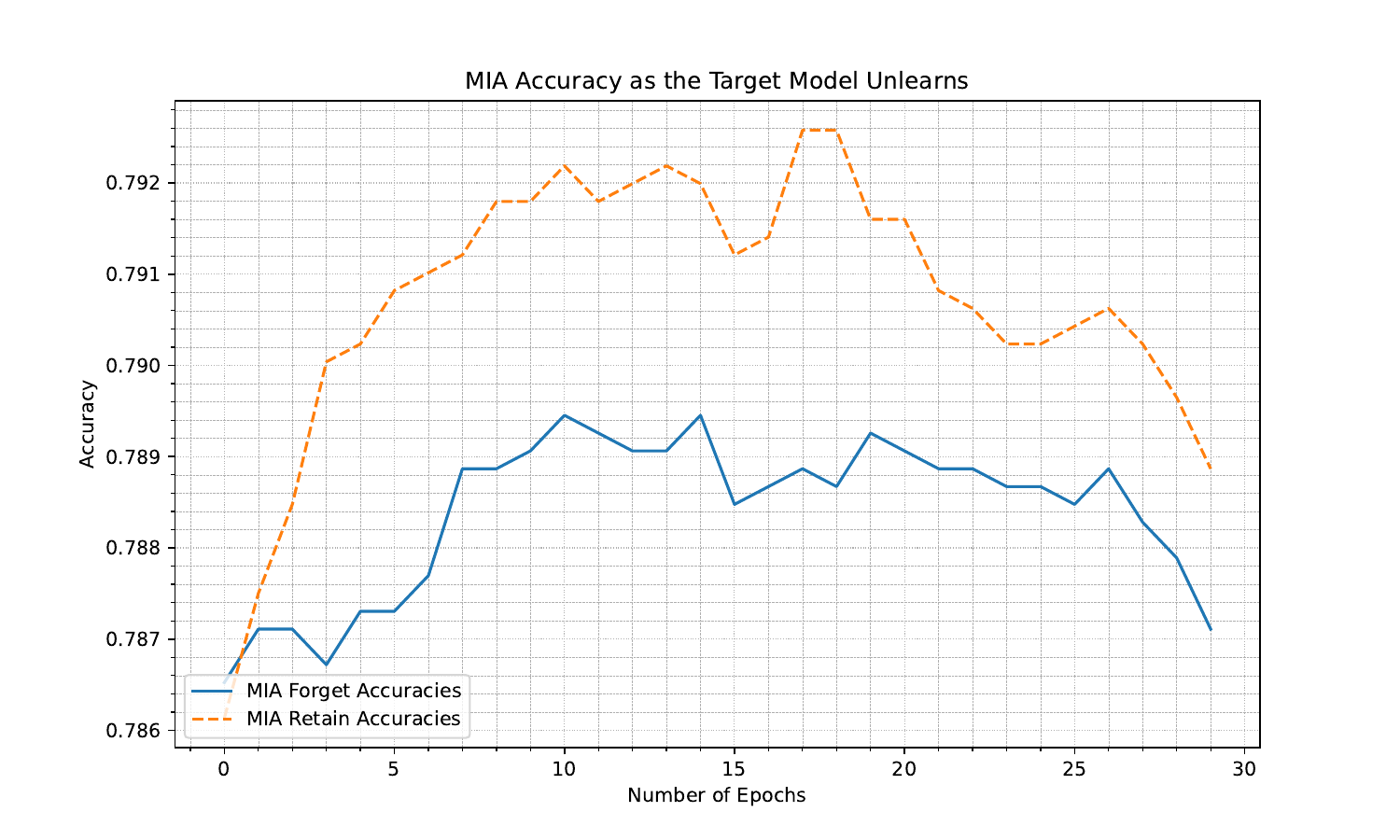}
        \caption{MIA Forget and Retain Accuracy \textbf{scrub} for \textbf{Texas}}
        \label{fig:mia_accuracy_3d_scrub}
    \end{subfigure}
    \caption{These graphs illustrate the Membership Inference Attack (MIA) performance at each epoch of the unlearning process. This specific set of results corresponds to the \textbf{third group} of experiments, utilizing the scrub unlearning algorithm.}
    \label{fig:scrub_3_group}
\end{figure*}

For the CIFAR-10 (Figure~\ref{fig:mia_accuracy_3a_scrub}) dataset using complex architectures, the Scrub algorithm demonstrates a highly effective and stable unlearning profile. While the MIA Retain Accuracy remains remarkably consistent, fluctuating minimally around 0.88, the MIA Forget Accuracy exhibits a steady and significant decline from 0.875 down to approximately 0.835 by epoch 30. This clear divergence indicates that for high-capacity models, Scrub can successfully target and diminish the membership signal of the forget set without compromising the privacy profile of the retained data.

In the MUFAC experiment (Figure~\ref{fig:mia_accuracy_3b_scrub}), we observe a similar but more pronounced downward trajectory for the forget set. The MIA Forget Accuracy starts near 0.86 and decreases consistently, reaching a low of approximately 0.79 by the end of the unlearning process. Meanwhile, the MIA Retain Accuracy maintains a high baseline around 0.86, showing only minor oscillations. This represents a successful unlearning scenario where the ``forgetting'' is surgical, specifically reducing the identifiable traces of the target samples in a high-dimensional model space.

The results for the Purchase-100 dataset (Figure~\ref{fig:mia_accuracy_3c_scrub}) reveal a high degree of volatility but a relatively stable overall privacy baseline. Both the forget and retain accuracies oscillate significantly between 0.790 and 0.806. Despite these fluctuations, there is no significant downward trend for either metric, with the accuracies ending near their initial starting points. This suggests that for complex architectures on Purchase-100, the Scrub algorithm struggles to find a consistent gradient to effectively reduce the membership signal, resulting in a model that remains equally susceptible to MIA throughout the process.

The Texas-100 dataset (Figure~\ref{fig:mia_accuracy_3d_scrub}) presents a paradoxical result. Following the start of unlearning, both MIA Forget and Retain accuracies actually increase during the first 15 epochs, rising from 0.786 to nearly 0.792. In the latter half of the process, both metrics begin to decline, eventually returning to near-baseline levels. This ``bell-curve behavior suggests that for complex Texas-100 models, the Scrub algorithm initially makes the data points more identifiable—likely due to model updates creating distinctive artifacts—before finally beginning to mask the membership signal in the final epochs.

In this final group, the Scrub algorithm shows its greatest strength in CIFAR-10 and MUFAC, where it achieves a clear separation between forget and retain membership signals. However, the Texas-100 results serve as a warning: in complex architectures, the unlearning process can initially increase privacy risk before reducing it. Furthermore, the resistance observed in Purchase-100 indicates that even advanced algorithms like Scrub are not universally effective and are highly dependent on the underlying data distribution and model capacity.

\subsection{SFTC Algorithm}

\subsubsection{First Group}

\begin{figure*}[t]
    \centering
    \begin{subfigure}[b]{0.49\textwidth}
        \centering
        \includegraphics[width=\textwidth]{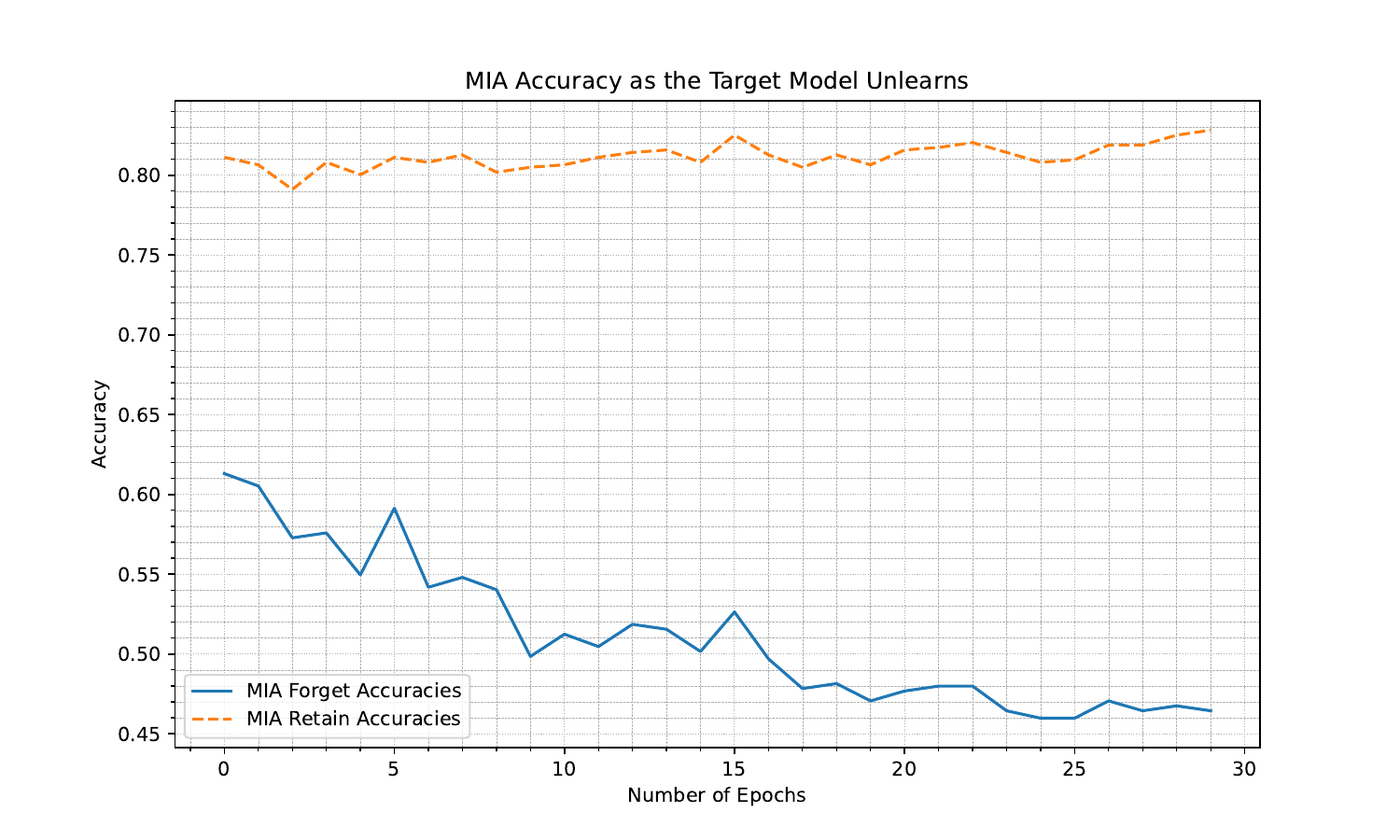}
        \caption{MIA Forget and Retain Accuracy \textbf{SFTC} for \textbf{Cifar}}
        \label{fig:mia_accuracy_1a_sftc}
    \end{subfigure}
    \hfill
    \begin{subfigure}[b]{0.49\textwidth}
        \centering
        \includegraphics[width=\textwidth]{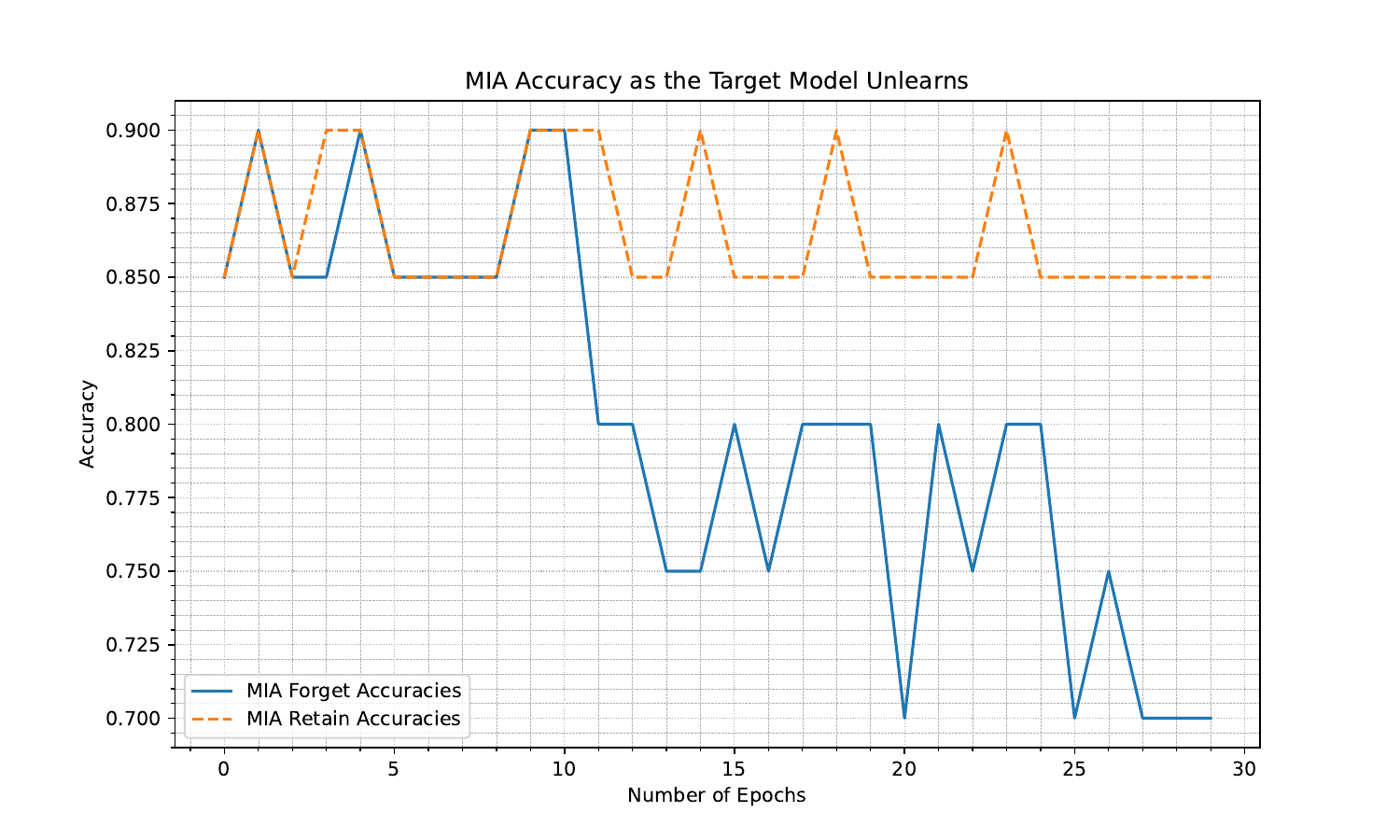}
        \caption{MIA Forget and Retain Accuracy \textbf{SFTC} for \textbf{Mufac}}
        \label{fig:mia_accuracy_1b_sftc}
    \end{subfigure}

    \medskip 
    
    \begin{subfigure}[b]{0.49\textwidth}
        \centering
        \includegraphics[width=\textwidth]{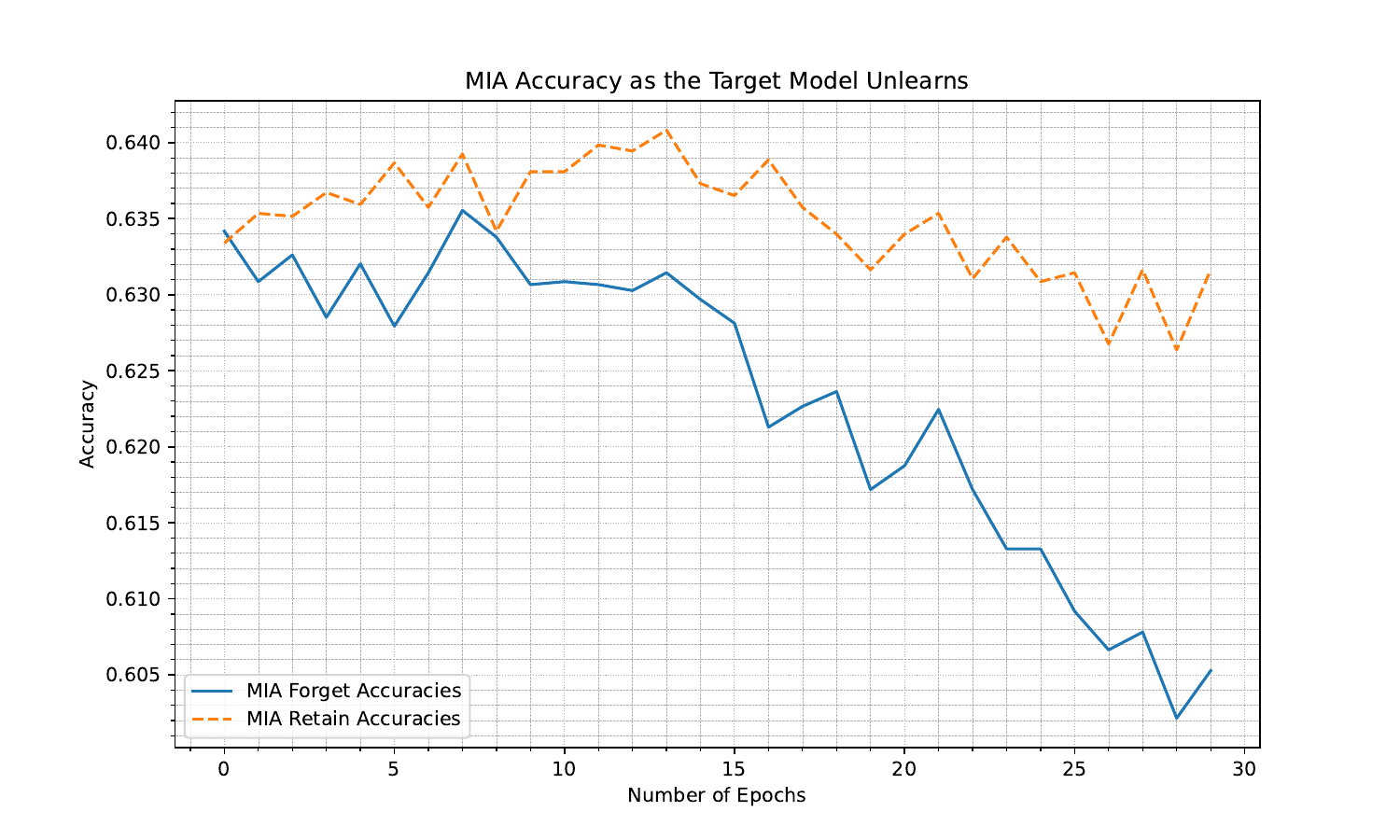}
        \caption{MIA Forget and Retain Accuracy \textbf{SFTC} for \textbf{Purchase}}
        \label{fig:mia_accuracy_1c_sftc}
    \end{subfigure}
    \hfill
    \begin{subfigure}[b]{0.49\textwidth}
        \centering
        \includegraphics[width=\textwidth]{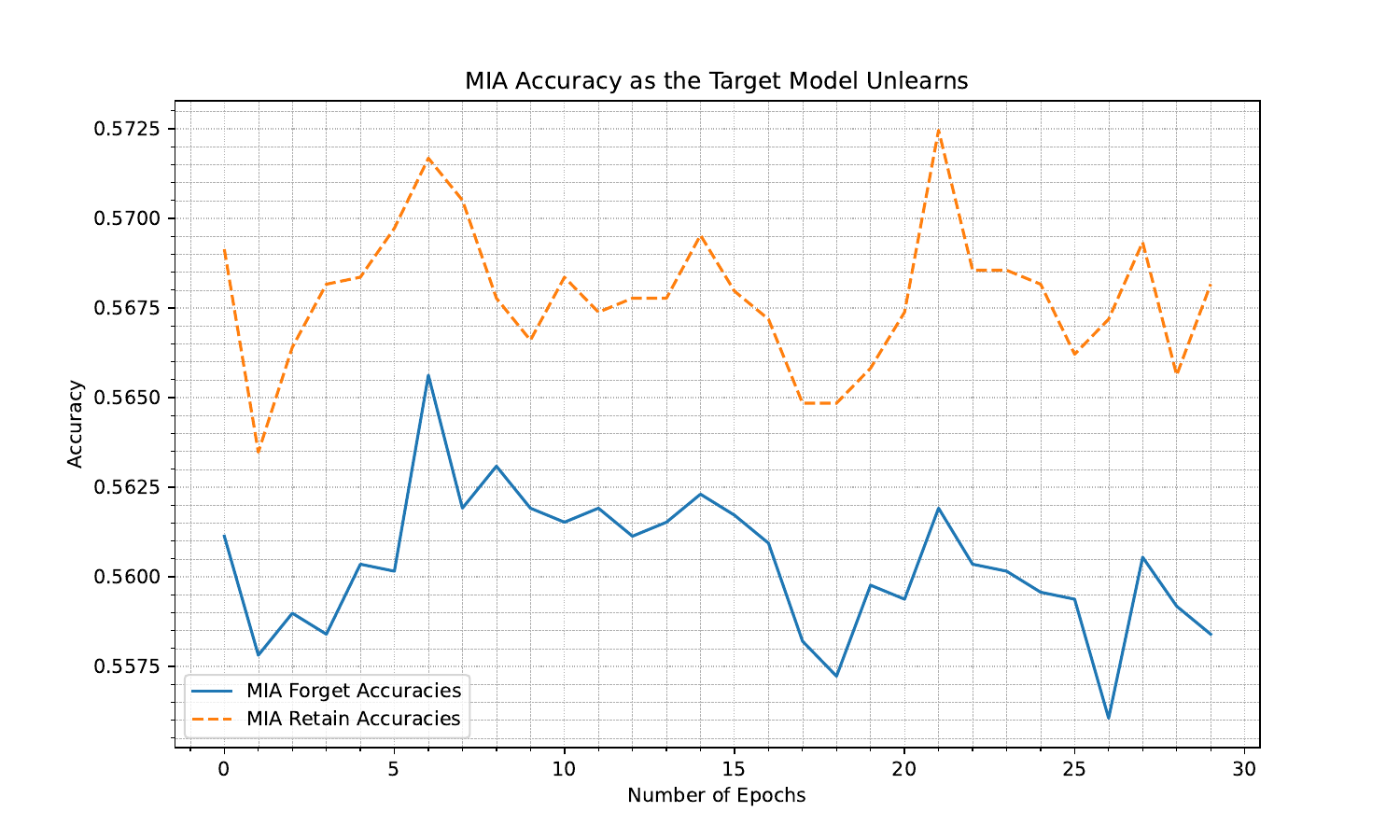}
        \caption{MIA Forget and Retain Accuracy \textbf{SFTC} for \textbf{Texas}}
        \label{fig:mia_accuracy_1d_sftc}
    \end{subfigure}
    
    \caption{These graphs illustrate the Membership Inference Attack (MIA) performance at each epoch of the unlearning process. This specific set of results corresponds to the \textbf{first group} of experiments, utilizing the sftc unlearning algorithm.}
    \label{fig:sftc_1_group}
\end{figure*}

For the CIFAR-10 dataset (Figure~\ref{fig:mia_accuracy_1a_sftc}), the SFTC algorithm demonstrates high precision in selectively targeting the forget set. Initially, the MIA Forget Accuracy starts near 0.62, but it experiences a sharp and consistent decline, reaching a low of approximately 0.46 by epoch 30—falling below the threshold of random guessing. Crucially, the MIA Retain Accuracy remains exceptionally stable throughout the process, maintaining a baseline above 0.80. This suggests that SFTC effectively removes membership traces of the forget set while preserving the model's structural knowledge of the retained data.

The MUFAC results (Figure~\ref{fig:mia_accuracy_1b_sftc}) highlight a highly volatile unlearning process characterized by significant oscillations. While both metrics start at a high baseline of 0.85, the MIA Forget Accuracy undergoes dramatic ``spikes,'' momentarily dropping to 0.70 before rebounding. By the final epochs, it stabilizes significantly lower than its starting point, around 0.70. The MIA Retain Accuracy also fluctuates but stays generally higher, around 0.85. This indicates that while SFTC manages to reduce privacy risk, the unlearning path for MUFAC in simple models is unstable, likely due to the sensitive nature of the dataset's features.

In the Purchase-100 experiment (Figure~\ref{fig:mia_accuracy_1c_sftc}), we observe a steady downward trend for both the forget and retain sets. The MIA Forget Accuracy starts at 0.635 and concludes at approximately 0.605. However, the MIA Retain Accuracy also mirrors this decline, falling from 0.635 to roughly 0.625. While a gap between the two metrics eventually emerges, the simultaneous decline suggests that the SFTC algorithm has a ``thinning'' effect on the model's overall confidence, affecting both sets rather than performing a purely surgical removal of the forget set's identity.

Regarding the Texas-100 dataset (Figure~\ref{fig:mia_accuracy_1d_sftc}), the SFTC algorithm appears to struggle with providing a significant defense. Both the forget and retain accuracies remain highly coupled and volatile throughout the 30 epochs, hovering between 0.557 and 0.573. There is no distinct downward trajectory for the forget set; in fact, both accuracies end near their starting points. This lack of impact suggests that for simple architectures on Texas-100, the teacher-student dynamics of SFTC fail to effectively distinguish or ``scrub'' the membership signals, resulting in negligible privacy gains.

Comparing the SFTC algorithm to the previous Scrub and Negative Gradient methods on simple architectures reveals a key strength: selectivity. In datasets like CIFAR-10, SFTC achieves the most surgical separation between forget and retain signals seen thus far. However, its performance on Texas-100 remains a point of concern, showing that gradient-based and teacher-student unlearning both face difficulties when the initial membership signal is weak or the data distribution is high-dimensional.

\subsubsection{Second Group}

\begin{figure*}[t]
    \centering
    \begin{subfigure}[b]{0.49\textwidth}
        \centering
        \includegraphics[width=\textwidth]{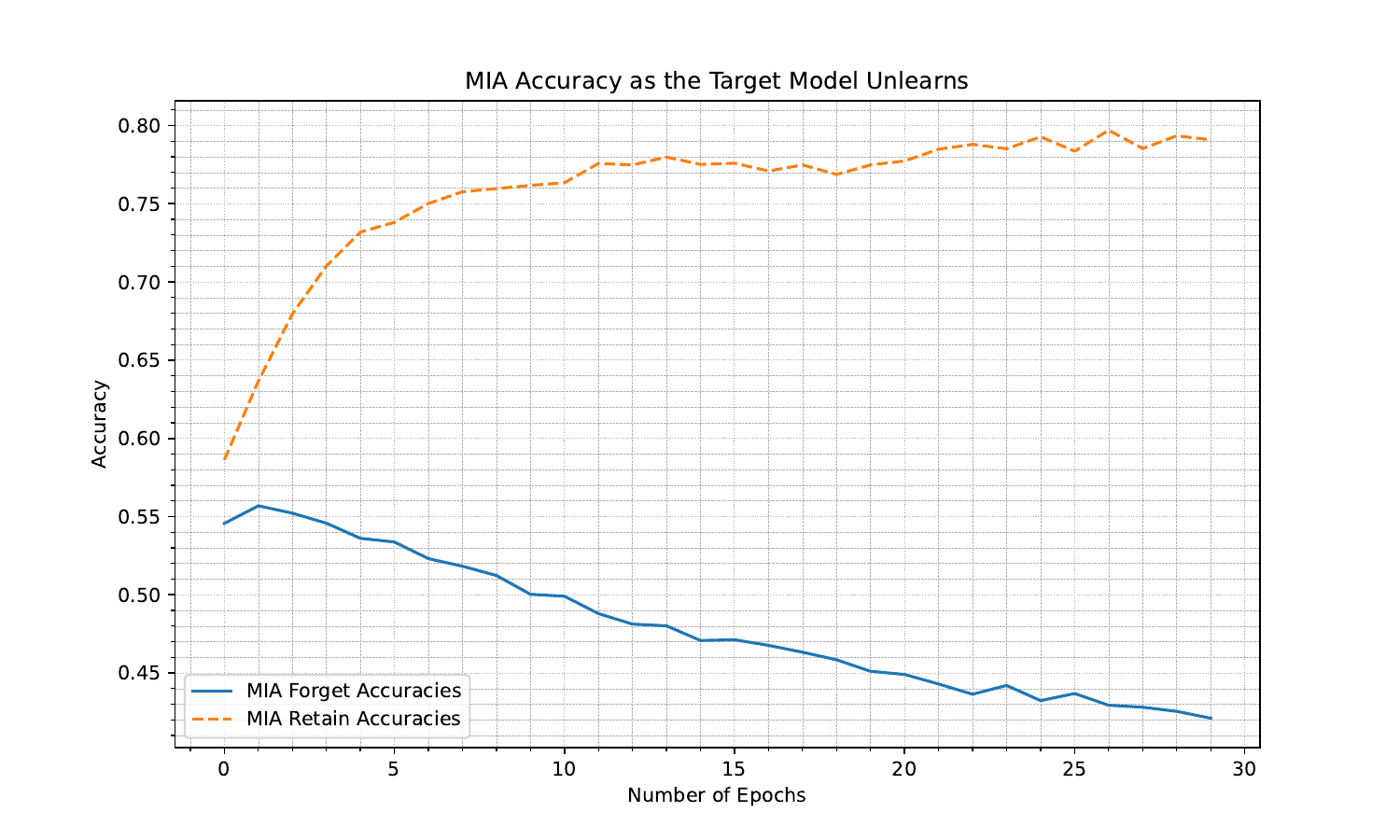}
        \caption{MIA Forget and Retain Accuracy \textbf{SFTC} for \textbf{Cifar}}
        \label{fig:mia_accuracy_2a_sftc}
    \end{subfigure}
    \hfill
    \begin{subfigure}[b]{0.49\textwidth}
        \centering
        \includegraphics[width=\textwidth]{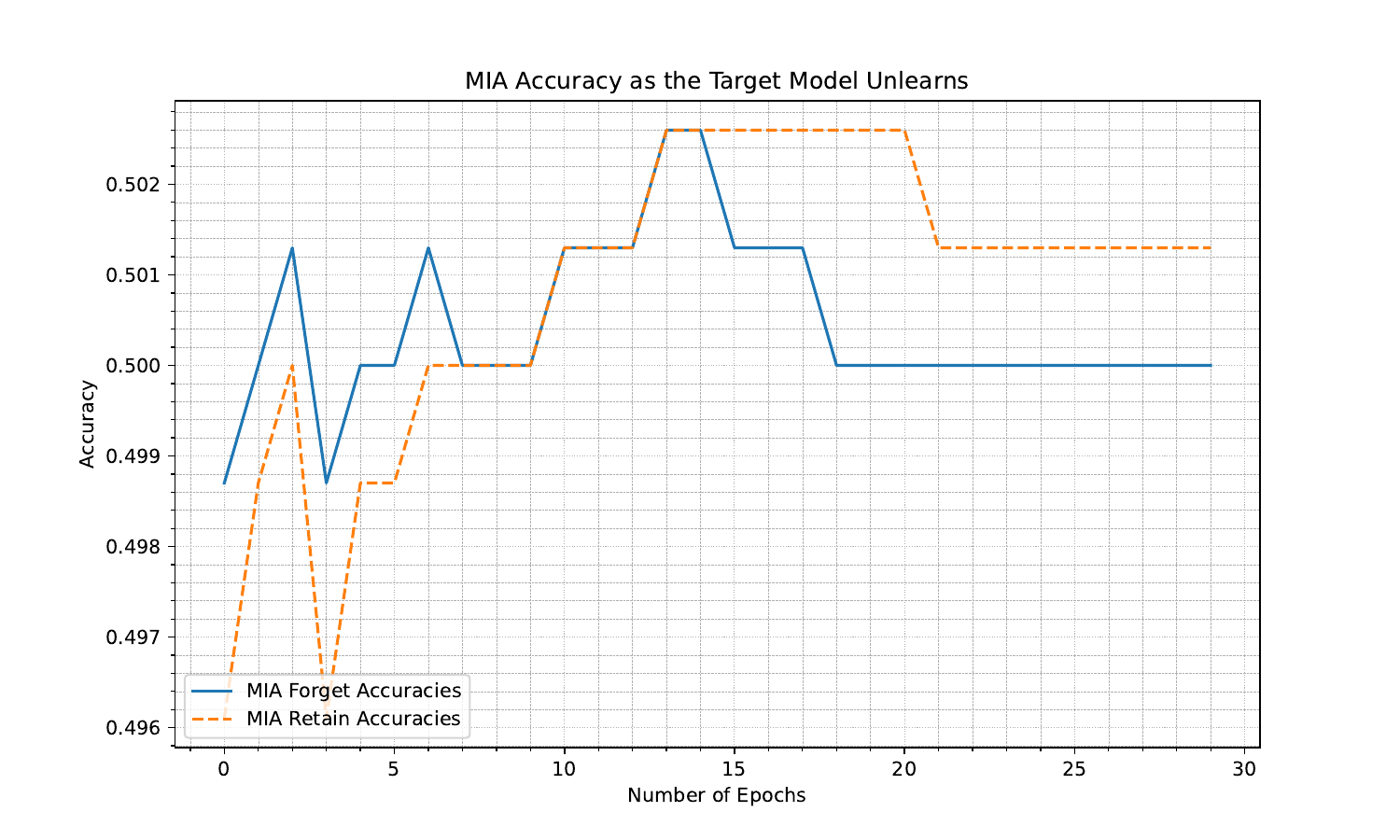}
        \caption{MIA Forget and Retain Accuracy \textbf{SFTC} for \textbf{Mufac}}
        \label{fig:mia_accuracy_2b_sftc}
    \end{subfigure}

    \medskip 
    
    \begin{subfigure}[b]{0.49\textwidth}
        \centering
        \includegraphics[width=\textwidth]{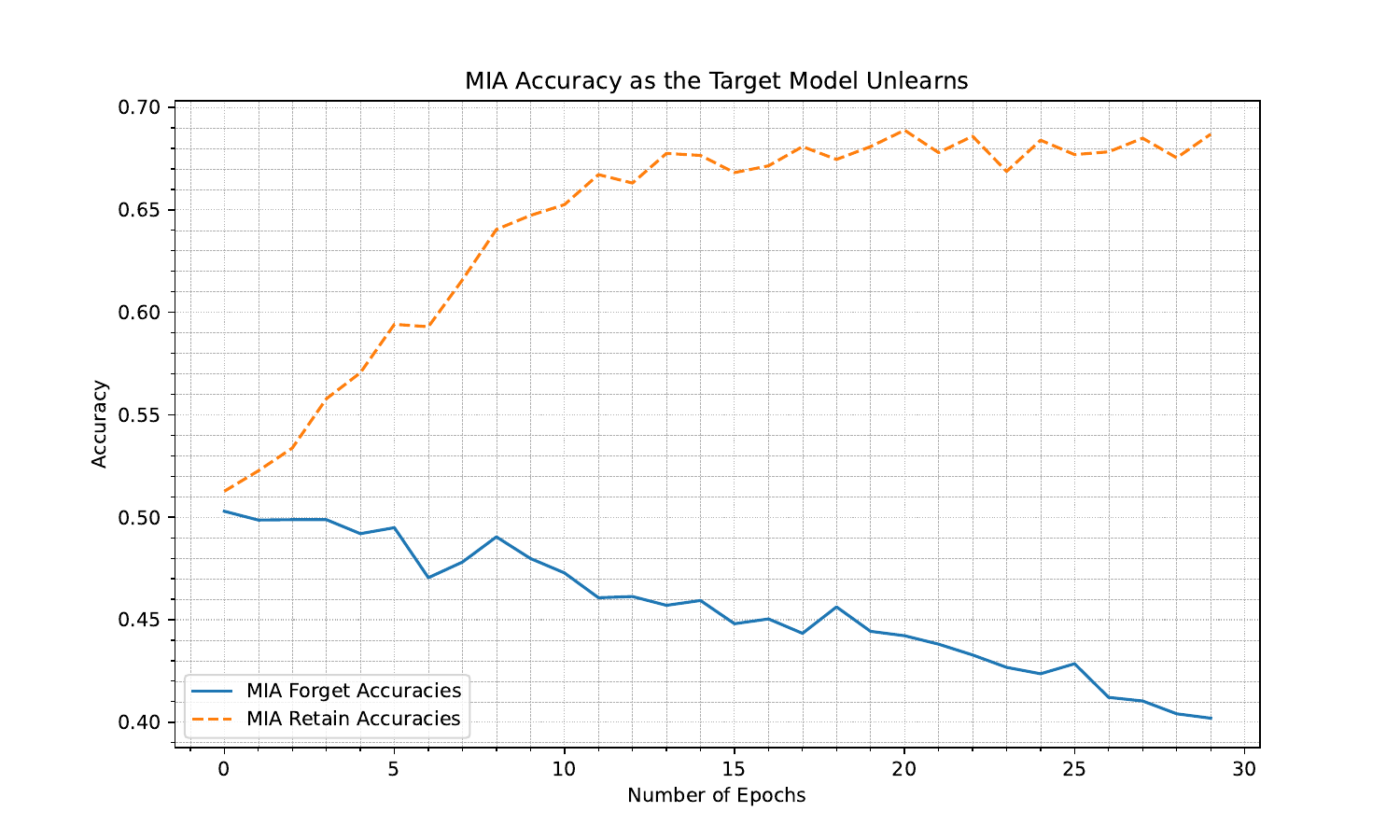}
        \caption{MIA Forget and Retain Accuracy \textbf{SFTC} for \textbf{Purchase}}
        \label{fig:mia_accuracy_2c_sftc}
    \end{subfigure}
    \hfill
    \begin{subfigure}[b]{0.49\textwidth}
        \centering
        \includegraphics[width=\textwidth]{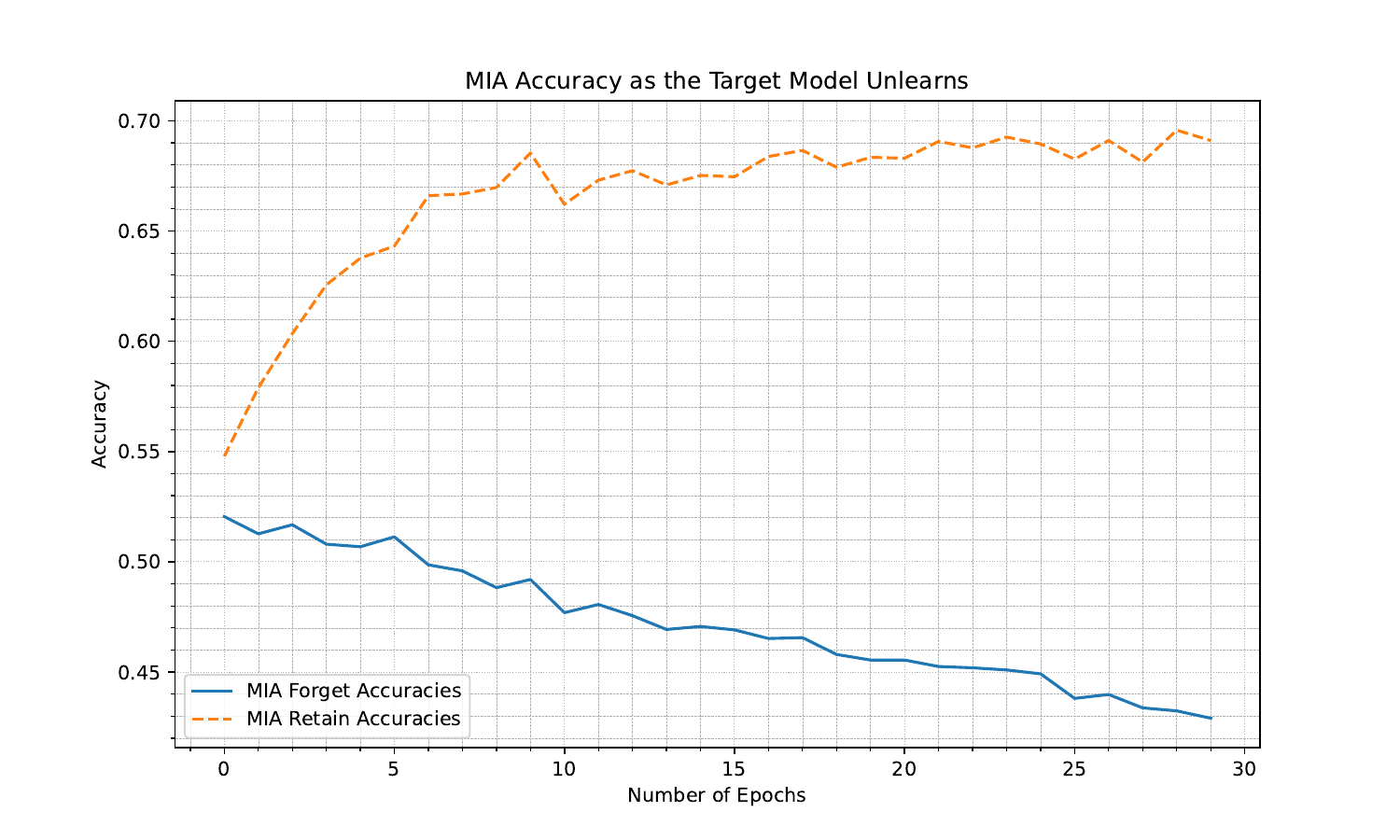}
        \caption{MIA Forget and Retain Accuracy \textbf{SFTC} for \textbf{Texas}}
        \label{fig:mia_accuracy_2d_sftc}
    \end{subfigure}
    
    \caption{These graphs illustrate the Membership Inference Attack (MIA) performance at each epoch of the unlearning process. This specific set of results corresponds to the \textbf{second group} of experiments, utilizing the sftc unlearning algorithm.}
    \label{fig:sftc_2_group}
\end{figure*}

In the CIFAR-10 scenario  (Figure~\ref{fig:mia_accuracy_2a_sftc}), the SFTC algorithm exhibits a highly effective unlearning trajectory. The MIA Forget Accuracy starts at approximately 0.55 and undergoes a steady, consistent decline, reaching a final accuracy of 0.42 by epoch 30. Simultaneously, the MIA Retain Accuracy shows a significant upward trend, rising from 0.57 to approximately 0.80. This suggests that as the model selectively forgets the target data, its generalization and membership signal for the retained data are strengthened. This raise the privacy for the retain data.

For the MUFAC dataset(Figure~\ref{fig:mia_accuracy_2b_sftc}), the MIA attack remains largely ineffective, with accuracies hovering near the random guessing baseline of 0.50. We observe minor fluctuations where both the forget and retain accuracies rise slightly to 0.502 before stabilizing back at 0.50. This indicates that for this intermediate architecture, the membership signals are naturally well-masked, and the SFTC unlearning process maintains this high level of privacy without introducing new identifiable artifacts.

The results for Purchase-100 (Figure~\ref{fig:mia_accuracy_2c_sftc}) demonstrate a clear divergent trend. The MIA Forget Accuracy begins at 0.50 and declines steadily to approximately 0.40. In contrast, the MIA Retain Accuracy increases from 0.50 to nearly 0.70. This pattern is highly desirable in unlearning applications; it demonstrates that the SFTC algorithm is successfully removing the specific membership traces of the forget set while the model simultaneously becomes more ``certain'' about its retained training members.

The Texas-100 dataset (Figure~\ref{fig:mia_accuracy_2d_sftc}) follows a nearly identical trajectory to Purchase-100. We observe the MIA Forget Accuracy dropping from an initial 0.52 down to 0.43, while the MIA Retain Accuracy climbs from 0.55 to roughly 0.70. This consistency across multiple datasets suggests that for intermediate-complexity models, the teacher-student dynamic of SFTC is robust at distinguishing between forget and retain samples, effectively ``pushing'' their membership probabilities in opposite directions.

In this second group, the SFTC algorithm it show better performance on the unlearning. Unlike the previous algorithms (Negative Gradient and Scrub), SFTC consistently achieves a widening gap between forget and retain membership signals. The fact that the MIA Retain Accuracy increases while the MIA Forget Accuracy decreases suggests that the ``teacher'' model in the SFTC framework is successfully guiding the ``student'' to retain high-confidence features for the correct data while actively suppressing them for the forgotten data.

\subsubsection{Third Group}

\begin{figure*}[t]
    \centering
    \begin{subfigure}[b]{0.49\textwidth}
        \centering
        \includegraphics[width=\textwidth]{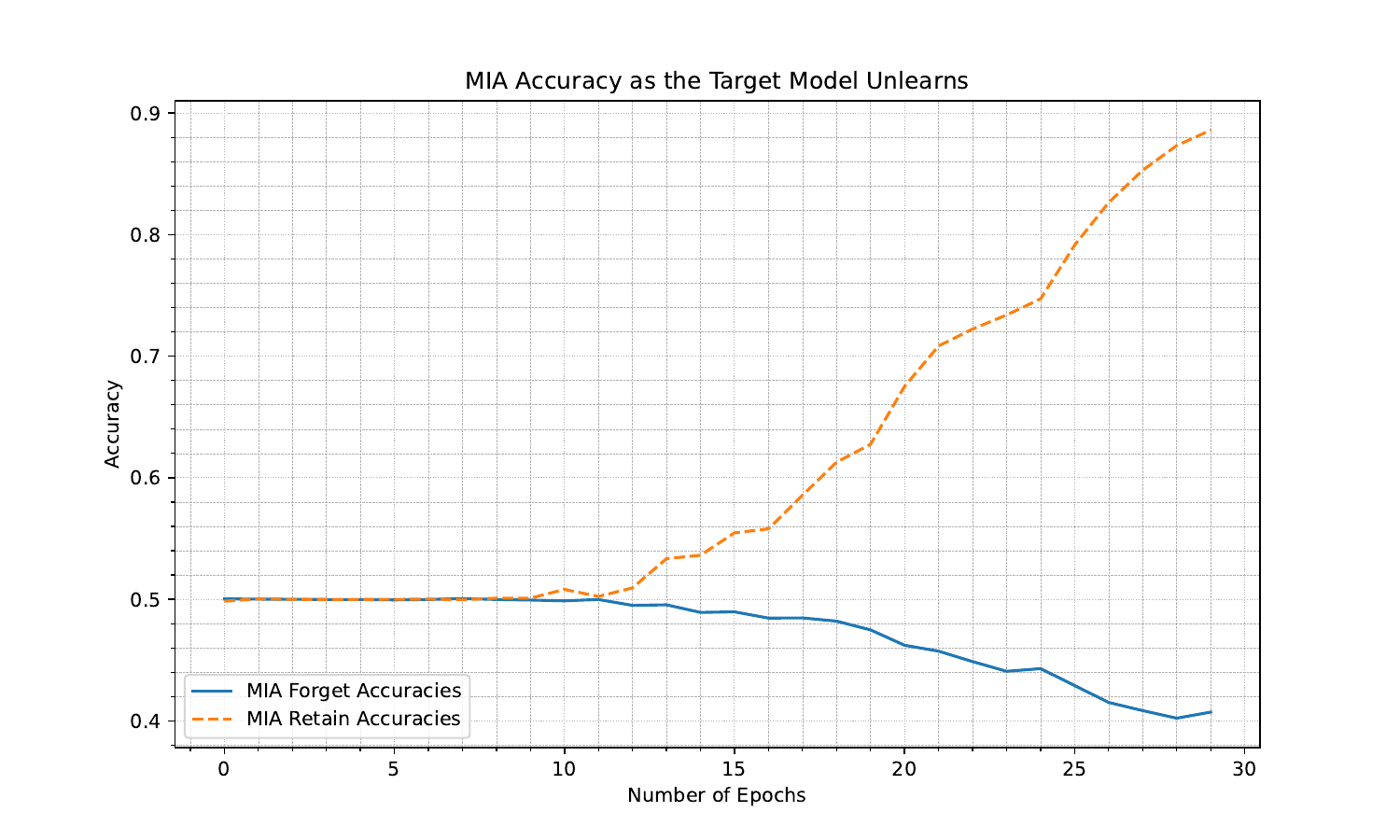}
        \caption{MIA Forget and Retain Accuracy \textbf{SFTC} for \textbf{Cifar}}
        \label{fig:mia_accuracy_3a_sftc}
    \end{subfigure}
    \hfill
    \begin{subfigure}[b]{0.49\textwidth}
        \centering
        \includegraphics[width=\textwidth]{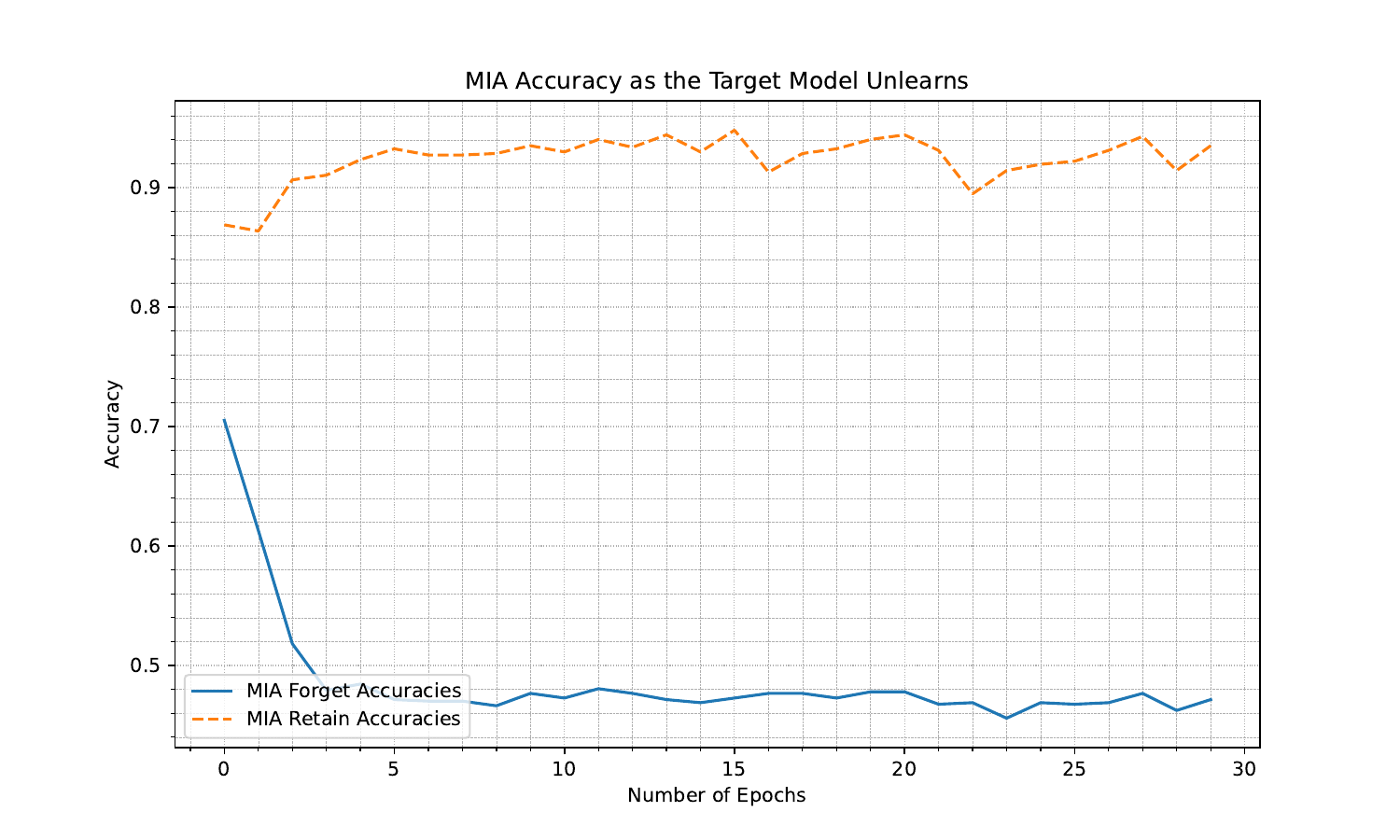}
        \caption{MIA Forget and Retain Accuracy \textbf{SFTC} for \textbf{Mufac}}
        \label{fig:mia_accuracy_3b_sftc}
    \end{subfigure}

    \medskip 
    
    \begin{subfigure}[b]{0.49\textwidth}
        \centering
        \includegraphics[width=\textwidth]{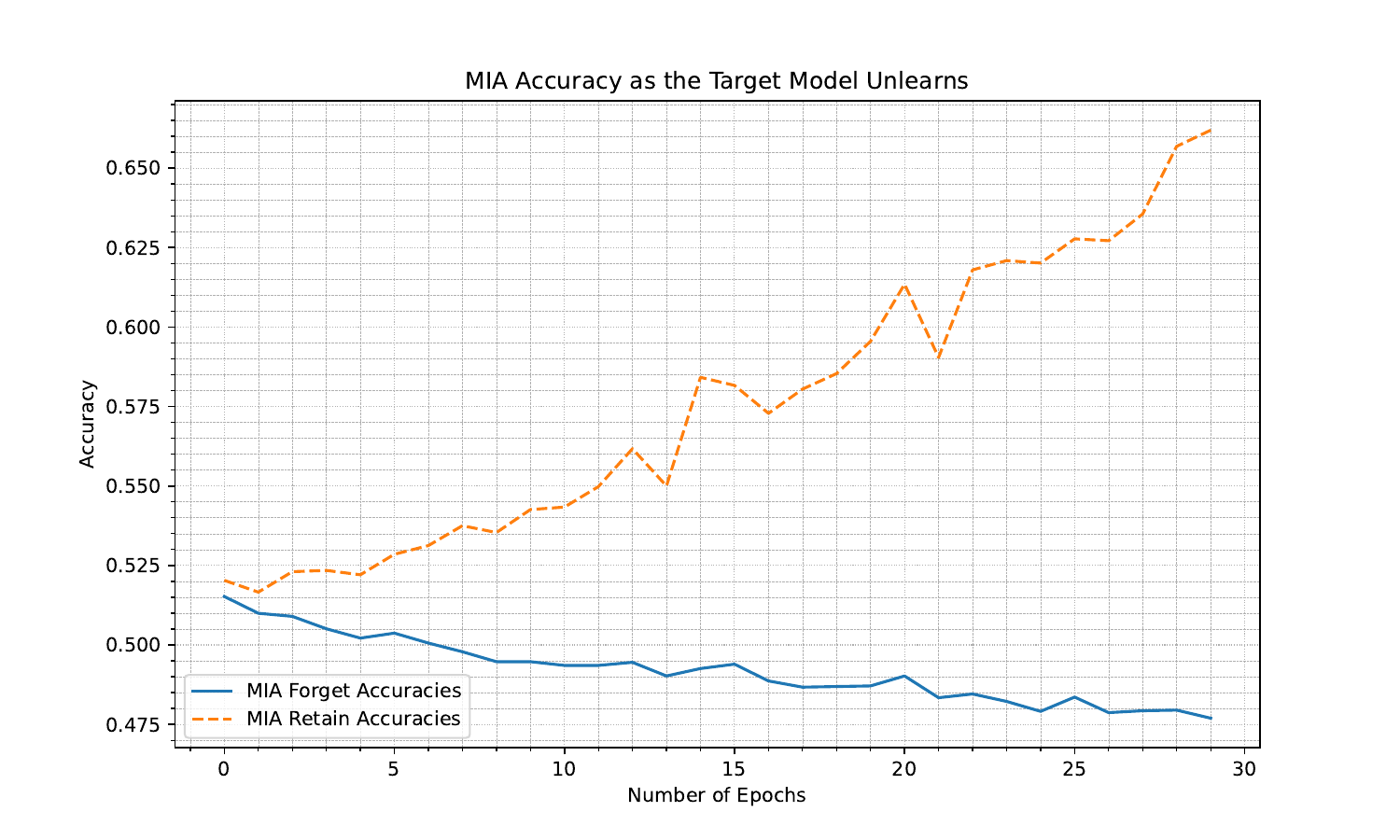}
        \caption{MIA Forget and Retain Accuracy \textbf{SFTC} for \textbf{Purchase}}
        \label{fig:mia_accuracy_3c_sftc}
    \end{subfigure}
    \hfill
    \begin{subfigure}[b]{0.49\textwidth}
        \centering
        \includegraphics[width=\textwidth]{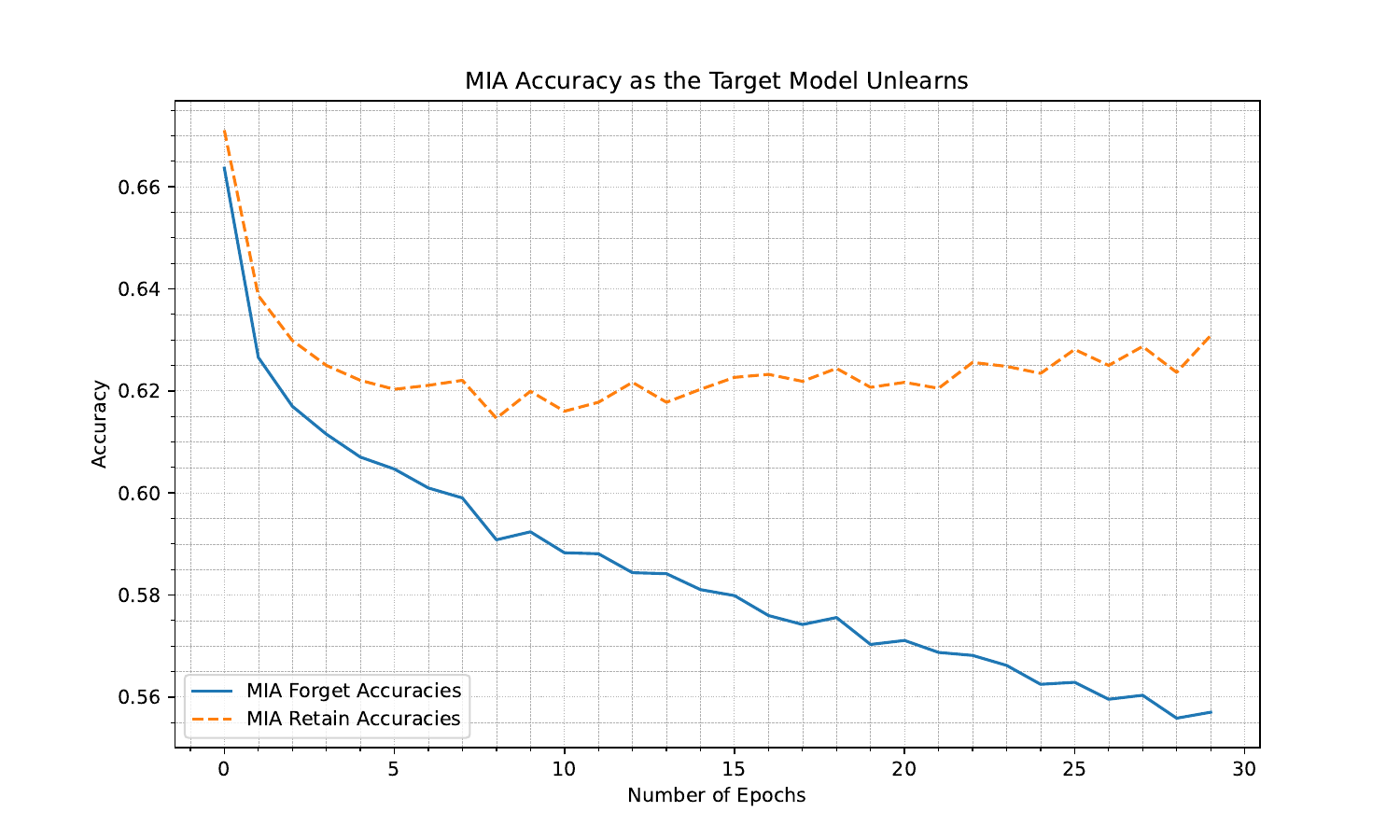}
        \caption{MIA Forget and Retain Accuracy \textbf{SFTC} for \textbf{Texas}}
        \label{fig:mia_accuracy_3d_sftc}
    \end{subfigure}
    \caption{These graphs illustrate the Membership Inference Attack (MIA) performance at each epoch of the unlearning process. This specific set of results corresponds to the \textbf{third group} of experiments, utilizing the sftc unlearning algorithm.}
    \label{fig:sftc_3_group}
\end{figure*}

In the high-complexity CIFAR-10 scenario (Figure~\ref{fig:mia_accuracy_3a_sftc}), the SFTC algorithm demonstrates a remarkable ``divergence'' effect. The MIA Forget Accuracy begins at approximately 0.50 and decreases steadily to roughly 0.40 by epoch 30. Simultaneously, the MIA Retain Accuracy experiences a significant surge, climbing from 0.42 to nearly 0.90. This suggests that while the algorithm successfully erases the membership traces of forgotten samples, it aggressively consolidates the membership signal of the retained data, creating the largest privacy-utility gap observed in this group.

For the MUFAC dataset (Figure~\ref{fig:mia_accuracy_3b_sftc}), SFTC achieves an immediate and sharp reduction in the forget set's identifiability. The MIA Forget Accuracy drops precipitously from 0.70 to a baseline of 0.42 within the first five epochs and remains stable thereafter. Meanwhile, the MIA Retain Accuracy shows a steady improvement, rising from 0.87 to 0.94. This rapid stabilization indicates that the teacher-student dynamic is highly efficient at ``de-identifying'' the forget set in complex MUFAC models almost instantly.

The results for Purchase-100 (Figure~\ref{fig:mia_accuracy_3c_sftc}) mirror the divergent trend seen in CIFAR-10 but with a more gradual progression. The MIA Forget Accuracy starts at 0.51 and slowly declines to 0.475. Conversely, the MIA Retain Accuracy increases from 0.51 to over 0.65. This consistent widening of the gap throughout the 30 epochs demonstrates that SFTC maintains its precision even in high-dimensional datasets when paired with complex architectures.

Regarding the Texas-100 dataset (Figure~\ref{fig:mia_accuracy_3d_sftc}), SFTC exhibits a highly controlled unlearning trajectory. We observe a clear downward trend in the MIA Forget Accuracy, which moves from 0.67 to 0.55. During the same period, the MIA Retain Accuracy shows high resilience, initially dropping slightly before stabilizing around 0.63. This outcome is significant as it shows SFTC can successfully mitigate privacy risks (the 0.12 drop in forget accuracy) while maintaining a stable and higher membership signal for the retained data.

\subsection{Sensitivity Analysis} 

\begin{figure*}[t]
    \centering
    \begin{subfigure}[b]{0.32\textwidth}
        \centering
        \includegraphics[width=\textwidth]{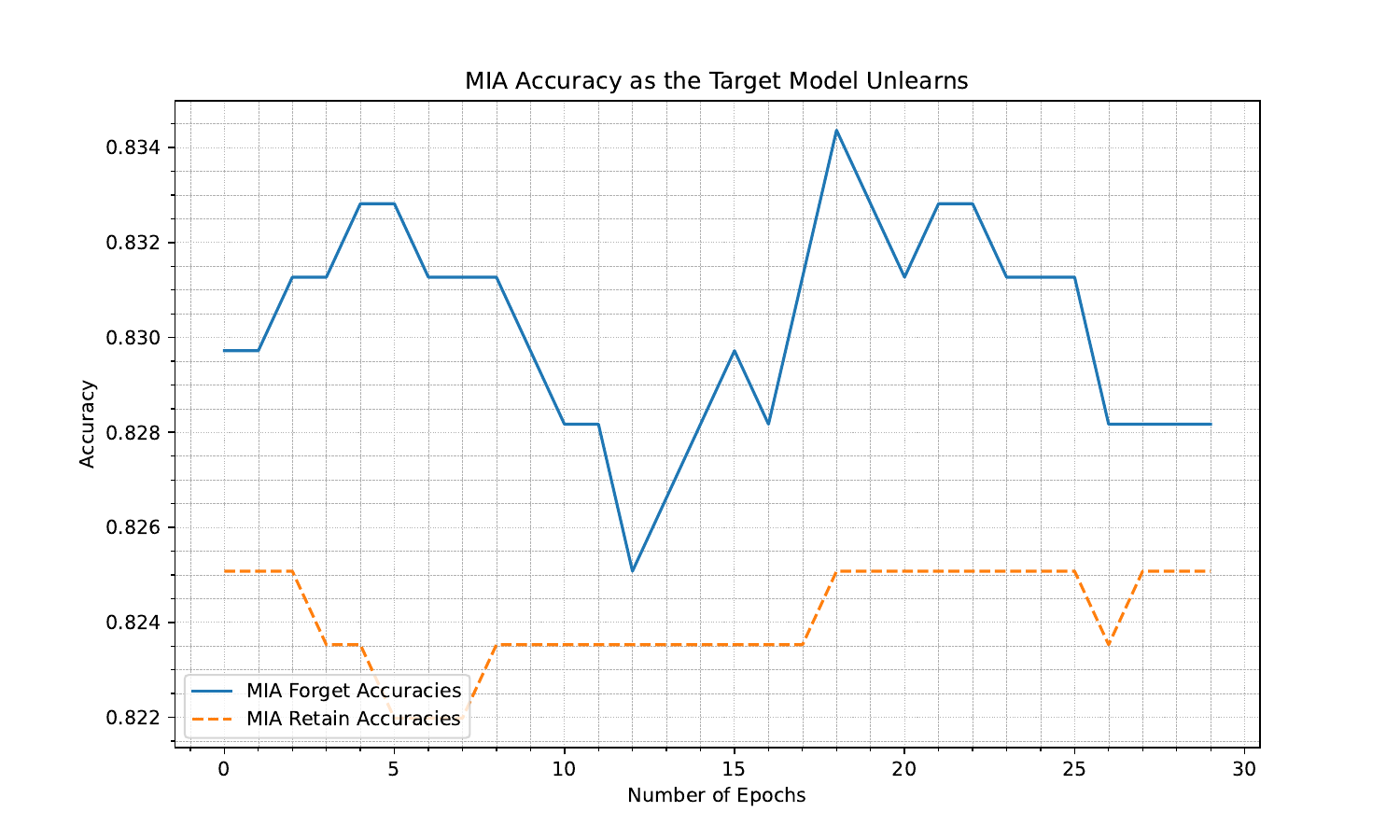}
        \caption{Neg Grad at cifar with lr 0.00001.}
        \label{fig:first}
    \end{subfigure}
    \hfill 
    \begin{subfigure}[b]{0.32\textwidth}
        \centering
        \includegraphics[width=\textwidth]{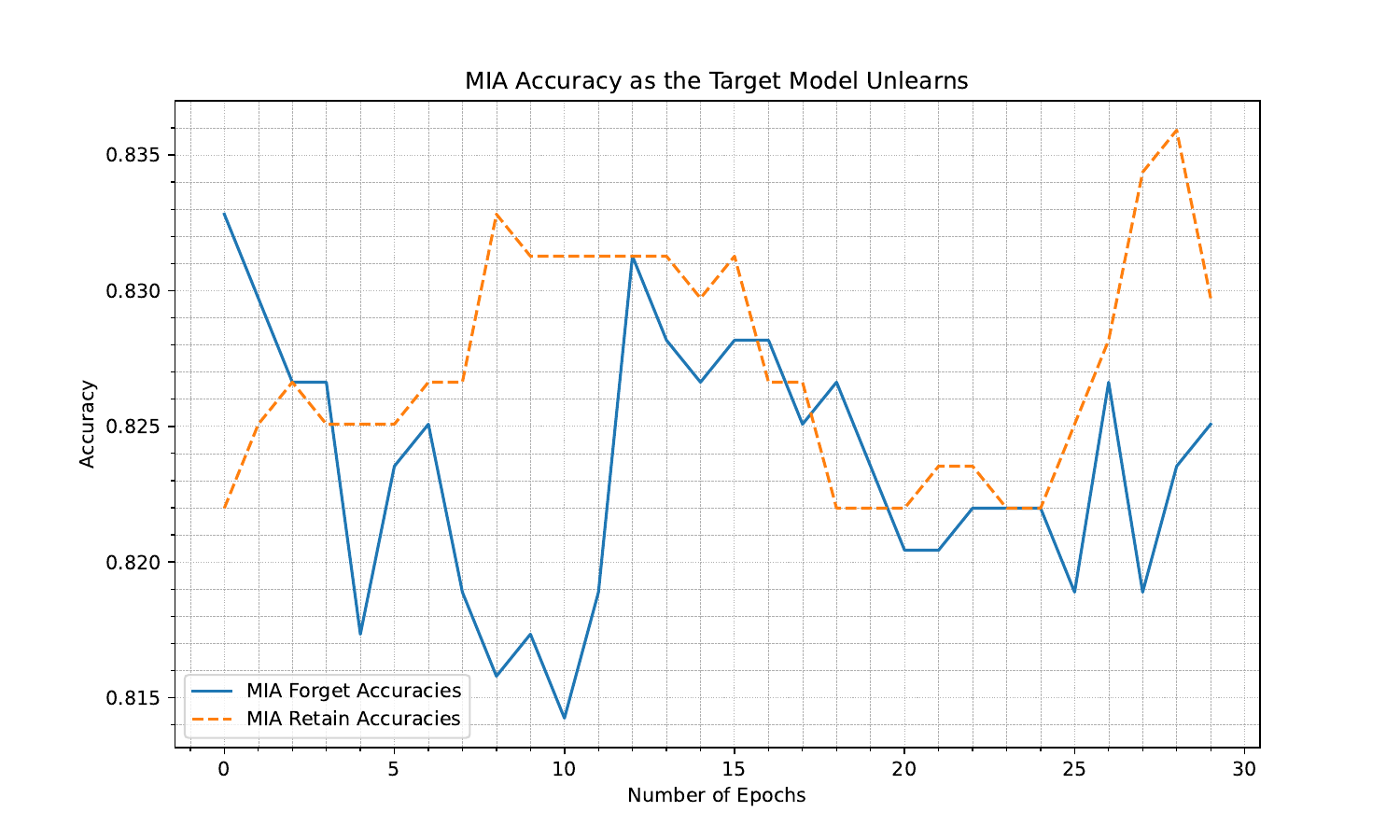}
        \caption{Neg Grad at cifar with lr 0.0001.}
        \label{fig:second}
    \end{subfigure}
    \hfill
    \begin{subfigure}[b]{0.32\textwidth}
        \centering
        \includegraphics[width=\textwidth]{fig/exp1a/1a_neg_grad_lr_0.0002.pdf}
        \caption{Neg Grad at cifar with lr 0.00001.}
        \label{fig:third}
    \end{subfigure}
    
    \caption{The unlearning curve through MIA for three different learning rates of Neg Grad for the cifar dataset and the architecture 1.}
    \label{fig:compare_lr}
\end{figure*}.

Throughout the implementation of our experiments, we observed a profound sensitivity regarding the learning rate of the unlearning algorithms. Minor adjustments to this hyperparameter resulted in substantial shifts in both the effectiveness of the MIA and the overall success of the unlearning process (see Figure~\ref{fig:compare_lr}). For instance, with a learning rate of 0.00001, the MIA Forget Accuracy reached 0.830, surpassing the Retain accuracy, which indicates a highly effective attack on the forget set. Consequently, we conducted extensive grid searches across various learning rates, ultimately selecting the values that provided the most robust defense against MIAs. This high sensitivity underscores a critical challenge in machine unlearning that warrants rigorous exploration in future research.

\section{Conclusion}
\label{sec:conclusions}

This paper investigates the efficacy of Machine Unlearning as a defense mechanism against Membership Inference Attacks (MIAs) across a diverse range of datasets and model architectures. Our experimental results demonstrate that Machine Unlearning can effectively mitigate a model's vulnerability to MIAs for the forget set, but this process requires a very careful consideration of the parameters. However, the success of this defense is critically contingent upon several factors: the selection of the unlearning algorithm, hyperparameter tuning (specifically the learning rate), the initial effectiveness of the MIA, the degree of model overfitting, and the number of unlearning epochs utilized.

In real-world scenarios, the application of 30 unlearning epochs typically results in the predictive collapse of the model. In our experiments, we deliberately extended the unlearning process to this duration to examine the unlearning dynamics through the perspective of MIA susceptibility.

The experimental results across the Negative Gradient, Scrub, and SFTC algorithms reveal that the effectiveness of machine unlearning as a defense against Membership Inference Attacks (MIA) is deeply tied to both model complexity and the specific unlearning mechanism utilized. While the Negative Gradient method provides a baseline level of protection, it often results in generalized degradation where the membership signals of both forget and retain sets decline simultaneously, particularly in datasets like Texas-100 and Purchase-100. The Scrub algorithm offers a more aggressive approach but is prone to significant instability in simple architectures and can paradoxically increase model transparency in certain scenarios, creating a ``bell-curve'' risk profile where data becomes more identifiable before being masked.

The SFTC algorithm is characterized by a distinct ``divergence'' effect, where the MIA Forget Accuracy decreases while the MIA Retain Accuracy is consolidated or even strengthened. This phenomenon is particularly evident in high-capacity models. The internal mechanics of SFTC involve intensive fine-tuning on the retain data alongside the selective ``confusion'' of the forget data. Because the student model is trained rigorously to mirror the teacher's high confidence on retain samples, it inadvertently reinforces their membership signals, making them more identifiable to an adversary. And that is obvious in our experiments where the MIA Retain accuracy is increasing.

In contrast, the SCRUB algorithm also employs a dual-phase approach: a maximize phase to induce forgetting and a minimize phase for fine-tuning on the retain data. However, our experiments show that the resulting increase in MIA Retain Accuracy is generally negligible compared to SFTC. Instead, SCRUB focuses on maintaining a significant drop in MIA Forget Accuracy. While SCRUB uses fine-tuning to recover model utility, it typically aims for stabilization rather than the aggressive reinforcement of the retain set's signal seen in SFTC.

Despite the promise of Machine Unlearning, it is clear that this technique is not a universal safeguard against MIAs. Residual vulnerabilities may still exist in models post-unlearning, as attackers could exploit subtle traces of the forget data.

While this study provides valuable insights into the defensive potential of machine unlearning against MIAs, there remain several areas for future exploration to further enhance its effectiveness. This includes investigating adaptive unlearning strategies, which dynamically adjust based on the model's vulnerability to MIAs, and evaluating the impact of unlearning on other types of adversarial attacks. Additionally, integrating Machine Unlearning with other privacy-preserving mechanisms, such as differential privacy, could enhance the overall defense against MIAs.

It will also be important to explore the long-term effects of repeated unlearning requests and their impact on model stability, as well as to develop standardized benchmarks for evaluating the robustness of unlearning algorithms. As MIAs evolve, unlearning methods must continue to adapt, ensuring comprehensive protection for user data in an increasingly adversarial landscape.

\begin{acknowledgements}
We thank the anonymous reviewers for their insightful and constructive comments, which have significantly improved the quality of this work.
\end{acknowledgements}

\section*{Declarations}

The authors declare that no funds, grants, or other support were received during the preparation of this manuscript. The authors have no competing interests to declare.

\bibliographystyle{spmpsci}
\bibliography{main.bib}   

@misc{murakonda2020mlprivacymeteraiding,
      title={ML Privacy Meter: Aiding Regulatory Compliance by Quantifying the Privacy Risks of Machine Learning}, 
      author={Sasi Kumar Murakonda and Reza Shokri},
      year={2020},
      eprint={2007.09339},
      archivePrefix={arXiv},
      primaryClass={cs.CR},
      url={https://arxiv.org/abs/2007.09339}, 
}

@misc{spartalis2025,
      title={LoTUS: Large-Scale Machine Unlearning with a Taste of Uncertainty}, 
      author={Christoforos N. Spartalis and Theodoros Semertzidis and Efstratios Gavves and Petros Daras},
      year={2025},
      eprint={2503.18314},
      archivePrefix={arXiv},
      primaryClass={cs.LG},
      url={https://arxiv.org/abs/2503.18314}, 
}

@misc{nasr2018machinelearningmembershipprivacy,
      title={Machine Learning with Membership Privacy using Adversarial Regularization}, 
      author={Milad Nasr and Reza Shokri and Amir Houmansadr},
      year={2018},
      eprint={1807.05852},
      archivePrefix={arXiv},
      primaryClass={stat.ML},
      url={https://arxiv.org/abs/1807.05852}, 
}

@misc{tu2025,
      title={A Reliable Cryptographic Framework for Empirical Machine Unlearning Evaluation}, 
      author={Yiwen Tu and Pingbang Hu and Jiaqi Ma},
      year={2025},
      eprint={2404.11577},
      archivePrefix={arXiv},
      primaryClass={cs.LG},
      url={https://arxiv.org/abs/2404.11577}, 
}

@misc{zou2020privacyanalysisdeeplearning,
      title={Privacy Analysis of Deep Learning in the Wild: Membership Inference Attacks against Transfer Learning}, 
      author={Yang Zou and Zhikun Zhang and Michael Backes and Yang Zhang},
      year={2020},
      eprint={2009.04872},
      archivePrefix={arXiv},
      primaryClass={cs.CR},
      url={https://arxiv.org/abs/2009.04872}, 
}

@INPROCEEDINGS{He2016resnet,
  author={He, Kaiming and Zhang, Xiangyu and Ren, Shaoqing and Sun, Jian},
  booktitle={2016 IEEE Conference on Computer Vision and Pattern Recognition (CVPR)}, 
  title={Deep Residual Learning for Image Recognition}, 
  year={2016},
  volume={},
  number={},
  pages={770-778},
  doi={10.1109/CVPR.2016.90}}

@inproceedings{akiba2019optuna,
author = {Akiba, Takuya and Sano, Shotaro and Yanase, Toshihiko and Ohta, Takeru and Koyama, Masanori},
title = {Optuna: A Next-generation Hyperparameter Optimization Framework},
year = {2019},
isbn = {9781450362016},
publisher = {Association for Computing Machinery},
address = {New York, NY, USA},
url = {https://doi.org/10.1145/3292500.3330701},
doi = {10.1145/3292500.3330701},
booktitle = {Proceedings of the 25th ACM SIGKDD International Conference on Knowledge Discovery \& Data Mining},
pages = {2623–2631},
numpages = {9},
location = {Anchorage, AK, USA},
series = {KDD '19}
}

@article{kurmanji2023towards,
  title={Towards unbounded machine unlearning},
  author={Kurmanji, Meghdad and Triantafillou, Peter and Hayes, Jamie and Triantafillou, Eleni},
  journal={Advances in neural information processing systems},
  volume={36},
  pages={1957--1987},
  year={2023}
}

@inproceedings{golatkar2020eternal,
  title={Eternal sunshine of the spotless net: Selective forgetting in deep networks},
  author={Golatkar, Aditya and Achille, Alessandro and Soatto, Stefano},
  booktitle={Proceedings of the IEEE/CVF conference on computer vision and pattern recognition},
  pages={9304--9312},
  year={2020}
}

@inproceedings{shokri2017membership,
  title={Membership inference attacks against machine learning models},
  author={Shokri, Reza and Stronati, Marco and Song, Congzheng and Shmatikov, Vitaly},
  booktitle={2017 IEEE Symposium on Security and Privacy (SP)},
  pages={3--18},
  year={2017},
  organization={IEEE}
}

@inproceedings{purchase100,
  title={The Tradeoff Between Privacy and Accuracy in Anonymized Big Data},
  author={V. Shokri and R. Póczos and A. Kosut and A. D. Sarwate},
  booktitle={Proceedings of the 18th ACM Conference on Computer and Communications Security (CCS)},
  year={2011}
}

@TECHREPORT{cifar10,
    author = {Alex Krizhevsky and Geoffrey Hinton},
    title = {Learning Multiple Layers of Features from Tiny Images},
    institution = {University of Toronto},
    year = {2009}
}

@article{Xu,
author = {Xu, Heng and Zhu, Tianqing and Zhang, Lefeng and Zhou, Wanlei and Yu, Philip},
year = {2023},
month = {06},
pages = {},
title = {Machine Unlearning: A Survey},
volume = {56},
journal = {ACM Computing Surveys},
doi = {10.1145/3603620}
}

@article{Baumhauer,
author = {Baumhauer, Thomas and Schöttle, Pascal and Zeppelzauer, Matthias},
year = {2022},
month = {07},
pages = {},
title = {Machine unlearning: linear filtration for logit-based classifiers},
volume = {111},
journal = {Machine Learning},
doi = {10.1007/s10994-022-06178-9}
}

@article{NIU2024404,
title = {A survey on membership inference attacks and defenses in machine learning},
journal = {Journal of Information and Intelligence},
volume = {2},
number = {5},
pages = {404-454},
year = {2024},
issn = {2949-7159},
doi = {https://doi.org/10.1016/j.jiixd.2024.02.001},
author = {Jun Niu and Peng Liu and Xiaoyan Zhu and Kuo Shen and Yuecong Wang and Haotian Chi and Yulong Shen and Xiaohong Jiang and Jianfeng Ma and Yuqing Zhang}
}

@article{Lu,
  title={Label-only membership inference attacks on machine unlearning without dependence of posteriors},
  author={Lu, Zhaobo and Liang, Hai and Zhao, Minghao and Lv, Qingzhe and Liang, Tiancai and Wang, Yilei},
  journal={International Journal of Intelligent Systems},
  volume={37},
  number={11},
  pages={9424--9441},
  year={2022},
  publisher={Wiley Online Library}
}

@misc{DoughanItani,
author="Doughan, Ziad
and Itani, Sari",
editor="Iliadis, Lazaros
and Maglogiannis, Ilias
and Papaleonidas, Antonios
and Pimenidis, Elias
and Jayne, Chrisina",
title="Machine Unlearning, A Comparative Analysis",
booktitle="Engineering Applications of Neural Networks",
year="2024",
publisher="Springer Nature Switzerland",
address="Cham",
pages="558--568",
isbn="978-3-031-62495-7"
}

@misc{Chen,
author = {Chen, Min and Zhang, Zhikun and Wang, Tianhao and Backes, Michael and Humbert, Mathias and Zhang, Yang},
title = {When Machine Unlearning Jeopardizes Privacy},
year = {2021},
isbn = {9781450384544},
publisher = {Association for Computing Machinery},
address = {New York, NY, USA},
doi = {10.1145/3460120.3484756},
booktitle = {Proceedings of the 2021 ACM SIGSAC Conference on Computer and Communications Security},
pages = {896–911},
numpages = {16},
location = {Virtual Event, Republic of Korea},
series = {CCS '21}
}

@article{Sommer,
author = {Sommer, David and Song, Liwei and Wagh, Sameer and Mittal, Prateek},
year = {2022},
month = {07},
pages = {268-290},
title = {Athena: Probabilistic Verification of Machine Unlearning},
volume = {2022},
journal = {Proceedings on Privacy Enhancing Technologies},
doi = {10.56553/popets-2022-0072}
}

@inproceedings{Bourtoule,
  author={Bourtoule, Lucas and Chandrasekaran, Varun and Choquette-Choo, Christopher A. and Jia, Hengrui and Travers, Adelin and Zhang, Baiwu and Lie, David and Papernot, Nicolas},
  booktitle={2021 IEEE Symposium on Security and Privacy (SP)}, 
  title={Machine Unlearning}, 
  year={2021},
  volume={},
  number={},
  pages={141-159},
  doi={10.1109/SP40001.2021.00019}}

@inbook{Ginart,
author = {Ginart, Antonio A. and Guan, Melody Y. and Valiant, Gregory and Zou, James},
title = {Making AI forget you: data deletion in machine learning},
year = {2019},
publisher = {Curran Associates Inc.},
address = {Red Hook, NY, USA},
booktitle = {Proceedings of the 33rd International Conference on Neural Information Processing Systems},
articleno = {316},
numpages = {14}
}

@inproceedings{Cao,
  author={Cao, Yinzhi and Yang, Junfeng},
  booktitle={2015 IEEE Symposium on Security and Privacy}, 
  title={Towards Making Systems Forget with Machine Unlearning}, 
  year={2015},
  volume={},
  number={},
  pages={463-480},
  doi={10.1109/SP.2015.35}}

@inproceedings{perifanis2024sftc,
  title={SFTC: Machine Unlearning via Selective Fine-tuning and Targeted Confusion},
  author={Perifanis, Vasileios and Karypidis, Efstathios and Komodakis, Nikos and Efraimidis, Pavlos},
  booktitle={European Interdisciplinary Cybersecurity Conference},
  pages={29--36},
  year={2024}
}

@article{hu2021membership,
  title={Membership Inference Attacks on Machine Learning: A Survey},
  author={Hu, Hongsheng and Salcic, Zoran and Sun, Lichao and Dobbie, Gill and Yu, Philip and Zhang, Xuyun},
  journal={ACM Computing Surveys (CSUR)},
  volume={54},
  number={37},
  pages={1--37},
  year={2021},
  doi={10.1145/3523273},
}

@article{banerjee2023miabad,
  title={MIA-BAD: An Approach for Enhancing Membership Inference Attack and its Mitigation with Federated Learning},
  author={Banerjee, Soumya and Roy, Sandip and Ahamed, Sayyed Farid and others},
  journal={ArXiv},
  volume={abs/2312.00051},
  year={2023},
  doi={10.48550/arXiv.2312.00051},
}

@article{shi2023scalemia,
  title={Scale-MIA: A Scalable Model Inversion Attack against Secure Federated Learning via Latent Space Reconstruction},
  author={Shi, Shanghao and Wang, Ning and Xiao, Yang and others},
  journal={ArXiv},
  volume={abs/2311.05808},
  year={2023},
  doi={10.48550/arXiv.2311.05808},
}

@article{Zhong2022Understanding,
title={Understanding Disparate Effects of Membership Inference Attacks and their Countermeasures},
author={Da Zhong and Haipei Sun and Jun Xu and N. Gong and Wendy Hui Wang},
journal={Proceedings of the 2022 ACM on Asia Conference on Computer and Communications Security},
year={2022},
doi={10.1145/3488932.3501279}
}

@misc{salem2018ml,
  title={ML-Leaks: Model and data independent membership inference attacks and defenses on machine learning models},
  author={Salem, Ahmed and Zhang, Yi and Humbert, Mathias and Backes, Michael and Zhang, Yang},
  booktitle={Proceedings of the 2018 Network and Distributed System Security Symposium (NDSS)},
  year={2018},
  organization={Internet Society}
}

@inproceedings{pavlidis2023extensive,
  title={An Extensive Overview of Feature Representation Techniques for Molecule Classification},
  author={Pavlidis, Nikolaos and Nikolaidis, Christos Chrysanthos and Perifanis, Vasileios and Papadopoulou, Anastasia and Efraimidis, Pavlos and Arampatzis, Avi},
  booktitle={Proceedings of the 27th Pan-Hellenic Conference on Progress in Computing and Informatics},
  pages={156--162},
  year={2023}
}

@inproceedings{tsiolakis2024carbon,
  title={Carbon-Aware Machine Learning: A Case Study on Cellular Traffic Forecasting with Spiking Neural Networks},
  author={Tsiolakis, Theodoros and Pavlidis, Nikolaos and Perifanis, Vasileios and Efraimidis, Pavlos S},
  booktitle={IFIP International Conference on Artificial Intelligence Applications and Innovations},
  pages={178--191},
  year={2024},
  organization={Springer}
}

@article{zhang2023review,
  title={A review on machine unlearning},
  author={Zhang, Haibo and Nakamura, Toru and Isohara, Takamasa and Sakurai, Kouichi},
  journal={SN Computer Science},
  volume={4},
  number={4},
  pages={337},
  year={2023},
  publisher={Springer}
}

@article{liu2024threats,
  title={Threats, attacks, and defenses in machine unlearning: A survey},
  author={Liu, Ziyao and Ye, Huanyi and Chen, Chen and Lam, Kwok-Yan},
  journal={arXiv preprint arXiv:2403.13682},
  year={2024}
}

@article{niu2024survey,
  title={A survey on membership inference attacks and defenses in Machine Learning},
  author={Niu, Jun and Liu, Peng and Zhu, Xiaoyan and Shen, Kuo and Wang, Yuecong and Chi, Haotian and Shen, Yulong and Jiang, Xiaohong and Ma, Jianfeng and Zhang, Yuqing},
  journal={Journal of Information and Intelligence},
  year={2024},
  publisher={Elsevier}
}

@article{choi2023towards,
  title={Towards machine unlearning benchmarks: Forgetting the personal identities in facial recognition systems},
  author={Choi, Dasol and Na, Dongbin},
  journal={arXiv preprint arXiv:2311.02240},
  year={2023}
}

@article{Nikolaidis2024,
author={Nikolaidis, Christos Chrysanthos
and Efraimidis, Pavlos S.},
title={Advancing elderly social care dropout prediction with federated learning: client selection and imbalanced data management},
journal={Cluster Computing},
year={2024},
month={Nov},
day={26},
volume={28},
number={2},
pages={114},
issn={1573-7543},
doi={10.1007/s10586-024-04850-4},
url={https://doi.org/10.1007/s10586-024-04850-4}
}

@Article{Nikolaidis2025,
author={Nikolaidis, Christos Chrysanthos
and Efraimidis, Pavlos S.},
title={A study of membership inference attacks on a federated health care application},
journal={Computing},
year={2025},
month={Jun},
day={16},
volume={107},
number={7},
pages={149},
issn={1436-5057},
doi={10.1007/s00607-025-01507-x},
url={https://doi.org/10.1007/s00607-025-01507-x}
}


\end{document}